\documentclass[12pt,a4]{article} 

\usepackage{amsfonts}
\usepackage{amstext}
\usepackage{amsmath}
\usepackage{amssymb}

\usepackage{graphicx}
\usepackage{color}
\usepackage{here}

\newcommand{\rf}[1]{(\ref{#1})}
\newcommand{\beq}{\begin{equation}}
\newcommand{\eeq}{\end{equation}}
\newcommand{\bea}{\begin{eqnarray}}
\newcommand{\eea}{\end{eqnarray}}
\newcommand{\e}{{\rm e}}
\newcommand{\g}{\gamma}
\renewcommand{\l}{\lambda}

\renewcommand{\a}{\alpha}
\newcommand{\n}{\nu}
\newcommand{\m}{\mu}
\newcommand{\del}{\delta}
\newcommand{\ep}{\varepsilon}
\newcommand{\om}{\omega}

\newcommand{\oh}{\frac{1}{2}}

\newcommand{\dg}{\dagger}
\newcommand{\tr}{{\rm tr}{\dbltinyspace}}
\newcommand{\ra}{\rangle}
\newcommand{\la}{\langle}
\newcommand{\rara}{\ra\!\ra}

\newcommand{\prt}{\partial}

\newcommand{\cF}{{\cal F}}
\newcommand{\cH}{{\cal H}}

\newcommand{\cJ}{{\cal J}}

\newcommand{\tH}{{\tilde{H}}}

\newcommand{\nn}{\nonumber \\}

\newcommand{\bra}[1]{ \langle #1 | }
\newcommand{\ket}[1]{ | #1 \rangle }
\newcommand{\vac}{ \bra{{\rm vac}} }
\newcommand{\cuum}{ \ket{{\rm vac}} }

\newcommand{\Hj}{{\cH^\star}}
\newcommand{\tHj}{\tilde{\cH}^{\star}}
\newcommand{\tcH}{\tilde{\cH}}
\newcommand{\Hdisk}{\cH_{\rm disk}}
\newcommand{\tHdisk}{\tilde{\cH}_{\rm disk}}

\newcommand{\cc}{\mu} 
\newcommand{\T}{T} 
\newcommand{\define}{ \stackrel{\rm def}{\equiv} }

\newcommand{\intdzeta}{ { \int\limits_{- i \infty}^{+ i \infty}
                        \frac{d \zeta}{2 \pi i} } }

\newcommand{\bz}{{\bar{z}}}

\font\twelvemsbm = msbm10 scaled\magstep1
\font\tenmsbm = msbm10
\font\eightmsbm = msbm8
\font\sixmsbm = msbm6
\newcommand{\Dbl}[1]{\leavevmode\raise-.10ex\hbox{\twelvemsbm #1}}
\newcommand{\dbl}[1]{\leavevmode\raise-.00ex\hbox{\tenmsbm #1}}
\newcommand{\dblsmall}[1]{\leavevmode\raise-.05ex\hbox{\eightmsbm #1}}
\newcommand{\dbltiny}[1]{\leavevmode\raise-.05ex\hbox{\sixmsbm #1}}

\newcommand{\dd}{{\rm d}}

\newcommand{\pder}[1]{\frac{\partial}{\partial #1}}
\newcommand{\pdder}[1]{\frac{\partial^2}{\partial #1^2}}

\newcommand{\halftinyspace}{\hspace{0.0278em}} 
\newcommand{\tinyspace}{\hspace{0.0556em}}
\newcommand{\trehalftinyspace}{\hspace{0.0834em}}
\newcommand{\dbltinyspace}{\hspace{0.1112em}}

\newcommand{\negtinyspace}{\hspace{-0.0556em}}
\newcommand{\negdbltinyspace}{\hspace{-0.1112em}}

\newcommand{\pictureX}{100}
\newcommand{\pictureY}{141.4}

\newcommand{\GG}{G}
\newcommand{\pp}{p}
\newcommand{\qq}{q}

\renewcommand{\dim}{K}

\def\theequation{\arabic{section}.\arabic{equation}}
\def\thefigure{\arabic{section}.\arabic{figure}}

\begin{document}
\topmargin 0pt
\oddsidemargin 5mm                              
\headheight 0pt
\headsep 0pt
\topskip 9mm


\hfill \today 

\begin{center}
  \vspace{24pt}
  {\large \bf
     Models of the Universe based on  Jordan algebras
   \\
     
   }

  \vspace{24pt}

  {\sl J. Ambj\o rn }

  \vspace{6pt}

{\small
  The Niels Bohr Institute, University of Copenhagen\\
  Blegdamsvej 17, DK-2100 Copenhagen \O , Denmark\\

\vspace{6pt}

and 

\vspace{6pt}

Institute for Mathematics, Astrophysics and Particle Physics (IMAPP)\\ 
Radbaud University Nijmegen, Heyendaalseweg 135,
6525 AJ, \\
Nijmegen, The Netherlands}

  \vspace{12pt}

  and

  \vspace{12pt}

  {\sl Y. Watabiki}

  \vspace{6pt}

{\small
  Department of Physics\\
  Tokyo Institute of Technology\\
  Oh-okayama, Meguro, Tokyo 152, Japan\\
}

\end{center}
\vspace{24pt}

\vfill

\begin{center}
  {\bf Abstract}
\end{center}

\vspace{12pt}

\noindent
We propose a model for the universe based on Jordan algebras. The action consists  of cubic terms with 
coefficients being the structure constants of a Jordan algebra. Coupling constants only enter the theory via  
symmetry breaking which also selects a physical vacuum. ``Before'' the symmetry breaking
the universe is in a pre-geometric state where it makes no sense to talk about space or time, but time
comes into existence with the symmetry breaking together with a Hamiltonian which can create space 
from ``nothing'' and in some cases can propagate the space to macroscopic size in a finite time. There 
exists symmetry breaking which results in macroscopic spacetime dimensions 3, 4, 6 and 10, based 
on the Jordan algebras of Hermitian 3x3 matrices with real, complex, quarternion and octonion entries,
respectively.

\vfill

\vspace{36pt}



\newpage

\section{Introduction}

Observations indicate that our present universe is expanding and that it at an earlier stage 
has been much smaller and hotter than today. A theory of our universe should preferable explain 
what happened before this early and hot time. One scenario which brings us further back in time 
is the concept of inflation. This scenario was invented to solve a number of issues related 
to a Big Bang theory which naively extrapolated the hot and small universe which once existed 
further backwards in time. However, the concept of inflation still leaves unanswered what happened before 
 inflation: can one at all speak of space and time when the universe is of ``Planck scale'', and if not, what kind of theory can one image 
leads to our present universe. The purpose of this article is to present a model with allows us to address 
such questions. Needless to say it will not answer the questions in a completely satisfactory way, but we nevertheless 
feel that it is important to address such questions if we believe that there is an underlying theory from which 
the behaviour of our universe can be derived in some way. One interesting aspect of the model is that 
it links properties of the universe at the shortest scales, namely the possibility of the universe to change 
topology by quantum fluctuations, to the acceleration of the universe observed today. Another interesting 
aspect of the model is that it presents a scenario where the ``universe'' is naturally in a pre-geometric, 
very symmetric state where it makes no sense to talk about time and spatial geometry. After  
symmetry breaking  there is then the possibility of  emergence of time and an associated Hamiltonian  
and after that of spatial geometry and thus the emergence of a ``universe'' which may resemble our universe. 
Different kinds of ``universes'' might emerge from different patterns  of symmetry breaking, and one intriguing possibility is 
a universe with 9 extended and 16 small, compact  spatial dimensions, a situation similar to what occurs for the 
heterotic string. It is also possible to obtain a universe with three extended spatial dimensions, while the rest are 
small and compact.

In the pre-geometric phase there is no concept of time and thus no concept of unitary evolution. This has the 
consequence that the symmetry breaking results in the emergence of time and an associated  Hamiltonian which 
is not entirely Hermitian: it is possible to create spatial universes of infinitesimal extension. Such universes can 
grow with time and they can merge and split and eventually evolve to something resembling our universe. Under 
suitable assumptions one can then derive a modified Friedmann equation for such a universe, which, as already 
mentioned, links the future acceleration  to the coupling constant responsible for splitting and merging of universes.                                                                                                                                                                                                                                                                                                                                                                                                                                                                                                                                                                             

We were led to the present model by the study of causal dynamical triangulations (CDT). It is a non-perturbative
lattice model of quantum gravity. In four dimensions the model can be studied by Monte Carlo simulations
(see \cite{physrev,loll} for  reviews, \cite{original} for orginal articles  and \cite{recent} for recent results). 
However, in two dimensions (one space and one 
time dimension) the lattice model can be solved analytically and a continuum limit can be taken \cite{al}.
The continuum limit turns out to be a simple version \cite{ag} of Horava-Lifshitz gravity \cite{horava}.
However, there is an generalisation of two-dimensional CDT which allows for topology change of space-time
\cite{gcdt}. This generalisation can be formulated as a string field theory \cite{cdtsft} which is quite similar to
the standard  non-critical string field theory \cite{ncsft,aw}, just simpler since the kinetic term for string propagation is not 
singular in CDT string field theory, contrary to the situation for standard non-critical string field theory. Both
CDT string field theory and non-critical string field theory describes how the string can split and joining, or in the
context of two-dimensional quantum gravity how a  spatial (one-dimensional) universe can split and join.
However, these theories tell us surprisingly little about the creation of a universe from ``nothing'', the confusing
question we are confronted with in cosmology when we extrapolate backwards in time. It is well known that there 
is a $W_3$ symmetry underlying the standard  non-critical string theory \cite{w3,aw}. 
In \cite{aw1} we showed that one can also 
associate a $W_3$ symmetry with CDT string field theory, but the difference to non-critical string field theory 
was that insisting on a $W_3$ symmetric action from the beginning leads to a CDT string field theory which 
allows for the creation of spatial universes of infinitesimal size from the vacuum. In \cite{aw2} this approach 
was generalized to include models with exended $W_3$ algebra symmetry, various space dimensions being 
assigned different ``flavours''. In this way one can create higher dimensional space from ``nothing'', and 
potentially created space-times which resemble our present universe. In our model the concept of inflation 
is somewhat different from the usual concept.
Our model does not need an inflaton field, but is nevertheless able to solve the horizon problem, which is the 
most serious problem that the inflation models solve. It is also solves the flatness problem. It has nothing to 
say about the monopole problem, but there is of course only a monopole problem if there exists a suitable 
grand unified quantum field theory, which is maybe not so obvious any longer.

The purpose of this article 
is to provide some of the details of this model. Part of the present work has been published in short articles 
earlier \cite{aw1,aw2}.
The rest of the article is organized as follows:  In Sec.\ \ref{sec2} we describe CDT string field theory, its relation to 
the $W^{(3)}$ algebra and how this gives rise to the emergence of time and space. In Sec.\ \ref{sec3} we 
we describe Jordan algebras and their relation to extended $W^{(3)}$ algebras. In Sec.\ \ref{sec4}  we discuss 
some concrete models. Sec.\ \ref{sec5} deals with the dynamics which can create higher dimensional universes  from 
two-dimensional CDT universes with different ``flavours''.
Finally Sec.\ \ref{sec6} contains a discussion of the results obtained so far.

\section{Two-dimensional  CDT  gravity}\label{sec2}
\setcounter{equation}{0}

\subsection{Non-interacting CDT}

\begin{figure}[h]
\vspace{10pt}
 \begin{center}
  \begin{picture}(\pictureX,\pictureY)
    \put(10,15){\includegraphics[width=100pt]{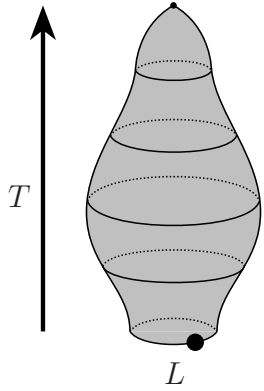}}
    \put(60,  3){$L$}
    \put( 1, 70){$T$}
  \end{picture}
 \end{center}
\vspace*{-20pt}
\caption[figDiskAmp]{{\small
A typical configuration of disk amplitude. 
The arrow describes the direction of time $T$.
The entrance loop has length $L$ with a marked point.
}}
\label{figDiskAmp}
\end{figure}

The theory of causal dynamical triangulations in two dimensions was first formulated in \cite{al} where the lattice 
theory was solved explicitly. The continuum limit of the lattice theory was found. It describes (in continuum notation)
the amplitude for   a spatial universe of size $L$ to propagate to a spatial universe of size $L'$, 
the propagation taking a proper  time $T$. This amplitude, or propagator, $G(L,L',T)$ satisfies the follow  
differential equation
\beq\label{GreenDiffEq}
\pder{\T} G(L,L';\T)=-H_0 G(L,L';\T), \qquad H_0 = 
L {\tinyspace}\bigg(\!- \pdder{L} + \cc \bigg).
\eeq
%
%
\begin{figure}[h]
\vspace{10pt}
 \begin{center}
  \begin{picture}(\pictureX,\pictureY)
    \put(10,15){\includegraphics[width=100pt]{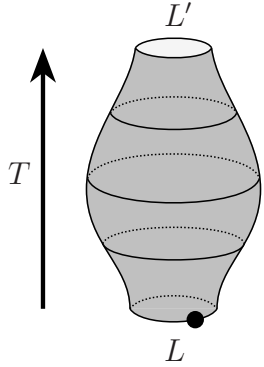}}
    \put(60,130){$L'$}
    \put(60,  3){$L$}
    \put( 1, 70){$T$}
  \end{picture}
 \end{center}
\vspace*{-20pt}
\caption[figGreenFun]{{\footnotesize
A typical configuration appearing in the path integral of the Green function. 
The arrow describes the direction of time $T$.
The entrance loop has length $L$ and a marked point 
and 
the exit loop has length $L'$ and no marked points.
}}
\label{figGreenFun}
\end{figure}%
In the path integral used for calculating the propagator $G(L,L',T)$ one integrates over all 
two-dimensional geometries where the two space-like boundaries of lengths $L$ and $L'$ are 
separated by proper time $T$ and where the topology of spacetime is that of a cylinder. Finally 
one assumes that the entrance loop with length $L$ has a marked point and the the initial condition
when $T \to 0$ is  
\beq\label{GreenInitialCond}
G(L,L';0)
\,=\,
\delta(L - L').
\eeq
A typical geometry associated with the propagator or Green function is illustrated in
Fig.\ \ref{figGreenFun}.
 The solution to this differential equation is\footnote{%
Details are provided in Appendix \ref{app:GreenFun}. 
}
\beq\label{GreenFun}
G(L,L';\T)
\,=\,
\frac{e^{- \sqrt{\cc}\,(L+L') \coth(\sqrt{\cc}\, T)}}
     {\sinh(\sqrt{\cc}\, T)}
\sqrt{\frac{\cc L}{L'}}
I_1\!\!\;\bigg(\frac{2\sqrt{\cc L L'}}{\sinh(\sqrt{\cc}\, T)}\bigg).
\eeq
Using the Green function we can obtain various amplitudes of interest, which we will 
now discuss. The first is the so-called   disk amplitude $F_1^{(0)}(L)$ obtained 
by contracting the exit loop to a point, i.e.\ taking $L' \to 0$, and integrating over $T$ 
\beq\label{DiskAmpDef}
F_1^{(0)}(L) \,=\,
\int\limits_0^\infty \dd \T \, G(L,0{\tinyspace};\T)
\,=\,
e^{- \sqrt{\cc}{\tinyspace} L}
\, .
\eeq
(See Figure \ref{figDiskAmp}.) 

Let us denote the size (the length) of the entrance loop $L_0$ and the size of the exit loop $L$. 
For $L_0 \to 0$ one obtains the amplitude of the universe after 
what we denote ``Big Bang'', i.e.\ the amplitude as a function of $T$ after the 
universe starts out with zero extension at $T=0$, 
\begin{eqnarray}\label{BigBangAmplitude1}
G_{\rm BB}(L;T) \!&:=&\!
\lim_{L_0 \to 0} \frac{1}{L_0}\!\; G(L_0,L;T)
\,=\,
\frac{\cc {\dbltinyspace} e^{- \sqrt{\cc}\,L \coth(\sqrt{\cc}\, T)}}
     {\sinh^2(\sqrt{\cc}\, T)}.
\end{eqnarray}
The reason we divide by $L_0$ in the definition \rf{BigBangAmplitude1} is that
a mark on the entrance boundary of length $L_0$ as shown in Fig.\ \ref{figGreenFun}
has introduced such a factor in the first place and we have done that to be in accordance 
with the standard definition of the disk amplitude.

If we have a universe which at time $T=0$ has length $L_0$, we can calculate the 
expectation value of the length (i.e.\ the expectation value of the size of the universe) at $T$:
\begin{equation}\label{AverageLengthUnderExpansion1}
\langle L \rangle_{(L_0,T)}
\,:=\,
\frac{\int_0^\infty d L {\dbltinyspace} L {\tinyspace} G(L_0,L;T)}
     {\int_0^\infty d L {\dbltinyspace} G(L_0,L;T)}
\,,
\end{equation}
As shown in Appendix A we obtain
\begin{eqnarray}\label{AverageLengthOfSpace1}
\langle L \rangle_{(L_0,T)}
\!&=&\!
\frac{L_0}
     {\cosh^2(\sqrt{\cc}\, T) \big(
        1
        -
        e^{- 2 \sqrt{\cc}\,L_0 / \sinh(2 \sqrt{\cc}\, T)}
      \big)}
\,,
\end{eqnarray}
and especially, for $L_0 \!=\! 0$, we have
\begin{eqnarray}\label{AverageLengthOfSpaceZero1}
\langle L \rangle_{(0,T)}
\!&=&\!
\frac{\tanh(\sqrt{\cc}\, T)}
     {\sqrt{\cc}}
\,.
\end{eqnarray}
This gives the average size of space at time $T$  after the Big Bang. 
For short time  $T$, \rf{AverageLengthOfSpaceZero1} behaves as 
$\langle L \rangle_{(0,T)} = T + O(T^3)$.
This implies that {\it the curvature at the birth of the universe is finite}.
It is seen that already when  $T$ is of the order of $\frac{1}{\sqrt{\cc}}$ after the Big Bang, the size of space is close 
to its maximum value $L_{\rm cs} := \frac{1}{\sqrt{\cc}}$.
Finally we can ask for the amplitude that such a Big Bang universe vanishes at time $T$:
\begin{equation}\label{BubbleProbability1}
\lim_{L_0 \to 0} \lim_{L \to 0}
\frac{1}{L_0} {\dbltinyspace} G(L_0,L;T)
\,=\,
\frac{\cc}{\sinh^2(\sqrt{\cc}\, T)}
\,.
\end{equation}

Below we will also need the corresponding results when $\cc$ is negative. The details are
provided in Appendix A.  Here we just state the results:
\begin{eqnarray}\label{CylinderAmplitudeNegativeMu1}
&&
 G(L_0,L;T)
\,=\,
\frac{e^{- \sqrt{-\cc}\,(L_0+L) \cot(\sqrt{-\cc}\, T)} }{\sin(\sqrt{-\cc}\, T)}
\,
\sqrt{\frac{-\cc L_0 }{L}}\;
I_1\!\!\;\bigg(\frac{2\sqrt{-\cc L_0 L}}{\sin(\sqrt{-\cc}\, T)}\bigg).
\end{eqnarray}
\begin{eqnarray}\label{BigBangAmplitudeNegativeMu1}
G_{\rm BB}(L;T) \!
&=&\!
\frac{-\cc {\dbltinyspace} e^{- \sqrt{-\cc}\,L \cot(\sqrt{-\cc}\, T)}}
     {\sin^2(\sqrt{-\cc}\, T)}
\,.
\end{eqnarray}
\begin{eqnarray}\label{AverageLengthOfSpaceInflation1}
\langle L \rangle_{(0,T)}
\!&=&\!
\frac{\tan(\sqrt{-\cc}\, T)}
     {\sqrt{-\cc}}
\,.
\end{eqnarray}
\begin{equation}
\lim_{L_0 \to 0} \lim_{L \to 0}
\frac{1}{L_0} {\dbltinyspace} G(L_0,L;T)
\,=\,
\frac{-\cc}{\sin^2(\sqrt{-\cc}\, T)}
\,.
\end{equation}
It is seen that a negative $\cc$ results in ``inflation'': the universe is expanding as 
a function of $T$ and for $T =  T_{\rm ps} :=\frac{\pi}{2\sqrt{-\cc}}$ it reaches infinite spatial size.
We will later argue that in an interacting theory it does not imply that the theory breaks down 
for values of $T > T_{\rm ps}$, but rather that Coleman's mechanism \cite{coleman} forces the effective value 
of $\mu$ to be close to zero.

\subsection{Second quantization}

We can now introduce a ``second quantized'' notation and the corresponding Fock space,
much like in many body theory. We let $H_0$ from \rf{GreenDiffEq} denote the ``single (spatial) universe'' 
Hamiltonian or single (closed) string Hamiltonian, in analogy with a single particle Hamiltonian in many body theory.
Let $\Psi^{\dagger}(L)$ and $\Psi(L)$ be operators
which create and annihilate one closed string
with length $L$, respectively.
More precisely  
$\Psi^{\dagger}(L)$ creates one closed string with a marked point, 
and 
$\Psi(L)$ annihilates one closed string with no marked points.
The commutation relations of the string operators are 
\bea
&&
\big[\, \Psi(L) \,, \Psi^\dagger(L') \,\big]
\,=\,
\delta(L-L')
\, ,
\label{CommutePsi}
\\
&&
\big[\, \Psi(L) \,, \Psi(L') \,\big]
\,=\,
\big[\, \Psi^\dagger(L) \,, \Psi^\dagger(L') \,\big]
\,=\,
0
\, .
\label{ZeroCommutePsi}
\eea
The vacuum states (no string states, in analogy with the no particle states in many body theory) $\cuum$ and $\vac$ 
are the Fock states defined by
\beq
\Psi(L) \cuum \,=\, 
\vac \Psi^\dagger(L) \,=\, 
0
\, .
\eeq
In this second quantized  Hamiltonian formalism the Green function
$G(L,L';\T)$ is  obtained by
\beq
G(L,L';\T)
\,=\,
\vac \Psi(L') {\dbltinyspace}e^{- \T \Hdisk} \Psi^\dagger(L) \cuum,
\, .
\eeq
where $\Hdisk$ consists of a tapole term and the kinetic term:
\beq\label{Hdisk}
\Hdisk \,=\, {\cal H}_{\rm tad} + {\cal H}_{0}
\eeq
\beq\label{Hkin}
{\cal H}_{\rm tad} =
- \int\limits_0^\infty\! d L {\dbltinyspace}
  \delta(L) {\tinyspace}
  \Psi(L),  \qquad 
 {\cal H}_{0} =
\int\limits_0^\infty\! \frac{d L}{L}  {\dbltinyspace}
  \Psi^\dag(L) {\tinyspace} H_0
  \,
  L\Psi(L)
\, .
\eeq
For the purpose of calculating $G(L,L',T)$ one does not need the tadpole term in \rf{Hdisk}. However,
if we want to reproduce the disk amplitude \rf{DiskAmpDef} directly in the second quantised formalism 
we need a closed string of length zero to have the possibility of disappearing into the vacuum. Including the 
tadpole term we can write 
\beq\label{DiskAmp}
F_1^{(0)}(L) \,=\,
\lim_{\T \rightarrow \infty}
\vac {\tinyspace} e^{- \T \Hdisk} \Psi^\dagger (L) \cuum
\,.
\eeq
Using that the right-hand side  of eq.\ \rf{DiskAmp} converges to the left-hand side of eq.\ \rf{DiskAmp}
which is  independent of $T$, we have  
\beq\label{jk90}
\lim_{\T \rightarrow \infty}
\pder{\T}
\vac {\tinyspace} e^{- \T \Hdisk} \Psi^\dagger (L) \cuum
\,=\,
0
\,.
\eeq
This equation is equivalent to the so-called Schwinger-Dyson equation derived directly from CDT 
when one allows a spatial universe to split in two \cite{gcdt}.

\subsection{Interacting CDT: String Field Theory}
\begin{figure}[h]
\vspace{50pt}
 \begin{center}
    {\includegraphics[width=250pt]{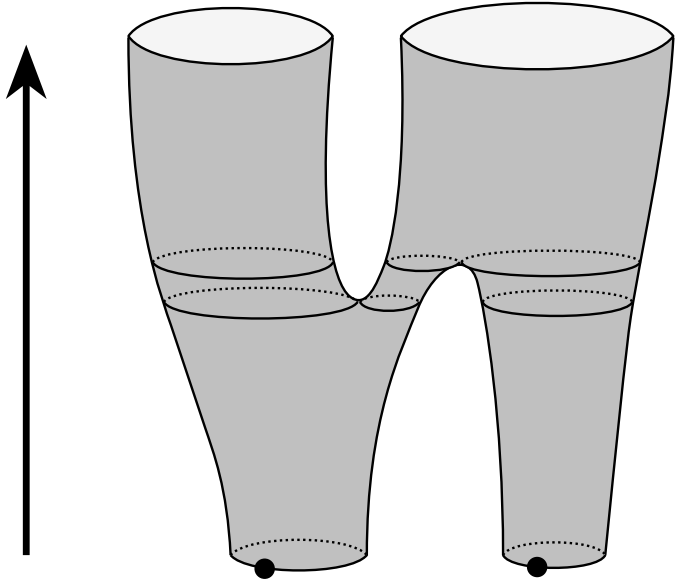}}
 \end{center}
\caption[hdynamics]{{\small
Dynamics of two spatial universes according to \rf{Hgeneral}. At time $T=0$ we have two 
spatial universes. Both propagate according to the first term in \rf{Hgeneral}. 
At a later time the left universe splits into two according to the second term in 
\rf{Hgeneral}. Then one of these universes merges with the universe at the right of the picture, according
to the third term in \rf{Hgeneral}, and after that the universes propagate according to the first term in \rf{Hgeneral}.
}}
\label{hdynamics}
\end{figure}

One advantage of  the second quantisation formalism in many body theory is that interaction between different 
particles is easily included in such a way that the statistics of many particles is automatically accounted for.
The same is true when the particles are replaced by closed strings.  If we assume that the interaction
is that one string can split in two or two strings can merge into one, we  can write  the complete
CDT string field Hamiltonian as
\bea\label{Hgeneral}
\cH \!&=&\!
\Hdisk
\nonumber\\&&\!
-\; g
\!\int\limits_0^\infty\! d L_1
\!\int\limits_0^\infty\! d L_2 {\dbltinyspace}
  \Psi^\dag(L_1) {\tinyspace}
  \Psi^\dag(L_2) {\tinyspace}
  (L_1{\negdbltinyspace}+{\negdbltinyspace}L_2)
  \Psi(L_1{\negdbltinyspace}+{\negdbltinyspace}L_2)
\nonumber\\&&\!
-\; \GG {\tinyspace} g
\!\int\limits_0^\infty\! d L_1
\!\int\limits_0^\infty\! d L_2 {\dbltinyspace}
  \Psi^\dag(L_1{\negdbltinyspace}+{\negdbltinyspace}L_2) {\tinyspace}
  L_1 \Psi(L_1) {\tinyspace}
  L_2 \Psi(L_2)
\,.
\eea
The coupling constant $g$ counts the combined number of times 
strings will split or join while $\GG$ counts 
the handles (genus) of the spacetime created by the string. 
The Hamiltonian \rf{Hgeneral} can be derived from so-called 
generalized two-dimensional CDT \cite{gcdt}. 
The dynamics of spacetime created by the Hamiltonian \rf{Hgeneral} 
is illustrated in Fig.\ \ref{hdynamics}.

The Hamiltonian $\cH$ is defined such that it  satisfies
\beq\label{StabilityOfVacuum}
\cH {\tinyspace} \cuum \,=\, 0
\,.
\eeq
This condition expresses that the vacuum is stable under 
the time evolution generated by $\cH$.

The model does not have time reversal symmetry. However,  under the transformation
\bea\label{TimeReverse}
\Psi^\dagger(L) \,\leftrightarrow\, G\,L \Psi(L)
\eea
the only term in ${\cal H}$ not compatible with time reversal symmetry is the 
the tadpole term in \rf{Hdisk}.

As a generalisation of the disk amplitude \rf{DiskAmp} one can define the amplitude for merging many entrance loops,
where the corresponding spacetime can have any topology by 
\bea\label{GeneralAmpDef}
\sum_{h=0}^\infty \GG^{{\tinyspace}h+N-1} \,
F_N^{(h)}(L_1,\ldots,L_N)
\!&=&\!
\lim_{\T \rightarrow \infty}
\vac {\tinyspace} e^{- \T \cH}
\Psi^\dagger (L_1) \ldots \Psi^\dagger (L_N) \cuum
\,.
\nn&&
\eea
In  eq.\ \rf{GeneralAmpDef} $N$ denotes 
the number of entrance loops and 
$h$ the number of handles (i.e.\ the genus) of the spacetime. 
The generalisation of   \rf{jk90} is then  
\beq\label{GeneralAmpLoopEq}
\lim_{\T \rightarrow \infty}
\pder{\T}
\vac {\tinyspace} e^{- \T \cH}
\Psi^\dagger (L_1) \ldots \Psi^\dagger (L_N) \cuum
\,=\,
0
\,.
\eeq

In the following it will be convenient to introduce  the Laplace transformation of 
$\Psi^\dagger(L)$ and $\Psi(L)$  as
\beq\label{LapPhiPsi}
\tilde\Psi^\dagger(\xi) \,=\,
\int\limits_0^\infty \! d L \, e^{- L \xi} {\tinyspace} \Psi^\dagger(L)
\,,
\qquad
\tilde\Psi(\eta) \,=\,
\int\limits_0^\infty \! d L \, e^{- L \eta} {\tinyspace} \Psi(L)
\,.
\eeq
Similarly, we also introduce the Laplace transformation 
for the other functions as 
\beq
\tilde{G}(\xi,\eta;\T) \,=\,
\int\limits_0^\infty \! d L  \!
\int\limits_0^\infty \! d L' \, e^{- L \xi - L' \eta} {\tinyspace}
G(L,L';\T)
\,,
\eeq
\beq
\tilde{F}_1^{(0)}(\xi) \,=\,
\int\limits_0^\infty \! d L \, e^{-L \xi} {\tinyspace} F_1^{(0)}(L)
\,,
\eeq
and
\beq
\tilde{F}_N^{(h)}(\xi_1,\ldots,\xi_N) \,=\,
\int\limits_0^\infty \! d L_1 \ldots\!
\int\limits_0^\infty \! d L_N \, 
e^{-L_1 \xi_1-\ldots-L_N \xi_N} {\tinyspace}
F_N^{(h)}(L_1,\ldots,L_N)
\,.
\eeq
Then, 
the differential equation \rf{GreenDiffEq} and 
the initial condition \rf{GreenInitialCond} become
\beq\label{GreenDiffEqLaplace}
\pder{\T} {\tinyspace} \tilde{G}(\xi,\eta;\T)
\,=\,
-( \xi^2 - \cc ) \pder{\xi} {\tinyspace} \tilde{G}(\xi,\eta;\T)
\,,
\eeq
and 
\beq\label{GreenInitialCondLaplace}
\tilde{G}(\xi,\eta;0)
\,=\,
\frac{1}{\xi + \eta}
\,.
\eeq
The disk amplitude expressions \rf{DiskAmpDef} and \rf{DiskAmp} transform to 
\beq\label{DiskAmpDefLaplace}
\tilde{F}_1^{(0)}(\xi)
\,=\,
\lim_{\T \rightarrow \infty}
\vac {\tinyspace} e^{- \T \tHdisk} \tilde\Psi^\dagger(\xi) \cuum
\,=\,
\frac{1}{\xi + \sqrt{\cc}}
\,,
\eeq
and the amplitudes with general topology \rf{GeneralAmpDef} read
\bea\label{GeneralAmp}
\sum_{h=0}^\infty \GG^{{\tinyspace}h+N-1} \,
\tilde{F}_N^{(h)}(\xi_1,\ldots,\xi_N)
\!&=&\!
\lim_{\T \rightarrow \infty}
\vac {\tinyspace} e^{- \T \tcH}
\tilde\Psi^\dagger(\xi_1) \ldots \tilde\Psi^\dagger(\xi_N) \cuum
\,,
\nn&&
\eea
where 
$\tHdisk = \tcH\big|_{g=0}$ 
and 
\bea\label{Hgen}
  \tcH  \!&=&\!\!
  \intdzeta \,\Bigl\{
    -\,   \tilde\Psi(-\zeta)
  \,-\,   ( \tilde\Psi^\dagger(\zeta) )
          \big( \zeta^2 -\cc \big) {\tinyspace}
          \frac{\prt}{\prt\zeta} \tilde\Psi(-\zeta)
\nonumber\\&&\!\!\phantom{%
  \intdzeta \,\Bigl\{
}\!\!{\dbltinyspace}
    -\, g {\tinyspace} ( \tilde\Psi^\dagger(\zeta) )^2
          \frac{\prt}{\prt\zeta} \tilde\Psi(-\zeta)
  \,-\, \GG {\tinyspace} g {\tinyspace} \tilde\Psi^\dagger(\zeta)
          \Bigl( \frac{\prt}{\prt\zeta} \tilde\Psi(-\zeta) \Bigr)^2
  \,\Bigr\}
\,.
\quad
\eea
This Hamiltonian can be obtained directly from \rf{Hgeneral}.

If we introduce the generating function
\beq\label{GeneratingFun}
 Z_F[\cJ]  \,\define \,
 \lim_{\T\rightarrow\infty} \vac \,
 e^{- \T \tcH}
 \exp\!\Big\{ \int\! \frac{d\zeta}{2 \pi i}
              \tilde\Psi^\dagger(\zeta) \cJ(-\zeta) \Big\}
 \cuum
\,,
\eeq
the amplitudes with general topology \rf{GeneralAmp} 
can be obtain in the standard way 
by functional differentiation after $\ln Z_F(\cJ)$:
\bea\label{GeneralAmp2}
\sum_{h=0}^\infty \GG^{{\tinyspace}h+N-1}
\tilde{F}_N^{(h)}(\xi_1,\ldots,\xi_N;\cc)
\!&=&\!
\frac{\delta^{N}}
     {\delta \cJ(\xi_1) \, \cdots \, \delta \cJ(\xi_N) } 
\ln Z_F[\cJ] \Bigg|_{\cJ=0}
\,.
\quad
\eea

Again, as in ordinary quantum field theory, 
the advantage of using the generating function $Z_F(\cJ)$ is that
operators $ A$ of $\Psi^\dagger$ and $\Psi$ acting $Z_F(J)$ can be expressed in terms of  $\frac{\delta}{\delta \cJ}$
and $\cJ$. To be more explicit we associate with 
an operator $A$ of $\Psi^\dagger$ and $\Psi$ the operator 
$A^\star$ of $\frac{\delta}{\delta J}$ and $J$ defined by  
\beq\label{StarOpDef}
 A^\star Z_F[\cJ]  \,=\,
 \lim_{\T\rightarrow\infty} \vac \,
 e^{- \T \tcH}
 A
 \exp\!\Big\{ \int\! \frac{d\zeta}{2 \pi i}
              \tilde\Psi^\dagger(\zeta) \cJ(-\zeta) \Big\}
 \cuum
\,.
\eeq
For example, 
\beq\label{StarOpPsi}
  \bigl( \tilde\Psi^\dagger(\xi) \bigr)^\star
  \,=\,
  \frac{\delta}{\delta \cJ(\xi)}
\,, \qquad
  \bigl( \tilde\Psi(\eta) \bigr)^\star
  \,=\,
  \cJ(\eta)
\,.
\eeq
Under this so-called star operation, one finds the following property:
\beq\label{StarOpProduct}
  \bigl( A_1 A_2 \cdots A_n \bigr)^\star
  \,=\,
  A_n^\star \cdots A_2^\star {\tinyspace} A_1^\star
\,.
\eeq
Using the star operation the  Schwinger-Dyson equation \rf{GeneralAmpLoopEq} 
can be  rewritten as 
\beq\label{SDGeneral2}
   \tHj Z_F[\cJ] \ = \  0
\,,
\eeq
where
\bea\label{Hgen2}
 \tHj \!&=&\!\!
 \intdzeta \Bigl\{ \,
   -\, \cJ(-\zeta)
 \,-\, \Bigl( \frac{\prt}{\prt\zeta} \cJ(-\zeta) \Bigr)
       \big( \zeta^2 - \cc \big)
       \Bigl( \frac{\delta}{\delta \cJ(\zeta)} \Bigr)
\nonumber\\&&\!\!\phantom{%
  \intdzeta \Bigl\{
}\!\!{\dbltinyspace}
   -\, g \Bigl( \frac{\prt}{\prt\zeta} \cJ(-\zeta) \Bigr)
         \Bigl( \frac{\delta}{\delta \cJ(\zeta)} \Bigr)^2
 \,-\, \GG {\tinyspace} g \Bigl( \frac{\prt}{\prt\zeta} \cJ(-\zeta) \Bigr)^2
         \frac{\delta}{\delta \cJ(\zeta)}
 \, \Bigr\}\,.
\nn&&
\eea

\subsection{The mode expansion of string field theory}

We now introduce the so-called mode expansion. It is the mode expansion which
allows us to relate CDT string field theory to $W$ symmetry in the most transparent way.
For large $\xi$ the disk amplitude \rf{DiskAmpDefLaplace} has the expansion 
\beq\label{DiskAmpModeExpansion}
\tilde{F}_1^{(0)}(\xi)
\,=\,
\xi^{-1}
+ \sum_{l=1}^\infty \xi^{-l-1} f_1^{(0)} (l)
\,.
\eeq
Similarly, 
the amplitudes with general 
topology 
have for large $\xi_i$ the expansion
\beq\label{GeneralAmpModeExpansion}
\tilde{F}_N^{(h)}(\xi_1,\ldots,\xi_N) \,=\,
\xi_1^{-1} \delta_{N,1} \delta_{h,0}
\,+
\sum_{l_1,\ldots,l_N=1}^\infty
\xi_1^{- l_1 - 1} \ldots \xi_N^{- l_N - 1}
f_N^{(h)}(l_1,\ldots,l_N)
\,.
\eeq
Note that the first term,  $\xi^{-1}$, 
on the right-hand  side of the last two equations does
not contribute to the finite spacetime volume since it does not contain any reference to $\mu$.
It can thus be considered as kind of non-universal term which we are free to choose in the 
mode expansion of the string fields. 
With this remark in mind we assume that the string fields 
have the following mode expansion   
\beq\label{StringModeExpansion}
\tilde{\Psi}^\dagger (\xi)
\,=\,
\xi^{-1}
+ \sum_{l=1}^\infty \xi^{-l-1} \phi^\dagger_l
\,,
\qquad
\tilde{\Psi}(\eta)= \sum_{l=1}^\infty (-\eta)^{l} \phi_l
\,.
\eeq
Using  this mode expansion  the commutation relations of string fields 
\rf{CommutePsi} and \rf{ZeroCommutePsi} become
\beq\label{Commutepsi}
\big[ \, \phi_l \, , \phi^\dagger_{l'} \, \big]  \ = \ \delta_{l,l'}
\,,
\qquad
\big[ \, \phi^\dagger_l \, , \phi^\dagger_{l'} \, \big] \ = \
\big[ \, \phi_l \, , \phi_{l'} \, \big] \ = \ 0
\,.
\eeq
Substituting the string mode expansion \rf{StringModeExpansion} 
into the Hamiltonian \rf{Hgeneral} we obtain
\bea\label{HgeneralModeExpansion}
 \cH \!&=&\!
 \cc {\halftinyspace} \phi_1
 -\, 2 g {\halftinyspace} \phi_2
 \,-\, \GG {\tinyspace} g {\halftinyspace} \phi_1 \phi_1
\nonumber\\&&\!
 -\> \sum_{l=1}^\infty \phi_{l+1}^\dagger l \phi_l
 \,+\, \cc \sum_{l=2}^\infty \phi_{l-1}^\dagger l \phi_l
 \,-\, 2 g \sum_{l=3}^\infty \phi_{l-2}^\dagger l \phi_l
\nonumber\\&&\!
 -\> g
  \sum_{l=4}^\infty \sum_{n=1}^{l-3}
  \phi_n^\dagger \phi_{l-n-2}^\dagger
  l \phi_l
\nonumber\\&&\!
 -\> \GG {\tinyspace} g
  \sum_{l=1}^\infty \sum_{n=\max(3-l,1)}^\infty
  \phi_{n+l-2}^\dagger
  n \phi_n l \phi_l
\,.
\eea
Note here that the term $2 g \sum_{l=3}^\infty \phi_{l-2}^\dagger l \phi_l$ originates from the 
production of baby universes.

In particular we have for the  non-interacting part of $\cH$
\beq\label{zj3}
 \cH_{0} =\int\limits_0^\infty\! \frac{d L}{L}  {\dbltinyspace}
  \Psi^\dag(L) {\tinyspace} H_0
 \,L \Psi(L) 
  =
  - \sum_{l=1}^\infty \phi_{l+1}^\dagger l \phi_l
+ \mu \sum_{l=2}^\infty \phi_{l-1}^\dagger l \phi_l.
\eeq
We have already seen that this part of the Hamiltonian is the one 
which  determines the amplitude $G(L,L',T)$ for a spatial universe of length $L$ to evolve
into a spatial  a universe of length $L'$ in time $T$ without the spatial 
universe splitting in two or being joined by another spatial universe.
From the explicit expression for $G(L,L',T)$ it is seen that there is 
a non-zero amplitude for even a spatial universe of  infinitesimal length  space to expand to 
macroscopic size $L'$ in a finite time $T$. Let us now consider a spatial universe in the 
quantum state $ \phi^\dg_l \cuum$. This is not a state where space has a specific length $L$.
However, the distribution of lengths can be expressed in terms of $\delta (L)$ and derivatives 
of $\delta(L)$, i.e.\ $ \phi^\dg_l \cuum$ has to be assigned an infinitesimal length (distribution) for any 
finite mode number $l$. This is seen by working out the relation between the operators $\phi_l,~\phi^\dg_l$ 
and the operators $\Psi(L),~\Psi^\dg(L)$ which  annihilate and create spatial universes of 
macroscopic length $L$. We have  
\beq\label{jj1}
\Psi^\dagger(L) = \sum_{l=0} \frac{L^l}{l!} \; \phi^\dg_l,\qquad \phi^\dg_l = \frac{d^{\, l}}{dL^l} \Psi^\dg(L) \Big|_{L=0}.
\eeq
The message to take along for the future discussion is that the kinetic term  $\cH_0$, 
given by the right hand side of eq.\ \rf{zj3}, is able to propagate a state of infinitesimal length like 
$| l \ra =\phi^\dg_l \cuum$ to a state of finite length $L'$ within finite time $T$.

Introducing the mode expansion for the source term $\cJ(\eta)$, 
in analogy with the mode expansion \rf{StringModeExpansion} for $\Psi(\eta)$, 
i.e.\ $\cJ(\eta) = \sum_{l=1}^\infty (-\eta)^l j_l$, 
the generating function $Z_F[\cJ]$ 
\rf{GeneratingFun} can be written as 
\beq\label{GeneratingFunModeExpansion}
 Z_f[j]  \,\define \,
 \lim_{\T\rightarrow\infty} \vac \,
 e^{- \T \cH}
 \exp\!\Big\{ \sum_{l=1}^\infty
              \phi^\dagger_l j_l \Big\}
 \cuum
\, .
\eeq
Then, 
the amplitudes $f_N^{(h)}$ defined in \rf{GeneralAmpModeExpansion}
can be obtain by differentiation after $\ln Z_f[j]$:
\bea\label{GeneralAmpModeExpansion2}
\sum_{h=0}^\infty \GG^{{\tinyspace}h+N-1}
f_N^{(h)}(l_1,\ldots,l_N)
\!&=&\!
\frac{\partial^{N}}
     {\partial j_{l_1} \, \cdots \, \partial j_{l_N} } 
\ln Z_f[j] \Bigg|_{j=0}
\,.
\quad
\eea
We can now apply the star operation  \rf{StarOpDef}   to  the string modes. 
Using the mode decomposition,  equation \rf{StarOpDef}  now reads
\beq\label{StarOpDefModeExpansion}
 A^\star Z_f[j]  \,=\,
 \lim_{\T\rightarrow\infty} \vac \,
 e^{- \T \cH}
 A
 \exp\!\Big\{ \sum_{l=1}^\infty
              \phi^\dagger_l j_l \Big\}
 \cuum
\, ,
\eeq
from which we obtain the analogy of  \rf{StarOpPsi} 
\beq\label{StarOpPsiModeExpansion}
  \bigl( \phi^\dagger_l \bigr)^\star \ = \ \pder{j_l}
\,,
\qquad
  \bigl( \phi_l \bigr)^\star \ = \ j_l
\,.
\eeq
Thus the star operation applied to the Hamiltonian \rf{HgeneralModeExpansion} 
leads to the following expression, which is the analogy of \rf{Hgen2},
\bea
 \Hj \!&=&\!
 \cc {\halftinyspace} j_1
 -\, 2 {\halftinyspace} g j_2
 \,-\, \GG {\tinyspace} g j_1 j_1
\nonumber\\&&\!
 -\, \sum_{l=1}^\infty l j_l {\dbltinyspace} \pder{j_{l+1}}
 \,+\, \cc \sum_{l=2}^\infty l j_l {\dbltinyspace} \pder{j_{l-1}}
 \,-\, 2 g \sum_{l=3}^\infty l j_l {\dbltinyspace} \pder{j_{l-2}}
\nonumber\\&&\!
 -\> g
  \sum_{l=4}^\infty \sum_{n=1}^{l-3}
  l j_l
  {\dbltinyspace} \pder{j_{n}}
  {\dbltinyspace} \pder{j_{l-n-2}}
\nonumber\\&&\!
 -\> \GG {\tinyspace} g
  \sum_{l=1}^\infty \sum_{n=\max(3-l,1)}^\infty
  n j_n {\tinyspace} l j_l
  {\dbltinyspace} \pder{j_{n+l-2}}
\,,
\label{HgeneneralModeExpansionStar}
\eea
Corresponding to \rf{SDGeneral2} we have
\beq\label{jk3}
\Hj \; Z_f[j] = 0.
\eeq
It should be noted that 
three parameters $g$, $\GG$ and $\cc$ appear in CDT string field theory,
while only  $\GG$ and $\cc$ appear in conventional, $c=0$ non-critical string theory.
The reason is that in $c=0$ non-critical string theory  
one can remove $g$ by a rescaling 
$T \to T/\sqrt{g}$, 
$\Psi^\dagger \to \Psi^\dagger/\sqrt{g}$, 
$\Psi \to \sqrt{g}\,\Psi$ and 
$\GG \to \GG/\sqrt{g}$ 
because $H_0 = 0$. 
This is impossible in CDT string field theory 
because $H_0 \neq 0$.

\subsection{The appearance of the W-algebra}

The role of the $W$-algebra in non-critical string theory is well known \cite{w3,aw}. For 
a description which is closest to the one we will use here, we refer to \cite{aw}. For completeness
let us define the $W^{(3)}$ algebra. Given a Fock space and operators $\a_n$ satisfying   
\beq\label{jx10} 
[\a_m,\a_n]= m\; \del_{0,n+m},
\eeq
we  define 
\beq\label{jx11}
J(z) = \sum_{n\in Z} \frac{\a_n}{z^{n+1}},~~~~
W^{(3)}(z)= \;\frac{1}{3}:J(z)^3: \;= \sum_{n\in Z} \frac{W^{(3)}_n}{z^{n+3}}.
\eeq
The normal ordering $:\!(\cdot)\!:$ 
refers to the $\a_n$ operators ($\a_n$ to the 
left of $\a_m$ for $n < m$). We then have 
\beq\label{jx12}
W^{(3)}_n = \frac{1}{3} \sum_{a+b+c=n} :\a_a\a_b\a_c:.
\eeq

In standard non-critical string theory for central charge $c=0$, i.e.\ the case of pure Euclidean two-dimensional 
quantum gravity, the Hamiltonian corresponding to our $\Hj$ can be written as 
\beq\label{jk1}
\Hj \propto  -\bar{W}^{(3)}_{-2} + Y ,
\eeq
where the $\a_n$ operators can be expressed in terms of $j_n$ and $\pder{j_n}$ and the operator $Y$ contains
a few $\a_n$ needed to ensure the equivalent of \rf{SDGeneral2} in the case 
of $c=0$ non-critical string theory. The bar over the $W$ in \rf{jk1} refers to the fact that $W$-algebra which 
appears in standard non-critical string theory is a so-called 2-reduced $W$ algebra, which (loosely speaking)
only involve the summation over even $n$ in formulas \rf{jx11} and \rf{jx12} (see \cite{aw} for precise 
definitions). The reason we recall these results from standard non-critical string theory is that we expect 
similar results for CDT string field theory, since the structure of this theory is quite similar to that of 
standard non-critical string theory (and, as already mentioned, in some sense simpler). Indeed, after a 
short calculation one finds  that \rf{HgeneneralModeExpansionStar} can be expressed  as
\bea\label{WalgebraHamiltonianStarAfterSSB}
\Hj
\!&=&\!
-\,
g \sqrt{\GG}\, W^{(3)}_{-2}
\,+\,
  \frac{1}{\GG} \bigg(
  \frac{\cc^2}{4 g}
+ \frac{1}{4 g} \pder{j_4}
- \frac{\cc}{2 g} \pder{j_2}
+ \pder{j_1}
\!\;\bigg)
\,,
\eea
where, according to \rf{jx12},
\beq\label{W3operator}
W^{(3)}_{-2} \ = \ 
\frac{1}{3} \sum_{k, l, m}\!
:\!\alpha_k \alpha_l \alpha_m\!: \delta_{k+l+m,-2}
= \mbox{res}_{z=0} \; W^{(3)}(z)
\,,
\eeq
and
\begin{eqnarray}\label{DefAlphaOperatorPureCDT}
\alpha_n :=
\left\{
\begin{array}{cl}
\displaystyle
\frac{1}{\sqrt{\GG}} {\dbltinyspace} \pder{j_n}
& \hbox{[\,$n \!>\! 0$\,]}
\rule[-12pt]{0pt}{30pt}\\
\displaystyle
\nu
& \hbox{[\,$n \!=\! 0$\,]}
\rule[-12pt]{0pt}{30pt}\\
\displaystyle
- n ( \lambda_{-n} + \sqrt{\GG} {\dbltinyspace} j_{-n} )
& \hbox{[\,$n \!<\! 0$\,]}
\end{array}
\right.
\end{eqnarray}
and
\begin{eqnarray}\label{CoherentEigenValuesPureCDT}
\nu = \frac{1}{\sqrt{\GG}}
\,,
\quad
\lambda_1 = -\,\frac{\cc}{2 g {\sqrt{\GG}}}
\,,
\quad
\lambda_3 = \frac{1}{6 g {\sqrt{\GG}}}
\,,
\quad
\lambda_n = 0,~~n \neq 1,3
\,.
\end{eqnarray}
The Fock space where the $\a_n$ operators act is formed by suitable functions of the sequence $j_n$ by 
a standard construction.
 
Let us now use the (inverse) star operation to return from $\Hj$ to $\cH$ which is 
expressed in terms of string modes. First we apply the (inverse) star operation to $\a_n$:
\begin{eqnarray}\label{DefAlphaOperatorPureCDT2}
\alpha_n =
\left\{
\begin{array}{cl}
\displaystyle
( a_n^\dagger )^\star
& \hbox{[\,$n \!>\! 0$\,]}
\rule[-12pt]{0pt}{30pt}\\
\displaystyle
( \pp )^\star
& \hbox{[\,$n \!=\! 0$\,]}
\rule[-12pt]{0pt}{30pt}\\
\displaystyle
- n ( a_{-n} )^\star
& \hbox{[\,$n \!<\! 0$\,]}
\end{array}
\right.
\end{eqnarray}
The $a_n$ and $a^\dagger_n$ operators are related to our string modes $\phi_n$ and $\phi_n^\dagger$ by
\beq\label{DefAlphaOperator}
a_n^\dagger \,=\, 
\frac{1}{\sqrt{\GG}} {\dbltinyspace} \phi^\dagger_n
\,,
\qquad
a_n \,=\,
\lambda_n + \sqrt{\GG} {\dbltinyspace} \phi_n,
\eeq
and with this notation the Hamiltonian \rf{HgeneralModeExpansion} can be written as 
\bea\label{WalgebraHamiltonianAfterSSB}
\cH
\!&=&\!
g \sqrt{\GG}\, \cH_{\rm W}
\,+\,
  \frac{1}{\GG} \bigg(
  \frac{\cc^2}{4 g}
+ \frac{1}{4 g} \phi_4^\dagger
- \frac{\cc}{2 g} \phi_2^\dagger
+ \phi_1^\dagger
\!\;\bigg)
\,,
\eea
where $\cH_{\rm W}$ is related to the  $W^{(3)}_{-2}$-operator by an (inverse) star operation:  
\beq\label{jy3}
\cH_{\rm W}^\star \,=\, - W^{(3)}_{-2} = - \mbox{res}_{z=0} W^{(3)} (z)
\,,
\eeq
and explicitly one has
\beq\label{WalgebraHamiltonian}
\cH_{\rm W} =
- \!\!\!\!\!\!\sum_{\stackrel{\scriptstyle l,\,m,\,n}
               {[n+m+2 = l\,]}}\!\!\!\!
  a_n^\dagger a_m^\dagger l a_l \; -  \!\!\!\!\!\!
\sum_{\stackrel{\scriptstyle n,\,m,\,l}{[n+2 = m+l\,]}}
 \!\!\! \!a_n^\dagger m a_m l a_l -
2 \!\!\!\sum_{\stackrel{\scriptstyle n,\,l}{[n+2 = l\,]}}\!\!
  p\!\> a_n^\dagger l a_l
 -  p\!\> a_1 a_1 -2 p^2 a_2
\,,
\eeq
where the commutation relations are 
\begin{eqnarray}
&&
\big[\, a_m \,,\, a_n^\dagger \,\big]
=
\delta_{m,n}
\,,
\qquad\quad
\big[\, \qq \,,\, \pp \,\big]
=
i
\,,
\\&&
\big[\, \pp \,,\, a_n^\dagger \,\big]
=
\big[\, \pp \,,\, a_n \,\big]
=
\big[\, \qq \,,\, a_n^\dagger \,\big]
=
\big[\, \qq \,,\, a_n \,\big]
=
0
\,,
\end{eqnarray}
and where we have introduced an operator $\qq$ which commutes with all $\a_n$ except $\a_0$. The pair 
$\a_n,\a_{-n}$ is related to the creation and annihilation operator $a^\dagger_n,a_n$ for $n > 0$, 
while $\a_0= \pp$ is an Hermitian 
operator. Introducing $\qq$ allow us to talk about annihilation and creation operators also for the mode
corresponding to $n=0$, precisely as for the ordinary harmonic oscillator if we define 
\beq\label{jk2}
a_0			\,=\, \frac{1}{\sqrt{2}} ( q + i p )
\,,
\qquad\quad
a_0^\dagger \,=\, \frac{1}{\sqrt{2}} ( q - i p )
\,.
\eeq
For furture reference we note that the eigenstates of $a_0$, 
i.e.\ states which satisfy $a_0 |\lambda_0\ra = \lambda_0 |\lambda_0\ra$, are related to the eigenstates 
of $p$, i.e.\ the states which satisfy $p |\nu \ra = \nu |\nu \ra$ by a standard formula from the theory
of coherent states:
\beq\label{jk6}
\la \nu | \lambda_0\ra = \exp \Big( - \oh ( \sqrt{2} \lambda_0^{(i)} -\nu)^2 + 
i (\sqrt{2} \lambda_0^{(i)} -\nu) \sqrt{2} \lambda_0^{(r)}\Big),
\eeq
where $\lambda_0 = \lambda_0^{(r)} + i \lambda_0^{(i)}$ and $\nu$ is real. 

{\it Note that $\cH_{\rm W}$ does not contain any coupling constants 
when expressed in terms of the operators $p$, $a_n$ and  $a_n^\dagger$ .}
Even the time $T$ placed in front of the Hamiltonian $\cH_{\rm W}$ 
can be absorbed by the rescaling $\alpha_n \to T^{n/2} \alpha_n$.

\subsection{Coherent states and the emergence of time and space}

When two-dimensional CDT was formulated as a theory of quantum gravity the assumption was that 
there existed a stable vacuum state, $\cuum$, such that $\cH \cuum =0$ (eq.\ \rf{StabilityOfVacuum}).
However, if we assume that the relevant Hamiltonian is proportional to $\cH_{\rm W}$, another state is 
the natural vacuum state, and we denote it {\it the absolute vacuum} $|0\rangle$. It satisfies 
\begin{equation}\label{DefVacuum}
a_n |0\rangle = \pp |0\rangle = 0
\,,
\qquad\quad\hbox{[\,$n \!=\!  1$, $2$, \ldots\,]}
\,,
\end{equation}
and we have 
\beq\label{jk4}
\cH_{\rm W} |0\rangle = 0.
\eeq

The states in the Fock space $\cF$ associated with $\cH_{\rm W}$ are obtained
by acting repeatedly on the absolute vacuum $|0\rangle$ 
with the operators $a_n^\dg$ and $\qq$. If our stating point is just the algebraic structure dictated 
by the $\a_n$'s, we have no geometric interpretation of these states, and we denote them as 
``pre-geometric''. However, the vacuum state for $\cH$ is a coherent state obtained from the absolute 
vacuum $| 0 \rangle$ in the following way
\beq\label{jz1a}
|\nu\rangle \,=\, \e^{i\nu q} |0\rangle
\,,
\qquad
\cuum \,=\, 
V(\lambda_1,\lambda_3) |\nu\rangle
\,,
\eeq
\beq\label{jz1}
V(\lambda_1,\lambda_3):=\,
\exp\!\bigg(\! - \frac{|\lambda_1|^2}{2} - \frac{|\lambda_3|^2}{2}
  + \lambda_1 a_1^\dagger + \lambda_3 a_3^\dagger \bigg)
\,,
\eeq
where $\lambda_1$ and $\lambda_3$ are given by \rf{CoherentEigenValuesPureCDT}, and where we have 
\begin{eqnarray}\label{jz2}
a_1 \cuum = \lambda_1 \cuum
\,,
\qquad
a_3 \cuum = \lambda_3 \cuum
\,,
\qquad
\pp \cuum = \nu \cuum
\,.
\end{eqnarray}
If we  now {\it assume} that our system prefers $\cuum$ as the {\it physical vacuum}, 
we can put a geometrical interpretation into the various terms appearing in $\cH_W$ via $\cH$.
$\cH$ is an ordinary Hamiltonian, and we have seen that the states in the Hilbert space associated with $\cH$
via its string field representation has a clear geometric interpretation. In addition we have a time evolution 
of these geometric states, governed by $\cH$.
The fact that $a_1$ and $a_3$ has an expectation value 
in the state $\cuum$, i.e.\ that $\cuum$ represents 
a {\it condensate} of $\phi_1$ and $\phi_3$ modes, implies that a quadratic term, namely the term 
$\cH_0$ defined in \rf{zj3}, is created among  the otherwise cubic interaction terms in $\cH_W$. We know 
already the $\cH_0$ can be associated with the propagation of states in $\cF$ if we view $T$ as a time.
The emergence of time is thus associated with the choice of a suitable coherent state in $\cF$ which 
we denote the physical vacuum. The important difference between $\cH$ and $\cH_{\rm W}$ in this 
context is that $\cH_{\rm W} \cuum \neq 0$: the last terms in the rhs of eq.\ \rf{WalgebraHamiltonianAfterSSB}
create universes of infinitesimal length, but the lengths of these universes 
will grow  to macroscopic size because of the presence of  $\cH_0$. {\em Thus we have 
the picture that a choice of $\cuum$ will first lead to the emergence of time, and successively to the 
emergence of space. }

\subsection{The interaction of  two-dimensional  W-universes}

\vspace{12pt}

Once coupling constants associated with a universe are assigned in the manner described above,
we are confronted with the situation that two universes with different coupling constants can 
merge into a third universe. This is a kind of generalised string field theory. Can we imagine 
some kind of interaction between states of different universes which governs such a merging?
In order to suggest a suitable ``vertex''  for such an interaction we first establish a suitable notation for coherent states.

\newcommand{\ketheart}{{{\hspace{-1.2pt}}{}_\heartsuit}}
\newcommand{\braheart}{{}_{{}_\heartsuit}\hspace{-2.4pt}}


In $\cF$ we can choose an (overcomplete) basis of coherent states as follows.
Let us generalise the vertex operator \rf{jz1} and introduce the corresponding  coherent states
\beq\label{jk7}
V[\lambda] :=V(\lambda_0,\lambda_1,\lambda_2,\ldots)
\,=\,
\exp\!\bigg( \sum_{n=0}^\infty \lambda_n a_n^\dagger  - \oh  \sum_{n=0}^\infty \lambda_n^2\bigg)
\eeq
and 
\beq\label{jk8}
| \lambda \rara = V[\lambda] \;| 0 \ra\!\ra
\,,
\eeq
where the state $| 0 \ra\!\ra $ is defined by 
\beq\label{jk9}
a_n | 0 \rara = 0 \quad {\rm for}\quad n=0,1,2,\ldots .
\eeq
From \rf{jk6} and the completeness relation for coherent states 
we have the following expression of the absolut vacuum $|0\ra$ 
in terms of the coherent  states $| \lambda \rara$:
\beq\label{jk10}
| 0 \ra = \int \frac{d^2 \lambda_0}{2 \pi} | \lambda_0, \lambda_1=0,\lambda_2 =0,\ldots \rara \; 
\e^{-\big(  \big( \lambda_0^{(i)}\big)^2 +\, i \lambda_0^{(i)} \lambda_0^{(r)}\big)}.
\eeq

By declaring $\cH_W$ the central object of our model, the absolut vacuum $| 0\ra$ as well 
as $|0\rangle\!\rangle$
become  natural selected states in $\cF$. From the properties of coherent states the overlap 
between states $| \lambda\rara$ and $| \lambda'\rara$ will in general 
be non-zero and inspired by string field theory where 
two strings with different quantum numbers can join 
and form a string with new quantum numbers, 
or a string can split into two, also with different quantum numbers, 
we can imagine something similar here, 
the splitting and joining mediated by interactions
 \begin{eqnarray}
V_{-\hspace{-2pt}\langle}
&=&
\int\!
\frac{{\rm d}^2 \lambda}{\pi}\!\>
\frac{{\rm d}^2 \lambda'}{\pi}\!\>
\bigg\{
  e^{(\lambda + \lambda') a^\dagger} |0\rangle\!\rangle
  \,
  \langle\!\langle 0| e^{\lambda^\ast a}
  \otimes
  \langle\!\langle 0| e^{\lambda^{\prime\ast} a}
\bigg\}
{\trehalftinyspace}
e^{- |\lambda|^2 - |\lambda'|^2}
\,,
\\
V_{{\dbltinyspace}\rangle\hspace{-2pt}-}
&=&
\int\!
\frac{{\rm d}^2 \lambda}{\pi}\!\>
\frac{{\rm d}^2 \lambda'}{\pi}\!\>
\bigg\{
  e^{(\lambda - \lambda') a^\dagger} |0\rangle\!\rangle
  {\negdbltinyspace}\otimes
  e^{\lambda^{\prime\ast} a^\dg} |0\rangle\!\rangle
  \,
  \langle\!\langle 0| e^{\lambda^{{\negtinyspace}\ast} a}
\bigg\}
{\trehalftinyspace}
e^{- |\lambda|^2 - |\lambda'|^2}
\,.
\end{eqnarray}
In these equations we have used the following notation
\bea
&&
\frac{{\rm d}^2 \lambda}{\pi}
\,=\,
\prod_{n=0}^\infty
\frac{{\rm d}^2 \lambda_n}{\pi}
\,,
\qquad\qquad
e^{-|\lambda|^2}
\,=\,
\exp\!\bigg(\!\!-\!\sum_{n=0}^\infty |\lambda_n|^2 \bigg)
\,,
\\
&&
e^{\lambda a^\dagger}
\,=\,
\exp\!\bigg( \sum_{n=0}^\infty \lambda_n a_n^\dagger \bigg)
\,,
\qquad\quad
e^{\lambda^{{\negtinyspace}\ast} a}
\,=\,
\exp\!\bigg( \sum_{n=0}^\infty \lambda_n^* a_n\bigg)
\,.
\eea
In these  interaction vertices the total value of $\lambda$ (i.e.\ the sum of all $\lambda$'s) is preserved:
\beq
V_{-\hspace{-2pt}\langle}\,
\big({\dbltinyspace} |\lambda_0,\lambda_1,\lambda_2,\ldots\rangle\!\rangle
  \otimes |\lambda_0',\lambda_1',\lambda_2',\ldots\rangle\!\rangle \big)
\ = \ 
|\lambda_0+\lambda_0',\lambda_1+\lambda_1',\lambda_2+\lambda_2',\ldots\rangle\!\rangle
\,.
\eeq
As an example 
the state $a^\dagger |0\rangle\!\rangle $  
can be changed into the state $a^\dagger \cuum$, representing an excitation of the physical 
vacuum by
 \beq
V_{-\hspace{-2pt}\langle}\,
\big({\dbltinyspace} a^\dagger |0\rangle\!\rangle  \otimes \cuum \big)
\ = \ 
a^\dagger \cuum
\,.
\eeq
``Before'' the interaction mediated by $V_{-\hspace{-2pt}\langle}$
the concept of time did not exist in 
the universe defined by the excited state $a^\dg |0\rangle\!\rangle$, 
but it existed in $\cuum$.
``After'' the interaction by $V_{-\hspace{-2pt}\langle}$, 
the new universe $a^\dg \cuum $ represents an excitation 
in the physical universe where time exists. 
In this sense we can say that ``time'' has survived the interaction. 
In this way, 
changing the value of $\lambda_n$ makes it possible to discuss some aspects of 
a so-called  ``pregeometric age'',  where time has come into existence but space not. 
This also makes it possible to discuss certain aspects of a possible  end  of our universe as we know it.

The formulation of ordinary CDT as a path integral over geometries of fixed spatial topology 
can be viewed as a ``first quantization'', like the path integral for a particle. The ``second quantization'' 
we introduced above allows several of these spatial universes to interact, to split and join while the 
spatial volume (the length) is conserved. It led in a natural way  to the concept of creation and 
annihilation operators for spatial universes of length $L$. Dealing with operators creating and destroying 
spatial universes has sometimes in quantum gravity been denoted ``third quantization'', to distinguish 
it from a standard second quantized matter theory defined on a fixed background geometry. In this sense,
what we have suggested above is a ``fourth quantization'', where two  universes with different  coupling 
constants can interact and form a new universe with new coupling constants (or vice versa, a universe 
can split in two universes, but with different coupling constants).

\section{Extension by Jordan symmetry}\label{sec3}
\setcounter{equation}{0}

Above we have described a model of a two-dimensional universes, where universes can be 
continuously created from a pre-geometric state, and where these universes when once 
created,  can split and merge. The emergence of the universes and with them the concept of time and space 
from a the pre-geometric state was related to a breaking of an underlying $W$ symmetry.
The breaking of the $W$ symmetry was signalled by the condensation of some of the $W$ modes and 
the accompanying  assignment of coupling constants to the universe, in the same way as the 
condensation of the Higgs field determines a number of the coupling constants in the Standard Model.
We now want to extend this scenario to higher dimensions, following the idea that in string theory 
different dimensions $X^\mu$ come with different flavors $\mu =1,2,\ldots,D$, but they are interrelated 
via the symmetry and the interaction governed of the string-action. Here we will thus look for extended 
$W^{(3)}$ symmetric Hamiltonians $H_W$, i.e.\ $W$ algebras where our operators $\a_n$ have 
an additional flavour index $a$, i.e.\ $\a_n \to \a_n^a$, $a =1,2,\ldots,N$. It turns out \cite{romans,hull} that the 
{\it classical} extended $W^{(3)}$ algebras are related to Jordan algebras,  
so for completeness the next section provides the definitions of Jordan algebras and how this relation 
comes about.

\subsection{Formally real Jordan algebras}

A Jordan algebra is an algebra where the multiplication $\circ$ satisfies 
\beq\label{jk20}
 X \circ Y = Y\circ X,\qquad (X\circ Y) \circ (X \circ  X) = X\circ (Y \circ (X \circ X))).
 \eeq
One of the most important properties of Jordan algebra 
is power-associativity which implies that the expression 
$X^n \define
 \underbrace{{\dbltinyspace}X \!\circ\! X \!\circ\! \ldots \!\circ\! X}_n$ 
[\,$n {\negdbltinyspace}\in{\negdbltinyspace} \Dbl{N}$\,] can be defined 
without specifying the order of multiplications. We consider here the so-called 
formally real Jordan algebras which are algebras over the real numbers which 
have the properties that a sum of squares is zero only if each individual term in the sum is zero
(the extension to non-real Jordan algebras is relatively straight forward). Every finite dimensional
formally real Jordan algebra can be written as a direct sum of simple formally real algebras
(where ``simple'' means that the algebra cannot be further decomposed in non-trivial direct sums), and 
these have been classified. Apart from the trivial algebra $\Dbl{R}$, the real numbers, they consist of 
four families of algebras and one ``exceptional'' algebra, called the  Albert algebra.
The first family is  constructed from the real vector space $V_n$ generated by the 
 $n$  Hermitian, traceless $\gamma$-matrices $\gamma_a$, $a =1,\ldots, n$ of dimension $2^{[n/2]}$ and the 
$2^{[n/2]}\times 2^{[n/2]}$ identity matrix $\gamma_0 =I$, $n \geq 2$. Thus 
\beq\label{jk21}
\gamma_a \gamma_b+ \gamma_b \gamma_a= 2 \del_{a b} I, \qquad a,b = 1,\ldots,n.
\eeq 
The next  three  families consist of the real vector spaces formed by $H_n (\Dbl{R})$, $H_n(\Dbl{C})$,
$H_n(\Dbl{H})$, i.e.\ the 
$n\times n$ Hermitian matrices, $n \geq 3$,  with real, complex and quaternion  
entries, respectively. Finally, the exceptional case\footnote{Given an associative algebra $A$ were 
the product of two elements $X$ and $Y$ is denoted $XY$, $A$ becomes a Jordan algebra if one 
defines a new product by \rf{JordanProduct}. If a Jordan algebra is a sub-algebra of such an $A$ it  is called {\it special}
and else it is called {\it exceptional}. The four families are all special Jordan algebras since they are 
sub-Jordan algebras in the associative algebras of $n\times n$ matrices $M_n(\Dbl{Q})$, with entries 
in $\Dbl{Q} = \Dbl{R}, \Dbl{C}, \Dbl{H}$, respectively, and in the $2^{[n/2]}$-dimensional associative Clifford algebra
generated by the $2^{[n/2]}$-dimensional $\g$-matrices. 
However, the multiplication is non-associative in the case of octonions $\Dbl{O}$
and it turns out that only for $3\times 3 $ matrices  \rf{jk20} will lead to a Jordan algebra and it 
belongs to the class of exceptional Jordan algebras. In fact it is the only finite dimensional exceptional simple
Jordan algebra.} 
 consists of the real vector space $H_3(\Dbl{O})$ of  Hermitian $3\times 3$ 
matrices  with octonion entries. Ordinary matrix multiplication is defined on all of  these vector spaces of matrices 
and for two elements $X$ and $Y$ we write $XY$ for this matrix multiplication  (the matrix 
$XY$ does not necessarily belong to the vector space). The vector spaces  become 
Jordan algebras with the following multiplication
 \beq\label{JordanProduct}
X \!\circ\! Y
\ \define \ 
\frac{1}{2} \{ X {\tinyspace},{\tinyspace} Y \}
\ \define \ 
\frac{1}{2} ( X Y + Y X )
\,.
\eeq
For all of these Jordan algebras we can defined a real scalar product by 
\beq\label{jk22}
\la X ,Y \ra  = c_1 \tr X\circ Y,
\eeq
where $c_1$ is a positive constant which can be chosen freely for each algebra.
Relative to this scalar product we can now make an orthogonal decomposition of the 
algebras (viewed as vector spaces)
\bea\label{jk24}
H_n (\Dbl{Q}) &=& \Dbl{R} \; I \oplus \tH_n (\Dbl{Q}),\qquad \Dbl{Q} = \Dbl{R},\,\Dbl{C},\, \Dbl{H},\\
H_3(\Dbl{O}) &=&  \Dbl{R} \; I \oplus \tH_n (\Dbl{O}), \label{jk25}\\
V_n &=&   \Dbl{R} \; I \oplus \tilde{V}_n,\label{jk26}
\eea
where $I$ is the identity matrix and $\tilde{W}$ denotes  a vector space of traceless matrices.
Let us now for each of the algebras choose an orthogonal  basis with respect to the scalar product
and let us denote the basis vectors $E_\mu$. We always assume a vector proportional to 
the unit vector is among the basic vectors and we denote it $E_0$. Further we assume that the 
basis vectors $E_a$, $a \neq 0$ are normalised such that 
\beq\label{jj2}
\tr E_\mu \circ E_\nu = g_{\mu\nu},\qquad \tr E_a \circ E_b= g_{ab} = d_2 \delta_{ab}.
\eeq
One defines the structure constants of the Jordan algebra with respect to this orthogonal basis  as:
\beq\label{jk23}
d_{\kappa \mu \nu } = c_2 \tr E_\kappa \circ (E_\mu \circ E_\nu),
\eeq 
where $c_2$ is also a constant which we can choose freely for each algebra. It is clear that 
$d_{000} = c_2 g_{00}^{3/2}$ and that $d_{0ab} = c_2 g_{00}^{1/2}g_{ab}$, i.e.\ diagonal in the orthogonal 
basis of $\tilde{W}$. The only non-trivial structure constants are the ones involving $E_a,E_b,E_c$
from $\tilde{W}$.  The automorphisms of a Jordan algebra  are the invertible linear mappings of 
the algebra onto itself  which respect the multiplication. The corresponding Lie algebras are isomorphic to the 
derivation algebras of the Jordan algebras. If we define $J = J^\mu E_\mu$ one can show
that the automorphisms leave invariant the forms
\beq\label{jj1}
\tr J^2 = \sum_{\mu,\nu}g_{\mu \nu} J^\mu J^\nu, \qquad 
\tr J^3 = c_2 \sum_{\lambda,\mu,\nu} d_{\lambda \mu\nu} J^\lambda J^\mu J^\nu.
\eeq
The automorphisms leave the identity element invariant and map $\tilde{W}$ onto $\tilde{W}$.  

The automorphism groups of $H_3(\Dbl{Q})$, $\Dbl{Q} =\Dbl{R},\Dbl{C},\Dbl{H},\Dbl{O}$, are 
$SO(3)$, $SU(3)$, ${ U\!sp}(6)$ and $F_4$ respectively, while it is $SO(n)$ for $V_n$.

\subsubsection{Spin factor type Jordan algebras}

The simplest family of Jordan algebras is the one generated by a set of  $2^{[n/2]}$-dimensional 
Hermitian $\g_\m$-matrices, $\mu =0,1,\ldots,n$ which constitute a basis $E_\m$. These 
Jordan algebras are sometimes denoted spin factor type Jordan algebras. The structure constants 
$d_{\m\n\rho}$  are defined by 
\beq
d_{\mu\nu\rho}
\ = \ 
\frac{1}{2^{{\tinyspace}[n/2]}}{\dbltinyspace}
 \tr\!
  \big(
    \gamma_\mu \!\circ\! ( \gamma_\nu \!\circ\! \gamma_\rho )
  \big)
\qquad\hbox{[\,$\mu$, $\nu$,
 $\rho {\negdbltinyspace}=\! 0$, $1$, \ldots, $n$\,]}
\,.
\eeq
They  are symmetric in the indices $\mu$, $\nu$, $\rho$ and we have   
\bea
&&
d_{000} \,=\, 1
\,,
\qquad
d_{0ab} \,=\, d_{a0b} \,=\, d_{ab0} \,=\, \delta_{ab}
\qquad\hbox{[\,$a$, $b {\negdbltinyspace}=\! 1$, \ldots, $n$\,]}
\,,
\nonumber\\
&&
\hbox{otherwise} \,=\, 0
\,.
\eea

\subsubsection{The Hermitian-matrix type Jordan algebras}

Let us now consider the simple Jordan algebras $H_3(\Dbl{Q})$, $\Dbl{Q} = \Dbl{R},\,  \Dbl{C},\,\Dbl{H},\,\Dbl{O}$
(we consider only the $3 \times 3$ matrices since these are the ones related to the $W^{(3)}$ algebras).
We denote an orthogonal basis $E_\mu$ by $\l_\m$, following convention, and we write 
\beq\label{jk30}
  \lambda_0, \;\lambda_a,
\quad \hbox{$a \!=\! 1$, \ldots, $N$},
\eeq
where $N$ denotes the dimension of the vector space $\tH_3(\Dbl{Q})$ and 
$\lambda_0$  is proportional to the three-dimensional unit matrix. Finally the $\lambda_a$ (\,$a \!=\! 1$, \ldots, $N$\,) 
are the three-dimensional Hermitian, 
traceless matrices with entries in $\Dbl{Q}$. 
which satisfy 
\beq\label{SUnAlgebra}
\oh \{\, \lambda_a \,,{\tinyspace} \lambda_b \,\}
\ = \ 
\frac{2}{3}{\dbltinyspace}
\delta_{ab}
\,+\,
\sum_c d_{abc} {\tinyspace} \lambda_c
\qquad\hbox{[\,$a$, $b$, $c {\negdbltinyspace}=\! 1$, \ldots, $N$\,]}
\,.
\eeq
The detailed structures of these matrices are described in Appendix C for the four  $\Dbl{Q}$, while 
the reader is reminded about properties of the $\Dbl{Q}$'s in Appendix B.
The  $d_{\mu\nu\rho}$ in \rf{SUnAlgebra} are the structure constants from   \rf{jk23} with $c_2 = 1/2$:
\beq
d_{\mu\nu\rho}
\ = \ 
\frac{1}{2}{\dbltinyspace}
 \tr\!
  \big(
    \lambda_\mu \!\circ\! ( \lambda_\nu \!\circ\! \lambda_\rho )
  \big)
\qquad\hbox{[\,$\mu$, $\nu$,
 $\rho {\negdbltinyspace}=\! 0$, $1$, \ldots, $N$\,]}
\,.
\eeq
Again, by construction, the structure constants $d_{\mu\nu\rho}$ are  totally symmetric in $\mu$, $\nu$, $\rho$, and 
it is convenient to choose $\lambda_0 = \sqrt{\frac{2}{3}} I$. Then we have 
\beq\label{dHermite3dim}
d_{000} \,=\, \sqrt{\frac{2}{3}}
\,,
\qquad
d_{0ab} \,=\, d_{a0b} \,=\, d_{ab0} \,=\, \sqrt{\frac{2}{3}}\,\delta_{ab}
\qquad\hbox{[\,$a$, $b {\negdbltinyspace}=\! 1$, \ldots, $N$\,]},
\eeq
while the $d_{abc}$ are listed in Appendix C for the various $\tH_3(\Dbl{Q})$. With our choice of $\lambda_0$ we can 
further write
\beq\label{structure}
\lambda_\mu \circ \lambda_\nu = \sum_\rho d_{\mu\nu\rho} \lambda_\rho,\qquad
 \frac{1}{2}\; \tr (\lambda_\mu \circ \lambda_\nu) = \delta_{\mu \nu}
\,.
\eeq

Finally, we have the following dimensions 
$N= \mbox{dim}(\tH_3(\Dbl{Q}))$ of the vector spaces $\tH_3(\Dbl{Q})$:
\beq\label{jk31}
 \mbox{dim}(\tH_3(\Dbl{R})) =  5,\quad
 \mbox{dim}(\tH_3(\Dbl{C})) =  8,\quad
 \mbox{dim}(\tH_3(\Dbl{H})) = 14,\quad
 \mbox{dim}(\tH_3(\Dbl{O})) = 26.
\eeq
We refer to appendix C for details.

\subsection{Classical $W^{(3)}$ algebras and Jordan Algebras}

Consider the free two-dimensional action
\beq\label{jp1}
S_0 = \int dz\,d\bz  \; \prt_z \phi \prt_\bz \phi
\,.
\eeq
The classical theory is invariant under the symmetries generated by the infinitesimal transformations
\beq\label{jp2}
\del \phi = \ep^{(1)}(z) \prt_z \phi  + \ep^{(2)}(z)  \prt_z \phi  \prt_z \phi  + 
\ep^{(3)}(z)  \prt_z \phi  \prt_z \phi  \prt_z \phi  
+ \cdots
\,.
\eeq
The conserved currents (viewing $\bz$ as a formal time) corresponding to these symmetries are 
\beq\label{jp3}
W^{(2)} (z) = \oh  \prt_z \phi  \prt_z \phi ,\quad W^{(3)} (z) = \frac{1}{3}  \prt_z \phi  \prt_z \phi  \prt_z \phi , \; \cdots
\,,
\eeq
where $W^{(2)} (z) = T(z)$ is the holomorphic part of the energy-momentum tensor corresponding 
to the free action \rf{jp1}. Classically there are no normal ordering issues related to the $W^{(n)}(z)$ and 
we are then free to write $W^{(4)}(z) = (W^{(2)}(z))^2$. Again, viewing $\bz$ as a formal time, the momentum
conjugate to $\phi(z,\bz)$ with the action \rf{jp1}, is $\pi =\prt \phi$. If we denote the Poisson bracket
for the system by $[\cdot, \cdot ]_{cl}$,  in general $[W^{(n)}(z),W^{(m)}(z')]_{cl}$ will contain $W^{n+m-2}(z)$,
and thus the algebra generated by $W^{(2)}(z), W^{(3)}(z)$ by $[\cdot,\cdot]_{cl}$ will only close 
if we make the identification  $W^{(4)}(z) = (W^{(2)}(z))^2$, and it thus closes in a non-linear way.
This is contrary to the situation where we only consider $W^{(2)}(z)$. 

The situation becomes less trivial if we consider extended, classical $W^{(3)}$ algebras. Again the simplest way to 
introduce these algebras is by introducing the ``flavour'' generalisation of the action \rf{jp1}
\beq\label{jp4}
 S_0 = \int dz\,d\bz  \; \sum_{a,b} g_{ab}\;\prt_z \phi^a \prt_\bz \phi^a.
 \eeq 
 The classical theory is now invariant under the symmetries generated by the infinitesimal transformations
\beq\label{jp2}
\del \phi^a = \ep^{(1)}(z) \prt_z \phi^a  + \ep^{(2)}(z) \sum_{b,c} d^a_{\;bc}  \prt_z \phi^b  \prt_z \phi^c  + \cdots
\,,
\eeq
where $d_{a_1\cdots a_i}$ are symmetric tensors and indices are lowered  and raised by $g_{ab}$ and  $g^{ab}$
($\sum_{b} g_{ab}g^{bc} = \del_a^c$). Here we will only need a trivial $g_{ab} = \del_{ab}$.
The conserved currents corresponding to these symmetries are 
\beq\label{jp3}
W^{(2)} (z) = \oh  \sum_{a,b} g_{ab}\prt_z \phi^a  \prt_z \phi^b ,\quad 
W^{(3)} (z) = \frac{1}{3}  \sum_{a,b,c} d_{abc}\,\prt_z \phi^a  \prt_z \phi^a  \prt_z \phi^c , \; \cdots
\,.
\eeq
However,  it now becomes non-trivial that the classical currents  $ W^{(2)} (z), W^{(3)} (z)$ 
form a closed algebra with respect to $[\cdot,\cdot]_{cl}$. In fact a necessary and sufficient condition 
for this is that the symmetric tensor $d_{abc}$ satisfies
\beq\label{jp6}
\sum_{e,f} g^{ef}d_{\underline{ab}e}d_{\underline{cd}f} \propto g_{\underline{ab}}g_{\underline{cd}}
\,,
\eeq
where underlined indices are symmetrized. 
It can be shown that the constants defined  by \rf{jk23} satisfy \rf{jp6} (for more relations satisfied 
by the $d_{abc}$ for the Hermitian-matrix like Jordan algebras, see appendix D). Thus 
if the symmetric tensor $d_{abc}$ can be associated with a Jordan algebra, 
the classical currents  $ W^{(2)} (z), W^{(3)} (z)$ form a closed algebra 
with respect to $[\cdot,\cdot]_{cl}$
\cite{romans,hull}..

\subsection{Quantum Jordan and $W^{(3)}$ algebras}

The moment we leave the realm of classical physics, e.g.\ by considering the free quantum field 
theory corresponding to \rf{jp4}, the condition that $d_{abc}$ can be associated with a Jordan algebra
is not  a sufficient condition for quantum operators $ W^{(2)} (z), W^{(3)} (z)$ to 
form a closed algebra, the problem being caused by the need for a normal ordering prescription of operators 
at coinciding points. In fact it is know for the cases of most interest for us that they will not form a
closed algebra \cite{non-closed}. Below, we will describe certain features of this enlarged algebra which we will 
still denote an ``extended'' $W^{(3)}$ algebra. However, one can ask a simpler question:
does the definition \rf{jk20} of a Jordan algebra survive extending $X,Y$ to suitable operators acting in a Hilbert space.

As already mentioned we  want to add ``flavours'' to our CDT string field theory and we 
do that by writing  $\alpha^{\mu}_n$ instead of $\alpha_n$. In this way the $W^{(3)}$ algebra associated
with the CDT string field theory in a natural way becomes an ``extended'' $W^{(3)}$ algebra in the 
way described above. 
  
Let us first define the generalisations of eqs.\ \rf{jx11} and \rf{jx12}
\beq\label{jk45}
J(z)  \ \define \  \sum_\mu E_\mu J^{\mu}(z), \qquad J^{\mu}(z) \,\define\,
\sum_n \frac{\alpha^{\mu}_n}{z^{n+1}},
\qquad [J^\mu(z),J^\nu(\om)] = \frac{\del^{\mu \nu}}{(z-\om)^2}
\,,
\eeq
 \beq\label{WoperatorLaurentExpansion}
{\cal W}^{(3)}(z)
\ = \frac{c_2}{3} \! :\!  \tr{\negdbltinyspace} \big( J(z) \big)^3 \!: \;\; =
\frac{1}{3} \sum_{\mu,\nu,\rho} d_{\mu\nu\rho}
:\!
 J^{\mu}(z)
 J^{\nu}(z)
 J^{\rho}(z)\!:\, ,
\eeq
which promotes 
 $J(z)$ to a quantum operator which lives in the tensor product of the Jordan algebra and the free field Fock space,
 and is used in \rf{WoperatorLaurentExpansion} to define the quantum ${\cal W}^{(3)}(z)$ operator. It is interesting 
to check  if the operators $J(z)$ themselves constitute a Jordan algebra, or if the quantum nature of $J(z)$ provides an 
 obstruction to this,  i.e.\   do $J(z)$'s defined by \rf{jk45} satisfy 
\beq\label{quantJ1}
(J(z)\circ J(\om)) \circ (J(z) \circ  J(z)) = J(z)\circ (J(\om) \circ (J(z) \circ J(z)))).
\eeq
Clearly this relation needs to be regularised since the product of two $J$'s at the same $z$ is singular.
For instance, using point splitting, i.e.\ 
\beq\label{quantJ2}
J^\mu (z) J^{\nu} (z) \to J^{\mu} (z+\ep) J^\nu(z) - \frac{\del^{\mu \nu}}{\ep^2},
\eeq 
one verifies that \rf{quantJ1} is satisfied provided 
\beq\label{quantJ3X}
\sum_\rho (E^\mu \circ E^\nu) \circ (E^\rho\circ E^\rho) = \sum_\rho E^\mu \circ (E^\nu \circ (E^\rho\circ E^\rho)),
\eeq
\beq\label{quantJ3}
\sum_\rho (E^\rho \circ E^\mu) \circ (E^\rho\circ E^\nu) = \sum_\rho E^\rho \circ (E^\mu \circ (E^\rho\circ E^\nu)).
\eeq
While these relations are trivial when the Jordan algebra is associative, they  can also easily be
 proven  in the case $H_3(\Dbl{O})$ using \rf{structure}. 
 We thus conclude that the regularised version of \rf{quantJ1} is true,
 and one can show that it can be written as  
 \beq\label{quantJ4}
 : (J(z)\circ J(\om)) \circ (J(z) \circ  J(z)) : ~ =  ~ :J(z)\circ (J(\om) \circ (J(z) \circ J(z)))):.
\eeq

\vspace{12pt}

With these definitions the ${\cal W}_n$-operators from eq.\ \rf{jx12} are generalised to:
\beq\label{WoperatorMode}
{\cal W}^{(3)}(z) \,\define\,
\sum_n \frac{{\cal W}^{(3)}_n}{z^{n+3}} \qquad
{\cal W}^{(3)}_n
\,=
\frac{1}{3} \sum_{\mu,\nu,\rho} d_{\mu\nu\rho} \!
\sum_{k, l, m}\!
:\! \alpha^{\mu}_k \alpha^{\nu}_l \alpha^{\rho}_m \!:
\delta_{k+l+m,n}
\,.
\eeq
We now define the Hamiltonian as a generalisation of eq.\ \rf{jy3}
\beq\label{jy3a}
\cH_{\rm W}^\star \,=\, - {\cal W}^{(3)}_{-2} = - \mbox{res}_{z=0} {\cal W}^{(3)} (z).
\eeq
Before discussing the physics associated with $\cH_{\rm W}^\star$ we will discuss the algebraic 
structure of the various $W$-operators \rf{WoperatorMode}.

\subsubsection{Spin factor type algebras}

In the case of a spin factor type Jordan algebra, 
the $W$-operator \rf{WoperatorMode} has the following structure 
\bea
{\cal W}^{(3)}_n
\!&=&\!
\frac{1}{3}
\sum_{k, l, m}\!
:\!\alpha^{0}_k \alpha^{0}_l \alpha^{0}_m \!: \delta_{k+l+m,n}
\,+\,
\sum_a
\sum_{k, l, m}\!
:\!\alpha^{0}_k \alpha^{a}_l \alpha^{a}_m \!: \delta_{k+l+m,n}
\nn
&=&\!
W^{(3)}_n
\,+\,
2
\sum_{k,m}
\alpha^{0}_k {\dbltinyspace} \widetilde{W}^{(2)}_m
{\dbltinyspace} \delta_{k+m,n}
\,,
\eea
where 
\bea
&&
W^{(3)}_n
\,\define\,
\frac{1}{3}
\sum_{k, l, m}\!
:\!\alpha^{0}_k \alpha^{0}_l \alpha^{0}_m \!: \delta_{k+l+m,n}
\,,
\label{W3opmode}
\\
&&
\widetilde{W}^{(2)}_n
\,\define\,
\frac{1}{2} \sum_a
\sum_{k,m}\!
:\!\alpha^{a}_k \alpha^{a}_m \!: \delta_{k+m,n}
\,.
\label{Wf2opmode}
\eea
$W^{(3)}_n$ is the $W$-operator of {\it the singlet mode} $\a^0$ while 
$\widetilde{W}^{(2)}_n$ is the Virasoro operator of {\it the flavour modes} $\a^a$, i.e.\ the $\a^\m$ with $\m \neq 0$.
If we introduce the Laurent expansion as usual 
\beq\label{W3opLaurent}
W^{(3)}(z) \,\define\,
\sum_n \frac{W^{(3)}_n}{z^{n+3}}
\,,
\qquad\quad
\widetilde{W}^{(2)}(z) \,\define\,
\sum_n \frac{\widetilde{W}^{(2)}_n}{z^{n+2}}
\, ,
\eeq
we can (as above) express $W^{(3)}(z)$ and $\widetilde{W}^{(2)}(z)$ in terms of the currents $J^\m(z)$ 
as follows
\bea
W^{(3)}(z) \!&=&\!
\frac{1}{3}
:\!
 J^{{\tinyspace}0}(z)
 J^{{\tinyspace}0}(z)
 J^{{\tinyspace}0}(z)
\!:
\,,
\label{W3op}
\\
\widetilde{W}^{(2)}(z) \!&=&\!
\frac{1}{2} \sum_{a}
:\!
 J^{{\tinyspace}a}(z)
 J^{{\tinyspace}a}(z)
\!:
\,.
\label{Wf2op}
\eea
The $W$-operator ${\cal W}^{(3)}(z)$ can finally be written as 
\beq\label{jy16}
{\cal W}^{(3)}(z) \,=\,
W^{(3)}(z)
+
2 {\tinyspace} J^{{\tinyspace}0}(z) {\tinyspace}
 \widetilde{W}^{(2)}(z)
\,.
\eeq

\subsubsection{Hermitian matrix type algebras}

In the case of $3\times 3$ Hermitian-matrix type Jordan algebra, 
the $W$-operator \rf{WoperatorMode} becomes 
\bea
{\cal W}^{(3)}_n
\!&=&\!
\frac{1}{3}
\sum_{k, l, m}\!
d_{000} :\!\alpha^{0}_k \alpha^{0}_l \alpha^{0}_m \!: \delta_{k+l+m,n}
\,+\,
\sum_a
\sum_{k, l, m}\!
d_{0aa} :\!\alpha^{0}_k \alpha^{a}_l \alpha^{a}_m \!: \delta_{k+l+m,n}
\nn
&&\!
+\;
\frac{1}{3} \sum_{a,b,c} d_{abc} \!
\sum_{k, l, m}\!
:\!\alpha^{a}_k \alpha^{b}_l \alpha^{c}_m \!: \delta_{k+l+m,n}
\nn
&=&\!
\sqrt{\frac{2}{3}}{\tinyspace}
W^{(3)}_n
\,+\,
2\sqrt{\frac{2}{3}}
\sum_{k,m}
\alpha^{0}_k {\tinyspace} \widetilde{W}^{(2)}_m
{\dbltinyspace} \delta_{k+m,n}
\,+\,
\widetilde{W}^{(3)}_n
\,,
\eea
where $W^{(3)}_n$ and $\widetilde{W}^{(2)}_m$ are defined in eqs.\ 
\rf{W3opmode}, \rf{Wf2opmode} and where  
\beq\label{Wf3opmode}
\widetilde{W}^{(3)}_n
\,\define\,
\frac{1}{3} \sum_{a,b,c} d_{abc} \!
\sum_{k, l, m}\!
:\!\alpha^{a}_k \alpha^{b}_l \alpha^{c}_m \!: \delta_{k+l+m,n}
\,.
\eeq
$\widetilde{W}^{(3)}_n$ is the $W$-operator of {\it the flavour modes} and will be the operator
which has our primary interest. $W^{(3)}_n$ and $\widetilde{W}^{(2)}_m$ modes had the 
Laurent expansion \rf{W3opLaurent}, and we introduce a corresponding expansion for  
the $W$-operator corresponding to the flavour modes: 
\beq\label{Wf3opLaurent}
\widetilde{W}^{(3)}(z) \,\define\,
\sum_n \frac{\widetilde{W}^{(3)}_n}{z^{n+3}}
\,.
\eeq
We find that $W^{(3)}(z)$ and $\widetilde{W}^{(2)}(z)$ can still be expressed via the currents $J^0(z)$ and
$J^a(z)$ by formulas \rf{W3op} and \rf{Wf2op}, respectively, and that  
$\widetilde{W}^{(3)}(z)$ in analogy is given by 
\bea
\widetilde{W}^{(3)}(z) \!&=&\!
\frac{1}{3} \sum_{a, b, c} d_{abc}
 :\!
  J^{{\tinyspace}a}(z)
  J^{{\tinyspace}b}(z)
  J^{{\tinyspace}c}(z)
 \!:
\,.
\label{Wf3op}
\eea
Finally we find that for the $W$-operator ${\cal W}^{(3)}(z)$ an expression similar to eq.\ \rf{jy16}
for the spin factor  type algebra:
\beq
{\cal W}^{(3)}(z) \,=\,
 \sqrt{\frac{2}{3}}{\tinyspace} 
W^{(3)}(z)
+
2 \sqrt{\frac{2}{3}}{\tinyspace} J^{{\tinyspace}0}(z) {\tinyspace}
 \widetilde{W}^{(2)}(z)
+
\widetilde{W}^{(3)}(z)
\,.
\eeq

\subsubsection{The operator product expansions of the  singlet mode}

First, let us consider the closure of the algebra generated by  $W^{(3)}_n$.
This algebra is well understood. 
The commutation relation of  $W^{(3)}(z)$ operators is 
\bea\label{W3W3commutation}
\big[\, W^{(3)}(z) \,, W^{(3)}(w) \,\big]
\!\!&\sim&\!\!
\frac{4{\tinyspace}W^{(4)}(w)}{(z - w)^2}
+
\frac{2{\tinyspace}\partial W^{(4)}(w)}{z - w}
\nonumber\\
&&\!\!
+\,
\frac{4{\tinyspace}W^{(2)}(w)}{(z - w)^4}
+
\frac{2{\tinyspace}\partial W^{(2)}(w)}{(z - w)^3}
+
\frac{3{\tinyspace}\partial^2 W^{(2)}(w)}{5{\tinyspace}(z - w)^2}
\nonumber\\
&&\!\!
+\,
\frac{2{\tinyspace}\partial^3 W^{(2)}(w)}{15{\tinyspace}(z - w)}
+
\frac{2}{3{\tinyspace}(z - w)^6}
\,,
\qquad
\eea
where
\bea
W^{(2)}(z) \!&\define&\!
\frac{1}{2}{\negdbltinyspace}
:\! J^{{\tinyspace}0}(z) J^{{\tinyspace}0}(z) \!:
\,,
\label{W2op}
\\
W^{(4)}(z) \!&\define&\!
\frac{1}{4}{\negdbltinyspace}
 :\!
   J^{{\tinyspace}0}(z) J^{{\tinyspace}0}(z)
   J^{{\tinyspace}0}(z) J^{{\tinyspace}0}(z)
 \!:
\nonumber\\
&&\!
   +\>
\frac{1}{20}\! :\! \big({\tinyspace}
   2
   J^{{\tinyspace}0}(z) {\tinyspace}
   \partial^2{\negdbltinyspace} J^{{\tinyspace}0}(z)
   -
   3{\tinyspace}
   \partial J^{{\tinyspace}0}(z) {\tinyspace} \partial J^{{\tinyspace}0}(z)
\big) \!:
\,.
\label{W4op}
\eea
The notation \lq\lq$\sim$" means 
the equality up to non-singular terms. 

The commutation relation \rf{W3W3commutation} implies that 
the operators $W^{(2)}(z)$ and $W^{(4)}(z)$ 
should be incorporated in order to close the algebra. 
The commutation relations involving only  $W^{(2)}(z)$  are  closed and one 
obtain the Virasoro algebra with central charge $1$, 
\bea
\big[\, W^{(2)}(z) \,, W^{(2)}(w) \,\big]
\!\!&\sim&\!\!
\frac{2{\tinyspace}W^{(2)}(w)}{(z - w)^2}
+
\frac{\partial W^{(2)}(w)}{z - w}
+
\frac{1}{2{\tinyspace}(z - w)^4}
\,.
\label{W2W2commutation}
\eea
The commutation relations involving  
$W^{(2)}(z)$, $W^{(3)}(z)$ and $W^{(4)}(z)$ are 
\begin{eqnarray}
\big[\, W^{(2)}(z) \,, W^{(3)}(w) \,\big]
\!\!&\sim&\!\!
\frac{3{\tinyspace}W^{(3)}(w)}{(z - w)^2}
+
\frac{\partial W^{(3)}(w)}{z - w}
+
\frac{J^{{\tinyspace}0}(w)}{(z - w)^4}
\,,
\label{W2W3commutation}
\\
\big[\, W^{(2)}(z) \,, W^{(4)}(w) \,\big]
\!\!&\sim&\!\!
\frac{4{\tinyspace}W^{(4)}(w)}{(z - w)^2}
+
\frac{\partial W^{(4)}(w)}{z - w}
+
\frac{21{\tinyspace}W^{(2)}(w)}{5{\tinyspace}(z - w)^4}
\,,
\label{W2W4commutation}
\\
\big[\, W^{(3)}(z) \,, W^{(4)}(w) \,\big]
\!\!&\sim&\!\!
\frac{5{\tinyspace}W^{(5)}(w)}{(z - w)^2}
+
\frac{2{\tinyspace}\partial W^{(5)}(w)}{z - w}
\nonumber\\
&&\!\!
+\,
\frac{54{\tinyspace}W^{(3)}(w)}{5{\tinyspace}(z - w)^4}
+
\frac{18{\tinyspace}\partial W^{(3)}(w)}{5{\tinyspace}(z - w)^3}
+
\frac{27{\tinyspace}\partial^2 W^{(3)}(w)}{35{\tinyspace}(z - w)^2}
\nonumber\\
&&\!\!
+\,
\frac{9{\tinyspace}\partial^3 W^{(3)}(w)}{70{\tinyspace}(z - w)}
+
\frac{2{\tinyspace}J^{{\tinyspace}0}(w)}{(z - w)^6}
\,,
\qquad
\label{W3W4commutation}
\end{eqnarray}
where
\bea
W^{(5)}(z) \!&\define&\!
\frac{1}{5}{\negdbltinyspace}
 :\!
   J^{{\tinyspace}0}(z) J^{{\tinyspace}0}(z) J^{{\tinyspace}0}(z) J^{{\tinyspace}0}(z) J^{{\tinyspace}0}(z)
\nonumber\\
&&\!
   +\>
\frac{1}{7}\! :\! \big({\tinyspace}
   2
   J^{{\tinyspace}0}(z) J^{{\tinyspace}0}(z) {\tinyspace}
   \partial^2{\negdbltinyspace} J^{{\tinyspace}0}(z)
   -
   3
   J^{{\tinyspace}0}(z) {\tinyspace}
   \partial J^{{\tinyspace}0}(z) {\tinyspace} \partial J^{{\tinyspace}0}(z)
\big) \!:
\,.
\label{W5op}
\eea
The closure of algebra requires the operator $W^{(5)}(z)$. The standard 
definition of the $W^{(n)}(z)$ operator is:
\bea\label{jy17}
W^{(n)}(z)
\!&=&\!
\sum_{k=0}^{n-1}
\frac{(-1)^k {\tinyspace}
      ({\tinyspace} [n {\negdbltinyspace}-{\negdbltinyspace} 1]_k )^2}
     {(n {\negdbltinyspace}-{\negdbltinyspace} k){\tinyspace}
      k! {\dbltinyspace}
      [2n {\negdbltinyspace}-{\negdbltinyspace} 2]_k}{\dbltinyspace}
 \partial^{{\tinyspace}k}
\big(\!
 :\!
  e^{-\phi(z)} \partial^{{\tinyspace}n-k} e^{\phi(z)}
 \!:
\!\big)
\qquad\hbox{[\,$n \!=\! 1$, $2$, $3$, \ldots\,]}
\,,
\qquad
\eea
where $[n]_k \define \frac{n!}{(n-k)!}$ and where $ W^{(1)}(z)= \partial \phi(z)$. In our case 
$\phi(z)$ in \rf{jy17} should be identified with $\a^{0}$, and thus $ W^{(1)}(z)= \partial \a^0(z)=J^0(z)$.

With these definitions 
the commutator of $W^{(m)}(z)$ and $W^{(n)}(z)$ will include $W^{(m+n-2)}(z)$, 
so an  infinite number of $W^{(n)}(z)$ [\,$n \!=\! 4$, $5$, \ldots\,] will appear in order 
to close the $W^{(3)}(z)$ algebra. Further, the current operator 
$J^{{\tinyspace}0}(z)$ appears explicitly in the algebra. 
This kind of closed algebra is denoted  a $W^{(1+\infty)}$ algebra, the ``1'' refering to $J^{0}=W^{(1)}$ and 
``$\infty$'' to the $ W^{(n)}$, $n=2,3,\ldots$.

\subsubsection{The operator product expansions with flavour modes}

Next, let us consider the closure of the ``flavour'' algebra generated by $\widetilde{W}^{(3)}_n$. 
In order to derive the formulas below we need various identities for the $d_{abc}$ symbols.
They are listen in Appendix D.
The commutation relation of two  $\widetilde{W}^{(3)}(z)$ operators is 
\bea\label{Wf3Wf3commutation}
\big[\, \widetilde{W}^{(3)}(z) \,, \widetilde{W}^{(3)}(w) \,\big]
\!\!&\sim&\!\!
\frac{4{\tinyspace}\widetilde{W}^{(4)}(w)}{(z - w)^2}
+
\frac{2{\tinyspace}\partial \widetilde{W}^{(4)}(w)}{z - w}
\nonumber\\
&&\!\!
+\,
\frac{4{\tinyspace}C_{2}{\tinyspace}
      \widetilde{W}^{(2)}(w)}
     {(z - w)^4}
+
\frac{2{\tinyspace}C_{2}{\tinyspace}
      \partial \widetilde{W}^{(2)}(w)}
     {(z - w)^3}
+
\frac{3{\tinyspace}C_{2}{\tinyspace}
      \partial^2 \widetilde{W}^{(2)}(w)}{5{\tinyspace}(z - w)^2}
\nonumber\\
&&\!\!
+\,
\frac{2{\tinyspace}C_{2}{\tinyspace}
      \partial^3 \widetilde{W}^{(2)}(w)}{15{\tinyspace}(z - w)}
+
\frac{2{\tinyspace}N C_{2}}
     {3{\tinyspace}(z - w)^6}
\,,
\qquad
\eea
where $\widetilde{W}^{(2)}(z)$ is defined by \rf{Wf2op} and where  
\bea
\widetilde{W}^{(4)}(z) \!&\define&\!
\frac{1}{12} \sum_{a, b}\! 
 :\!
   J^{{\tinyspace}a}(z)
   J^{{\tinyspace}a}(z)
   J^{{\tinyspace}b}(z)
   J^{{\tinyspace}b}(z)
 \!:
\nonumber\\
&&\!
   +\>
\frac{C_{2}}{20} \sum_a\!
 :\! \big({\tinyspace}
   2
   J^{{\tinyspace}a}(z) {\tinyspace}
   \partial^2{\negdbltinyspace} J^{{\tinyspace}a}(z)
   -
   3{\tinyspace}
   \partial J^{{\tinyspace}a}(z) {\tinyspace}
   \partial J^{{\tinyspace}a}(z)
\big) \!:
\,,
\label{Wf4op}
\eea
and
\beq\label{jy18}
C_{2} {\tinyspace} \delta_{ab}
 {\negtinyspace}={\negdbltinyspace}
 \sum_{c, d} d_{acd} {\tinyspace} d_{bcd},\qquad C_{2} = 
\frac{N{\negdbltinyspace}+{\negtinyspace}2}
     {6}
\,.
\eeq
In \rf{jy18} $N$ denotes the number of flavours, and is given by \rf{jk31} for the 
various $3 \times 3$ matrices with entries in $\Dbl{Q} = \Dbl{R}, \Dbl{C}, \Dbl{H}$ and $\Dbl{O}$, respectively.


The commutation relation \rf{Wf3Wf3commutation} leads to the fact that 
the operators $\widetilde{W}^{(2)}(z)$ and $\widetilde{W}^{(4)}(z)$ 
should be incorporated in order to close the algebra. 
The commutation relation of $\widetilde{W}^{(2)}(z)$ is closed and is 
the Virasoro algebra with the central charge $N$, 
\bea
\big[\, \widetilde{W}^{(2)}(z) \,, \widetilde{W}^{(2)}(w) \,\big]
\!\!&\sim&\!\!
\frac{2{\tinyspace}\widetilde{W}^{(2)}(w)}{(z - w)^2}
+
\frac{\partial \widetilde{W}^{(2)}(w)}{z - w}
+
\frac{N}{2{\tinyspace}(z - w)^4}
\,.
\label{Wf2Wf2commutation}
\eea
The commutation relations of 
$\widetilde{W}^{(2)}(z)$, 
$\widetilde{W}^{(3)}(z)$ and 
$\widetilde{W}^{(4)}(z)$ are 
\begin{eqnarray}
\big[\, \widetilde{W}^{(2)}(z) \,, \widetilde{W}^{(3)}(w) \,\big]
\!\!&\sim&\!\!
\frac{3{\tinyspace}\widetilde{W}^{(3)}(w)}{(z - w)^2}
+
\frac{\partial \widetilde{W}^{(3)}(w)}{z - w}
\,,
\label{Wf2Wf3commutation}
\\
\big[\, \widetilde{W}^{(2)}(z) \,, \widetilde{W}^{(4)}(w) \,\big]
\!\!&\sim&\!\!
\frac{4{\tinyspace}\widetilde{W}^{(4)}(w)}{(z - w)^2}
+
\frac{\partial \widetilde{W}^{(4)}(w)}{z - w}
+
\frac{16{\tinyspace}C_{2}\widetilde{W}^{(2)}(w)}
     {5{\tinyspace}(z - w)^4}
\,,
\label{Wf2Wf4commutation}
\\
\big[\, \widetilde{W}^{(3)}(z) \,, \widetilde{W}^{(4)}(w) \,\big]
\!\!&\sim&\!\!
\frac{5{\tinyspace}\widetilde{W}^{(5)}(w)}{(z - w)^2}
+
\frac{2{\tinyspace}\partial \widetilde{W}^{(5)}(w)}
     {z - w}
+
\frac{\widetilde{W}^{(5,1)}(w)}
     {z - w}
\nonumber\\
&&\!\!
+\,
\frac{54{\tinyspace}C_{3}{\tinyspace}\widetilde{W}^{(3)}(w)}
     {5{\tinyspace}(z - w)^4}
+
\frac{18{\tinyspace}C_{3}{\tinyspace}\partial \widetilde{W}^{(3)}(w)}
     {5{\tinyspace}(z - w)^3}
+
\frac{27{\tinyspace}C_{3}{\tinyspace}\partial^2 \widetilde{W}^{(3)}(w)}
     {35{\tinyspace}(z - w)^2}
\nonumber\\
&&\!\!
+\,
\frac{9{\tinyspace}C_{3}{\tinyspace}\partial^3 \widetilde{W}^{(3)}(w)}
     {70{\tinyspace}(z - w)}
\,,
\qquad
\label{Wf3Wf4commutation}
\end{eqnarray}
where
\bea
\widetilde{W}^{(5)}(z) \!&\define&\!
\frac{1}{15} \sum_{a, b, c, k}\! d_{abc} \!
 :\!
   J^{{\tinyspace}a}(z)
   J^{{\tinyspace}b}(z)
   J^{{\tinyspace}c}(z)
   J^{{\tinyspace}k}(z)
   J^{{\tinyspace}k}(z)
 \!:
\nonumber\\
&&\!\hspace{-0pt}
   +\>
\frac{N{\negdbltinyspace}+{\negdbltinyspace}4}{105} \sum_{a, b, c} d_{abc} \!
 :\! \big({\tinyspace}
   2
   J^{{\tinyspace}a}(z) J^{{\tinyspace}b}(z) {\tinyspace}
   \partial^2{\negdbltinyspace} J^{{\tinyspace}c}(z)
   -
   3
   J^{{\tinyspace}a}(z) {\tinyspace}
   \partial J^{{\tinyspace}b}(z) {\tinyspace}
   \partial J^{{\tinyspace}c}(z)
\big) \!:
\,,
\qquad
\\
\widetilde{W}^{(5,1)}(z) \!&\define&\!
\frac{4}{15} \sum_{a, b, c, k}\! d_{abc} \!
 :\!
   J^{{\tinyspace}a}(z)
   J^{{\tinyspace}b}(z)
   \partial J^{{\tinyspace}c}(z)
   J^{{\tinyspace}k}(z)
   J^{{\tinyspace}k}(z)
\nn&&\!
\phantom{%
\frac{1}{15} \sum_{a, b, c, k}\! d_{abc} \!
 :\!
}
   -
   J^{{\tinyspace}a}(z)
   J^{{\tinyspace}b}(z)
   J^{{\tinyspace}c}(z)
   J^{{\tinyspace}k}(z)
   \partial J^{{\tinyspace}k}(z)
 \!:
\nonumber\\
&&\!\hspace{-0pt}
   +\>
\frac{3{\tinyspace}N{\negdbltinyspace}+{\negdbltinyspace}2}{360}
 \sum_{a, b, c} d_{abc} \!
 :\! \big({\tinyspace}
   6{\tinyspace}
   J^{{\tinyspace}a}(z) {\tinyspace}
   \partial J^{{\tinyspace}b}(z) {\tinyspace}
   \partial^2{\negdbltinyspace} J^{{\tinyspace}c}(z)
   -
   J^{{\tinyspace}a}(z) {\tinyspace}
   J^{{\tinyspace}b}(z) {\tinyspace}
   \partial^3{\negdbltinyspace} J^{{\tinyspace}c}(z)
\nn&&\!\hspace{-0pt}
\phantom{%
   +\>
\frac{3{\tinyspace}N{\negdbltinyspace}+{\negdbltinyspace}2}{360}
 \sum_{a, b, c} d_{abc} \!
 :\! \big({\tinyspace}
}
   -
   6{\tinyspace}
   \partial J^{{\tinyspace}a}(z) {\tinyspace}
   \partial J^{{\tinyspace}b}(z) {\tinyspace}
   \partial J^{{\tinyspace}c}(z)
\big) \!:
\,,
\qquad
\label{Wf5op}
\eea
and
\beq
C_{3} = 
\frac{3N{\negdbltinyspace}+{\negdbltinyspace}26}
     {108}
\,.
\eeq
The closure of the algebra requires the operators 
$\widetilde{W}^{(5)}(z)$ and $\widetilde{W}^{(5,1)}(z)$. 
In general, 
the commutator of $\widetilde{W}^{(m)}(z)$ and $\widetilde{W}^{(n)}(z)$ 
includes $\widetilde{W}^{(m+n-2)}(z)$, 
so an infinite number of $\widetilde{W}^{(n)}$ [\,$n \!=\! 4$, $5$, \ldots\,] 
appears in order to close the algebra. Note however that the  
the current operator $J^{{\tinyspace}a}(z)$ 
does not appear explicitly in this algebra because $\tr \lambda_a \!=\! 0$. 
Therefore, this closed algebra is a kind of $W^{(\infty)}$ algebra rather than  a $W^{(1+\infty)}$ algebra.

\section{Several concrete models}\label{sec4}
\setcounter{equation}{0}

In the case of the CDT with Jordan algebra, 
we assume nonzero expectation values for 
$\alpha^\mu_{0}$, $\alpha^\mu_{-1}$, and $\alpha^\mu_{-3}$, 
in the same way as we did for the simple non-extended CDT model. 
We denote these  
$\nu^\mu$, $\lambda^\mu_1$, $\lambda^\mu_3$ 
and, as before,  they are are related to  
$\omega^\mu$, $\mu^\mu$, $\sigma^\mu$, $g$ and $\GG$ by
\begin{eqnarray}\label{JordanCDTparameters}
\nu^\mu \,=\, \frac{\omega^\mu}{\sqrt{\GG}}
\,,
\qquad
\lambda^\mu_1 \,=\, -\,\frac{\mu^\mu}{2 g \sqrt{\GG}}
\,,
\qquad
\lambda^\mu_3 \,=\, \frac{\sigma^\mu}{6 g \sqrt{\GG}}
\,.
\end{eqnarray}

Then, the Hamiltonian $\cH_{\rm W}$ becomes
\begin{equation}
\cH_{\rm W}^\star \,=\, - {W}^{(3)}_{-2}
\,,
\end{equation}
\begin{eqnarray}\label{JordanWalgebraHamiltonianAfterSSB}
g \sqrt{\GG}\, \cH_{\rm W}
\!&=&\!
\cH
\,+\,
 \frac{1}{\GG}
 \sum_{\mu,\nu,\rho} d_{\mu\nu\rho} {\dbltinyspace}
  \sigma^\mu
\bigg(\!\!
  - \frac{\sigma^\nu}{4 g} {\tinyspace} \phi^{\rho\dagger}_4
  + \frac{\mu^\nu}{2 g} {\tinyspace} \phi^{\rho\dagger}_2
  - \omega^\nu \phi^{\rho\dagger}_1
\!\;\bigg)
\nonumber\\&&\!\phantom{%
\cH
}
\,-
 \frac{1}{4 g {\tinyspace} \GG}
 \sum_{\mu,\nu,\rho} d_{\mu\nu\rho} {\dbltinyspace}
  \mu^\mu \mu^\nu \omega^\rho
\,,
\end{eqnarray}
where
\begin{eqnarray}\label{GeneralJordanHamiltonian_modeexpansion}
\cH
\ &=& \ 
  \sum_{\mu,\nu,\rho} d_{\mu\nu\rho} {\dbltinyspace}
    \omega^\rho \big(
      \mu^\mu {\tinyspace} \phi^\nu_1
    - 2 g {\tinyspace} \omega^\mu \!\> \phi^\nu_2
    - g {\tinyspace} \GG {\tinyspace} \phi^\mu_1 \phi^\nu_1
    {\tinyspace}\big)
\nonumber\\&& \ 
-\> 
  \sum_{\mu,\nu,\rho} d_{\mu\nu\rho} {\dbltinyspace}
    \sigma^\rho \sum_{l=1}^\infty \phi^{\mu\dagger}_{l+1} l \phi^\nu_l
+
  \sum_{\mu,\nu,\rho} d_{\mu\nu\rho} {\dbltinyspace}
    \mu^\rho \sum_{l=2}^\infty \phi_{l-1}^{\mu\dagger} l \phi^\nu_l
\nonumber\\&& \ 
-\> 2 g
  \sum_{\mu,\nu,\rho} d_{\mu\nu\rho} {\dbltinyspace}
    \omega^\rho \sum_{l=3}^\infty \phi^{\mu\dagger}_{l-2} l \phi^\nu_l
\nonumber\\&& \ 
-\> g
  \sum_{\mu,\nu,\rho} d_{\mu\nu\rho}
    \sum_{l=4}^\infty \sum_{n=1}^{l-3}
      \phi^{\mu\dagger}_n \phi^{\nu\dagger}_{l-n-2}
      l \phi^\rho_l
\nonumber\\&& \ 
-\> g 
  \sum_{\mu,\nu,\rho} d_{\mu\nu\rho}
    \sum_{l=1}^\infty \sum_{m=\max(3-l,1)}^\infty
      \phi^{\rho\dagger}_{m+l-2}
      m \phi^\mu_m l \phi^\nu_l
\,.
\end{eqnarray}

Now, let us consider only the kinetic terms. These will determine the possibilities of 
propagation from a pre-geometric state to a universe of macroscopic size. 
We first neglect the term which vanishes for $g \!=\! 0$.
\begin{equation}\label{W3_KineticHamiltonian}
{\cal H}_{\rm kin}\Big|_{g=0}
\,=\,
-
  \sum_{\mu,\nu,\rho} d_{\mu\nu\rho} {\dbltinyspace}
    \sigma^\rho
    \sum_{l=1}^\infty \phi^{\mu\dagger}_{l+1}\!\; l \phi^\nu_l
+
  \sum_{\mu,\nu,\rho} d_{\mu\nu\rho} {\dbltinyspace}
    \mu^\rho
    \sum_{l=2}^\infty
      \phi^{\mu\dagger}_{l-1}\!\; l \phi^\nu_l.
\end{equation}
The notation ${\cal H}_{\rm kin}\big|_{g=0}$ is a little bit of a misnormer since it is 
the first term which is the real kinetic term since it, when translated to $L$ variables, 
contains the second derivative of $L$, while the last term is a ``mass'' or cosmological term.
 
In the following we will only consider the extensions of CDT based on 
the Jordan algebras $H_3(\Dbl{Q})$, $\Dbl{Q} =\Dbl{R},\Dbl{C},\Dbl{H},\Dbl{O}$, with main 
emphasis on $H_3(\Dbl{O})$.
As explained in Appendix B and C  it is then natural to divide the indices $\mu$ into four groups.
\begin{eqnarray}
&&
\{ \mu \}
\,=\,
\{ 0,\, a \}
\qquad
\{ a \}
\,=\,
\{ 8,\, 3,\, i,\, I,\, \tilde{I} {\dbltinyspace}\}
\nonumber\\
&&
\{ i \} =
\{ 1,\, 2,\, 2',\, 2'',\, \ldots {\tinyspace}\}
\qquad
\{ I {\dbltinyspace}\} =
\{ 4,\, 5,\, 5',\, 5'',\, \ldots {\tinyspace}\}
\qquad
\{ \tilde{I} {\dbltinyspace}\} =
\{ 6,\, 7,\, 7',\, 7'',\, \ldots {\tinyspace}\}
\nonumber\\
&&
\end{eqnarray}
Then, as explained in Appendix C, the non-zero coefficients $d_{abc}$ are classified as 
\begin{eqnarray}
d_{888},\,
d_{833},\,
d_{8ii},\,
d_{8II},\,
d_{8\tilde{I}\tilde{I}},\,
d_{3II},\,
d_{3\tilde{I}\tilde{I}},\,
d_{iI\tilde{I}}.
\end{eqnarray}
The rest of the  $d_{abc}$ are all zero.

\subsection{Hermitian-matrix $0$-$8$-$3$ type}

This model has the following vacumm expectation values:
\begin{eqnarray}
&&
\sigma^0 = 1\,,
\qquad
\sigma^\mu =0\,,
\quad
\mu \neq 0
\,,
\label{jk2}
\\
&&
\mu^0 = \m_0
\,,
\quad
\mu^8 = \sqrt{3} {\dbltinyspace} \mu
\,,
\quad
\mu^3 = \sqrt{3} {\dbltinyspace} \mu'
\,,
\qquad
\mu^\mu = 0
\,,
\quad
\mu^\mu \neq 0,3,8
\,.
\label{jk3}
\end{eqnarray}
This choice is actually the most general assignment one can make for $\mu^\mu$ when 
we have made the choice \rf{jk2}.
Recall from $SU(3)$ that the matrices of form $\mu' \lambda_3 +\mu \lambda_8$ span a Cartan subalgebra
of the  Lie algebra $su(3)$ of $SU(3)$, i.e.\ the traceless Hermitian $3 \times 3$ matrices, and further that
any element in $su(3)$ can be mapped to this Cartan subalgebra by an automorphism. The same 
is true for the Jordan algebras $H_3(\Dbl{Q})$, $\Dbl{Q} =\Dbl{R},\Dbl{C},\Dbl{H},\Dbl{O}$ where the 
automorphism groups are  $SO(3)$, $SU(3)$, $U \!sp(6)$ and $F_4$ respectively. The automophism group
leaves the identity matrix invariant, and a traceless Hermitian matrix can be rotated to belong to the subspace
spanned by $\lambda_3$ and $\lambda_8$.

\begin{eqnarray}\label{W3_KineticHamiltonianZeroType}
{\cal H}_{\rm kin}\Big|_{g=0}
\!&=&\!
-\,
 \sum_\mu  d_{0\mu\mu} {\dbltinyspace}
    \sigma^0
    \sum_{l=1}^\infty \phi^{\mu\dagger}_{l+1}\!\; l \phi^\mu_l + 
 \sum_\mu  d_{0\mu\mu} {\dbltinyspace}
    \mu^0
    \sum_{l=1}^\infty \phi^{\mu\dagger}_{l-1}\!\; l \phi^\mu_l    
 \nonumber\\&&\!
+
  \sum_{a} d_{8aa} {\dbltinyspace}
    \mu^8
    \sum_{l=1}^\infty \phi^{a\dagger}_{l-1}\!\; l \phi^a_l
+
  \sum_{I} d_{3II} {\dbltinyspace}
    \mu^3
    \sum_{l=1}^\infty \phi^{I\dagger}_{l-1}\!\; l \phi^I_l
+
  \sum_{\tilde{I}} d_{3\tilde{I}\tilde{I}} {\dbltinyspace}
    \mu^3
    \sum_{l=1}^\infty \phi^{\tilde{I}\dagger}_{l-1}\!\; l \phi^{\tilde{I}}_l
\nonumber\\&&\!    
+\,
  d_{808} {\dbltinyspace}
    \mu^8
    \sum_{l=1}^\infty \phi^{0\dagger}_{l-1}\!\; l \phi^8_l
+
  d_{880} {\dbltinyspace}
    \mu^8
    \sum_{l=1}^\infty \phi^{8\dagger}_{l-1}\!\; l \phi^0_l
 \nonumber\\&&\!    
+\,
  d_{303} {\dbltinyspace}
    \mu^3
    \sum_{l=1}^\infty \phi^{0\dagger}_{l-1}\!\; l \phi^3_l
+
  d_{330} {\dbltinyspace}
    \mu^3
    \sum_{l=1}^\infty \phi^{3\dagger}_{l-1}\!\; l \phi^0_l   
\nonumber\\&&\!
    +\,
  d_{338} {\dbltinyspace}
    \mu^3
    \sum_{l=1}^\infty \phi^{3\dagger}_{l-1}\!\; l \phi^8_l
+
  d_{383} {\dbltinyspace}
    \mu^3
    \sum_{l=1}^\infty \phi^{8\dagger}_{l-1}\!\; l \phi^3_l 
%
\nonumber\\
\!&=&\!
{\cal H}_{\rm kin}^{(083)}
\,+\,
{\cal H}_{\rm kin}^{(12)}
\,+\,
{\cal H}_{\rm kin}^{(45)}
\,+\,
{\cal H}_{\rm kin}^{(67)}
\,,
\end{eqnarray}
where (using the notation $d_{0\mu\nu} = \sqrt{\frac{2}{3}}\, \delta_{\mu \nu}  = \kappa \, \delta_{\mu \nu}$)
\begin{eqnarray}
{\cal H}_{\rm kin}^{(083)}
\!&=&\!
-\,
  \kappa \sum_{l=1}^\infty \phi^{0\dagger}_{l+1}\!\; l \phi^0_l
-
  \kappa \sum_{l=1}^\infty \phi^{8\dagger}_{l+1}\!\; l \phi^8_l 
 -
  \kappa  \sum_{l=1}^\infty \phi^{3\dagger}_{l+1}\!\; l \phi^3_l
\nonumber\\&&\!
+\,
  \kappa\, \mu_0  \sum_{l=1}^\infty \phi^{0\dagger}_{l-1}\!\; l \phi^0_l
+\,
  (\kappa \mu_0-\mu)
    \sum_{l=1}^\infty \phi^{8\dagger}_{l-1}\!\; l \phi^8_l
+ (\kappa \mu_0+\mu)
    \sum_{l=1}^\infty \phi^{3\dagger}_{l-1}\!\; l \phi^3_l   
 \nonumber\\&&\!   
+\,
  \kappa \sqrt{3} {\dbltinyspace}
  \mu
    \sum_{l=1}^\infty \phi^{0\dagger}_{l-1}\!\; l \phi^8_l
+\,
  \kappa \sqrt{3} {\dbltinyspace}
  \mu
    \sum_{l=1}^\infty \phi^{8\dagger}_{l-1}\!\; l \phi^0_l
 \nonumber\\&&\!   
+
  \kappa \sqrt{3} {\dbltinyspace}
  \mu'
    \sum_{l=1}^\infty \phi^{0\dagger}_{l-1}\!\; l \phi^3_l
+
  \kappa  \sqrt{3} {\dbltinyspace}
  \mu'
    \sum_{l=1}^\infty \phi^{3\dagger}_{l-1}\!\; l \phi^0_l
\nonumber\\&&\! 
     +\,
  \mu' \sum_{l=1}^\infty \phi^{3\dagger}_{l-1}\!\; l \phi^8_l
+
  \mu'  \sum_{l=1}^\infty \phi^{8\dagger}_{l-1}\!\; l \phi^3_l
\,,
\label{h083}%
\\
{\cal H}_{\rm kin}^{(12)}
\!&=&\!
-\,
 \kappa \sum_{i}
    \sum_{l=1}^\infty \phi^{i\dagger}_{l+1}\!\; l \phi^i_l
+
 (\kappa \mu_0+ \mu)
  \sum_{i}
    \sum_{l=1}^\infty \phi^{i\dagger}_{l-1}\!\; l \phi^i_l
\,,
\label{h12}%
\\
{\cal H}_{\rm kin}^{(45)}
\!&=&\!
-\,
 \kappa \sum_{I}
    \sum_{l=1}^\infty \phi^{I\dagger}_{l+1}\!\; l \phi^I_l
+
  \Big(\kappa \mu_0- \frac{\mu- \sqrt{3} \mu'}{2}\Big)
  \sum_{I}
    \sum_{l=1}^\infty \phi^{I\dagger}_{l-1}\!\; l \phi^I_l
\,,
\label{h45}%
\\
{\cal H}_{\rm kin}^{(67)}
\!&=&\!
-\,
 \kappa \sum_{\tilde{I}}
    \sum_{l=1}^\infty \phi^{\tilde{I}\dagger}_{l+1}\!\; l \phi^{\tilde{I}}_l
+
 \Big(\kappa \mu_0- \frac{\mu+\sqrt{3} \mu'}{2}\Big)
  \sum_{\tilde{I}}
    \sum_{l=1}^\infty \phi^{\tilde{I}\dagger}_{l-1}\!\; l \phi^{\tilde{I}}_l
\,.
\label{h67}%
\end{eqnarray}
Introducing new fields
\bea\label{CCsingletfield}
\hspace{-24pt}
\phi^{[0]\dagger}
\!\!&\define&\!\!
-\,\frac{1}{\sqrt{3}}\,\phi^{0\dagger}_l
+ \sqrt{\frac{2}{3}}\,\phi^{8\dagger}_l
\,,
\quad\hspace{54pt}
\phi^{[0]}
\,\define\,
-\,\frac{1}{\sqrt{3}}\,\phi^0_l
+ \sqrt{\frac{2}{3}}\,\phi^8_l
\,,
\nonumber\\
\hspace{-24pt}
\phi^{[\pm]\dagger}
\!\!&\define&\!\!
\pm\,\frac{1}{\sqrt{3}}\,\phi^{0\dagger}_l
\,\pm\,\frac{1}{\sqrt{6}}\,\phi^{8\dagger}_l
+ \frac{1}{\sqrt{2}}\,\phi^{3\dagger}_l
\,,
\quad
\phi^{[\pm]}
\,\define\,
\pm\,\frac{1}{\sqrt{3}}\,\phi^0_l
\,\pm\,\frac{1}{\sqrt{6}}\,\phi^8_l
+ \frac{1}{\sqrt{2}}\,\phi^3_l
\,,
\eea
the Hamiltonian \rf{h083} becomes
\bea\label{h083diag}
{\cal H}_{\rm kin}^{(083)}
\!&=&\!
-\,
 \kappa \sum_{l=1}^\infty \phi^{[0]\dagger}_{l+1}\!\; l \phi^{[0]}_l
+
 (\kappa \mu_0 - 2\mu)
    \sum_{l=1}^\infty \phi^{[0]\dagger}_{l-1}\!\; l \phi^{[0]}_l
 \nonumber\\&&\!
-\,
 \kappa \sum_{l=1}^\infty \phi^{[+]\dagger}_{l+1}\!\; l \phi^{[+]}_l
+
 (\kappa \mu_0 + \mu + \sqrt{3}\mu'\,)
    \sum_{l=1}^\infty \phi^{[+]\dagger}_{l-1}\!\; l \phi^{[+]}_l
 \nonumber\\&&\!
-\,
 \kappa \sum_{l=1}^\infty \phi^{[-]\dagger}_{l+1}\!\; l \phi^{[-]}_l
+
 (\kappa \mu_0 + \mu - \sqrt{3}\mu'\,)
    \sum_{l=1}^\infty \phi^{[-]\dagger}_{l-1}\!\; l \phi^{[-]}_l
\,.
\eea
Therefore, the $0$-$8$-$3$ system consists of three singlets 
with cosmological constants
\beq\label{CCsinglet}
\mu^{[0]}
  \,=\, \kappa \mu_0 - 2\mu
\,,
\qquad
\mu^{[\pm]}
  \,=\, \kappa \mu_0 + \mu \pm \sqrt{3}\mu'
\,,
\eeq
respectively.
On the other hand, for $H_3(\Dbl{O})$
the $1$-$2$, $4$-$5$, $6$-$7$ systems consist of three {\it octets}\footnote{ For $H_3(\Dbl{R})$ the octets are replaced by singlets, for $H_3(\Dbl{C})$ by doublets and for $H_3(\Dbl{H})$ by quartets (see the lists provided in Appendix C).} 
with cosmological constants
\beq\label{CCoctet}
\mu^{\prime[0]}
  \,=\, \kappa \mu_0 + \mu
\,,
\qquad
\mu^{\prime[\pm]}
  \,=\, \kappa \mu_0 + \frac{-{\dbltinyspace}\mu \pm \sqrt{3}\mu'}{2}
\,,
\eeq
respectively.
Since these cosmological constants satisfy the relations 
\beq
\mu^{\prime[0]} \,=\, \frac{\mu^{[+]} + \mu^{[-]}}{2}
\,,
\qquad
\mu^{\prime[\pm]} \,=\, \frac{\mu^{[0]} + \mu^{[\pm]}}{2}
\,,
\eeq
the descending order of cosmological constants is 
singlet--octet--singlet--octet--octet--singlet,  or the reflected order singlet--octet--octet--singlet--octet--singlet. 

For sufficient large $\mu_0$ 
all cosmological constants are positive,
and the spatial extension of a universe of any  flavour is  
of the order $1/\sqrt{\mu_0}$. 
The spaces created by these modes are thus 
finite (of ``Planckian size''). 
Similarly for sufficiently negative $\mu_0$ 
all spaces will expand to infinity 
in a time of order $1/\sqrt{-\mu_0}$. 
However, 
if $\mu_0$ is not  dominant  a number of scenarios can occur. 
In the case of $H_3(\Dbl{O})$ the possible number of infinite-expanding spaces are 
0, 1, 9, 10, 17, 18, 26, 27. 
Applying the same symmetry breaking pattern 
to the Jordan algebras $H_3(\Dbl{Q})$, 
$\Dbl{Q} =\Dbl{R}, \Dbl{C}, \Dbl{H}$, 
the octets are replaced by 
singlet, doublet, quartet, respectively, as mentioned above. 
Then the possible number of infinite-expanding spaces is 
0, 1, 2, 3, 4,  5,  6 for $\Dbl{R}$, 
0, 1, 3, 4, 5,  6,  8,  9 for $\Dbl{C}$ and  
0, 1, 5, 6, 9, 10, 14, 15 for $\Dbl{H}$. 
It should be noted that the so-called knitting mechanism, described below in sec.\ \ref{sec5}, 
does not work for any $H_3(\Dbl{Q})$ model if it  has 
the maximum possible number of infinite-expanding spaces. The reason is that the knitting mechanism requires
small wormholes.  

In the case of 
$\mu'\!=\!0$ or $\mu'\!=\!\pm\sqrt{3}\mu$, 
the discussion becomes particular simple. 
One interesting scenario is the following: 
For $\mu_0 \!=\! \mu'\!=\!0$ and $\mu \!<\! 0$ 
it is seen that $4$-$5$ and $6$-$7$ systems 
\rf{h45} and \rf{h67} have the same positive mass, and in
the case of  $H_3(\Dbl{O})$ we have 16 compact space dimensions, 
like in the heterotic string model. 
At the same time the $1$-$2$ system \rf{h12} has a negative mass and will expand
to infinity in a finite time of order $1/\sqrt{-\mu}$.
Finally 
the masses of $0$-$8$-$3$ system \rf{h083diag} will  in this case have  
one positive and two negative eigenvalues. 
A similar situation occurs 
for $\mu_0 \!=\! 0$, $\mu'\!=\!\pm\sqrt{3}\mu$ and $\mu \!>\! 0$. 
We thus have a scenario where we have the potential for creating 
$8+2$ extended spatial directions and $16+1$ compact, 
``Planckian size'', spatial directions. 
Including time $T$, 
this scenario thus leaves us with 11 extended spacetime dimensions, 
which  coincides with the critical dimensions $10+1$ of the supermembrane. 
For other values of $\mu'$, another interesting scenario is possible, where 
we have $8+1$ extended spatial directions and $16+2$ compact spatial dimensions.
Including time $T$, this leads to the extended spacetime dimension 10 which coincides with 
the critical dimensions of the superstring.  
In both cases the $+1$ and $+2$ compact spaces will play the  role of wormholes 
used in the knitting mechanism to be described in Sec.\ \ref{sec5}. Note also that the same symmetry breaking pattern in the 
case of Jordan algebras $H_3(\Dbl{Q})$, $\Dbl{Q} =\Dbl{R},\Dbl{C},\Dbl{H}$
results in  extended spacetime dimensions 3,4 and 6, i.e.\ the spacetime dimensions where one can define 
classical superstring theories.

Finally note that with the new fields \rf{CCsingletfield}, 
the interaction Hamiltonian for $0$-$8$-$3$ system becomes 
\bea
{\cal H}_{\rm int}^{(083)}
\!\!&=&\!\!
- \sqrt{2}\>\!g
    \sum_{l=4}^\infty \sum_{n=1}^{l-3}
     \big(
      \phi^{[+]\dagger}_n \phi^{[+]\dagger}_{l-n-2}
      l \phi^{[+]}_l
      +
      \phi^{[-]\dagger}_n \phi^{[-]\dagger}_{l-n-2}
      l \phi^{[-]}_l
      -
      \phi^{[0]\dagger}_n \phi^{[0]\dagger}_{l-n-2}
      l \phi^{[0]}_l
     \big)
\nonumber\\&&\!\!\!\!
- \sqrt{2}\>\!g 
    \sum_{l=1}^\infty \sum_{m=\max(3-l,1)}^\infty\!\!\!
     \big(
      \phi^{[+]\dagger}_{m+l-2}
      m \phi^{[+]}_m l \phi^{[+]}_l
      +
      \phi^{[-]\dagger}_{m+l-2}
      m \phi^{[-]}_m l \phi^{[-]}_l
      -
      \phi^{[0]\dagger}_{m+l-2}
      m \phi^{[0]}_m l \phi^{[0]}_l
     \big)
.
\nonumber\\&&
\eea
This result is consistent with the fact that 
each mode is singlet.

\subsection{Other models}

The modes $\alpha^{0}_n$, which classically related to the identity in Jordan algebras, has a 
special role in the above symmetry breaking. The choice $\sigma^0 =1$ is invariant under the 
symmetry of the model (the automorphisms of the algebra) and that is what allowed us to 
reduce the other symmetry breaking to the  relative simple case discussed above. However, it is 
of course possible to make other choices, e.g.\ choosing $\sigma^8 =1$ (and $\mu^8 \neq 0$, the 
rest of constants zero, say),
as an example. Such a choice comes with a price, namely that the sign of the kinetic term of the 
various modes will not be the same. Looking at the differential equations which govern the time evolution 
of the system it implies naively that some modes evolve from the length $L=0$ to negative values of $L$. 
Clearly, there is no physical interpretation of this if we insist that $L$ has the interpretation of length. However,
it is also possible to take a more formal point of view and consider the algebraic structure as the more fundamental
feature of the model. Then it is up to us if we can find another way to view these negative values of $L$. Such 
an interpretation indeed exists. The Hamiltonian in this case will be 
\beq
{\cal H}_{\rm kin} = {\cal H}_{\rm kin}^{(83)} + {\cal H}_{\rm kin}^{(12)}+{\cal H}_{\rm kin}^{(45)}+{\cal H}_{\rm kin}^{(67)},
\eeq
where 
\begin{eqnarray}
{\cal H}_{\rm kin}^{(83)}
\!&=&\!
    \sum_{l=1}^\infty \phi^{8\dagger}_{l+1}\!\; l \phi^8_l
-
    \sum_{l=1}^\infty \phi^{3\dagger}_{l+1}\!\; l \phi^3_l
-\,
    \mu
    \sum_{l=1}^\infty \phi^{8\dagger}_{l-1}\!\; l \phi^8_l
+
  \mu
    \sum_{l=1}^\infty \phi^{3\dagger}_{l-1}\!\; l \phi^3_l
\,,
%
\\
{\cal H}_{\rm kin}^{(12)}
\!&=&\!
-\,
  \sum_{i}
    \sum_{l=1}^\infty \phi^{i\dagger}_{l+1}\!\; l \phi^i_l
+
  \mu
  \sum_{i}
    \sum_{l=1}^\infty \phi^{i\dagger}_{l-1}\!\; l \phi^i_l
\,,
\\
{\cal H}_{\rm kin}^{(45)}
\!&=&\!
\frac{1}{2}
  \sum_{I}
    \sum_{l=1}^\infty \phi^{I\dagger}_{l+1}\!\; l \phi^I_l
-
  \frac{\mu}{2}
  \sum_{I}
    \sum_{l=1}^\infty \phi^{I\dagger}_{l-1}\!\; l \phi^I_l
\,,
\\
{\cal H}_{\rm kin}^{(67)}
\!&=&\!
\frac{1}{2}
  \sum_{\tilde{I}}
    \sum_{l=1}^\infty \phi^{\tilde{I}\dagger}_{l+1}\!\; l \phi^{\tilde{I}}_l
-
  \frac{\mu}{2}
  \sum_{\tilde{I}}
    \sum_{l=1}^\infty \phi^{\tilde{I}\dagger}_{l-1}\!\; l \phi^{\tilde{I}}_l
\,.
\end{eqnarray}
Here the kinetic terms related to the modes 8,4,5,6 and 7 have the wrong sign. However, we can make the 
following redefinition of the modes 
\begin{equation}
 \begin{array}{lll}
  \phi^{8\dagger}_l \,\to\,
  (-1)^l \phi^{8\dagger}_l
  \,,
  &\quad
  \phi^{I\dagger}_l \,\to\,
  (-2)^l \phi^{I\dagger}_l
  \,,
  &\quad
  \phi^{\tilde{I}\dagger}_l \,\to\,
  (-2)^l \phi^{\tilde{I}\dagger}_l
  \,,
  \\
  \phi^8_l \,\to\,
  (-1)^l \phi^8_l
  \,,
  &\quad
  \phi^I_l \,\to\,
  \big(\!{\negdbltinyspace}-{\negdbltinyspace}\frac{1}{2}
  \big)^{{\negdbltinyspace}l}
  \phi^I_l
  \,,
  &\quad
  \phi^{\tilde{I}}_l \,\to\,
  \big(\!{\negdbltinyspace}-{\negdbltinyspace}\frac{1}{2}
  \big)^{{\negdbltinyspace}l}
  \phi^{\tilde{I}}_l
  \,.
 \end{array}
\end{equation}
This redefinition will not change the commutation relations between the $\phi^\mu_l$'s, so from the algebraic  
point of view we are allowed to make such a change. However, with the new modes the 
Hamiltonians read 
\begin{eqnarray}
{\cal H}_{\rm kin}^{(83)}
\!&\to&\!
-\,
    \sum_{l=1}^\infty \phi^{8\dagger}_{l+1}\!\; l \phi^8_l
-
    \sum_{l=1}^\infty \phi^{3\dagger}_{l+1}\!\; l \phi^3_l
+\,
    \mu
    \sum_{l=1}^\infty \phi^{8\dagger}_{l-1}\!\; l \phi^8_l
+
  \mu
    \sum_{l=1}^\infty \phi^{3\dagger}_{l-1}\!\; l \phi^3_l
\,,
\\
{\cal H}_{\rm kin}^{(45)}
\!&\to&\!
-\,
  \sum_{I}
    \sum_{l=1}^\infty \phi^{I\dagger}_{l+1}\!\; l \phi^I_l
+
  \frac{\mu}{4}
  \sum_{I}
    \sum_{l=1}^\infty \phi^{I\dagger}_{l-1}\!\; l \phi^I_l
\,,
\\
{\cal H}_{\rm kin}^{(67)}
\!&\to&\!
-\,
  \sum_{\tilde{I}}
    \sum_{l=1}^\infty \phi^{\tilde{I}\dagger}_{l+1}\!\; l \phi^{\tilde{I}}_l
+
  \frac{\mu}{4}
  \sum_{\tilde{I}}
    \sum_{l=1}^\infty \phi^{\tilde{I}\dagger}_{l-1}\!\; l \phi^{\tilde{I}}_l
\,.
\end{eqnarray}
With the mode redefinition the model has an interpretation where all spatial directions have positive length
assignment and are 
either ``Planck size" compactified ($\mu \!>\! 0$) 
or expanding to infinity in a finite time ($\mu \!<\! 0$). However,
one price to pay for such a redefinition of modes is that while the interactions involving three modes
can readily be written down, it is not clear how to translate the interaction back to a geometric one
where a universe of length $L$ splits into two of lengths $L_1$ and $L_2$ with $L=L_1+L_2$, and the 
inverse process where two universes of lengths $L_1$ and $L_2$ merge to one universe of length $L_1+L_2$. 
For this reason we have not pursued this line of symmetry breaking further.

\section{High-dimensional CDT models}\label{sec5}
\setcounter{equation}{0}

\begin{figure}[t]
\vspace{-90pt}
 \begin{center}
 \includegraphics[width=300pt]{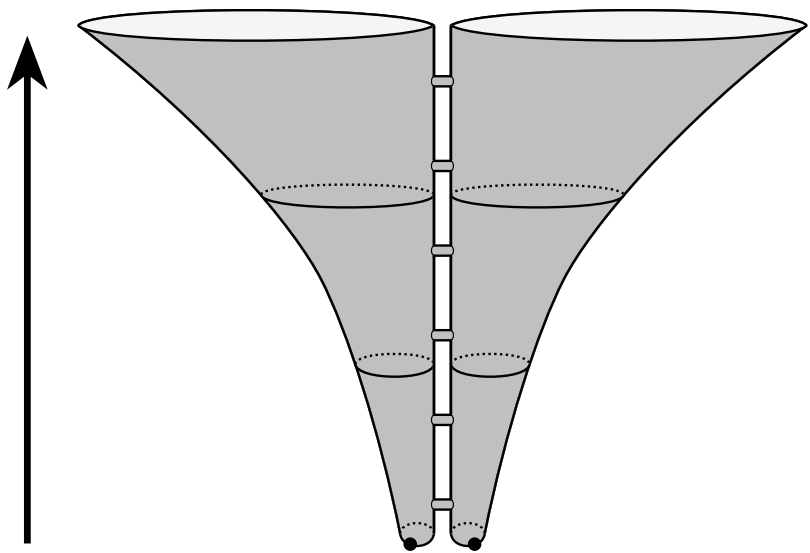}
 \end{center}
\vspace*{-10pt}
\caption[knitting]{{\footnotesize
Two universes of different flavours 1,2 or 3 knitted together via interaction with flavour 8.
}}
\label{knitting}
\end{figure}

In Sec.\ \ref{sec2} we described how a one-dimensional CDT universe can split in two and how two CDT universes
can join to one. This is described by CDT string field theory, which is very similar in structure to non-critical string 
field theory. When we generalise this model to include ``flavour''  a new situation arises. A given 
space now carries a flavour  index and the joining and splitting of space is dictated by the Jordan algebra 
structure constants $d_{abc}$.  As an example consider the model where the non-trivial part 
of the Jordan algebra is 
$\tH_3(\Dbl{C})$. The generators $\lambda_a$ are then the Gellmann matrices and the structure constants the 
corresponding $d_{abc}$ symbols of $SU(3)$. From the table of $d_{abc}$ symbols (see Appendix C)
it is seen that two universes of flavours $a = 1,2,3$ 
can only join to a universe of flavour $a=8$\footnote{
If we considered $H_3(\Dbl{C})$ rather than $\tH_3(\Dbl{C})$, then  universes with flavours 1,2 an 3 can also join to a 
universe with flavour $a=0$, as is clear from eq.\   \rf{dHermite3dim}. Flavours 0 and 8 will play identical roles and 
for simplicity we will omit  the flavour $a=0$ from the discussion here.}.
Which type of flavoured universe there can be created from ``nothing'' and 
how it develops before interacting with other universes,
depends on the symmetry breaking pattern. 
Examples were provided in the last section.
Let us first for simplicity consider the $0$-$8$ breaking 
(i.e.\ $\sigma_0 =1$, $\mu \neq 0$ and $\mu_0 = \mu' =0$) 
and further restrict ourselves to the sector 
of the Hilbert space where $(\phi^0)^\dg |v \ra =0$ 
(i.e.\ we allow symmetry breaking in the $0$ direction, but 
no further excitations).  
If we consider the situation where $\mu < 0$ it follows from
eqs.\ \rf{h083}-\rf{h67} that 
universes of flavours 1,2 and 3 will expand to infinite, 
once they are created 
(i.e.\ have   $\tan (\sqrt{|\mu |} T)$ behaviour) and as 
long as we ignore interaction with other modes during the expansion. 
On the other hand, 
universes with flavour 4,5,6 and 7 will only expand to 
finite (Planckian) size  (i.e.\ have a $\tanh (\sqrt{\mu} T)$ behaviour)
and the same is true for universes with flavour 8. 
Since the one-dimensional universes of type 1,2 or 3 
can only interact via universes of flavour 8, 
the one-dimensional universes with flavour 1,2,3 
will be ``knitted'' together via the point-like interactions 
with the 8-flavour and while one in the quadratic approximation 
has three independent spatial universes with flavours 1,2 and 3 
and each with the  spatial topology of the circle, $T^1$, we now have, 
via the cubic interact with flavour 8, 
effectively a three-dimensional universe 
with the topology of $T^3$. Including the time in 
the counting we thus have an extended four-dimension spacetime 
created from nothing. 
The ``knitting'' procedure which produces this dimension enhancement
is illustrated in Fig.\ \ref{knitting}. 

\begin{figure}[h]
 \begin{center}
 \includegraphics[width=100pt]{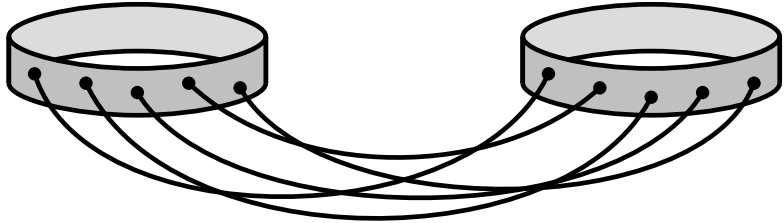}
 \end{center}
\vspace{20pt}
\caption[torusknitting]{{\footnotesize
Points of two one-dimensional universes with different flavours 1,2 or 3 are identified via knitting,
merging the two $T^1$ spatial universes into a a two-dimensional $T^2$ universe. 
}}
\label{torusknitting}
\end{figure}

\begin{figure}[h]
 \begin{center}
 \includegraphics[width=2.5cm]{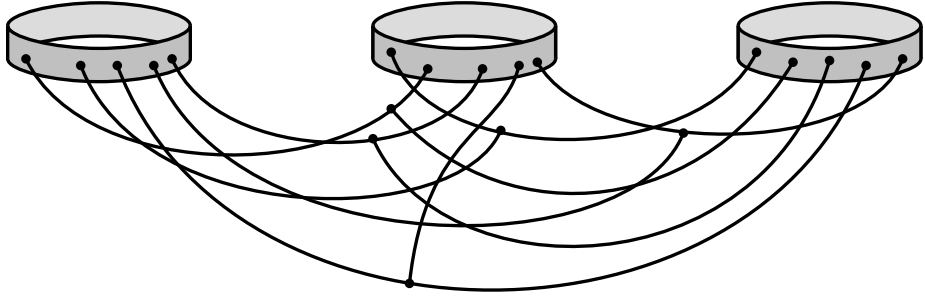}
 \end{center}
\vspace{2cm}
\caption[torusknitting3D]{{\footnotesize
Points of three one-dimensional universes with different flavours 1,2 or 3 are identified via knitting,
merging the three $T^1$ spatial universes into a three-dimensional $T^3$ universe. 
}}
\label{torusknitting3D}
\end{figure}

There are a number of aspects of such a knitting procedure which are not yet well understood. The basic process
is the one shown in Fig.\ \ref{hdynamics}, now only with flavours added. Two universes of different flavours 
1,2 or 3 propagate in time.  Then a ``wormhole'' of flavour 8 is created and connects them. 
This wormhole is less than  Planckian size. 
In this way two points in the different universes are connected. This is all fine and everything 
of this perturbative term (of order $g^2G$) can be explicitly calculated, and it corresponds to 
one of the connections shown in Fig.\ \ref{knitting}. {In particular one can from the explicit 
form of the propagator calculate how the wormhole of flavour 8 glues together two universes of 
flavour 1 and 2, say. If the wormhole exists for a very short time $\Delta T$, the free energy 
related to the process is proportional to  $1/2 \log (L \Delta T)$, where $L$ is the spatial extension
of the wormhole. Small wormholes with a short lifetime are thus favoured, as illustrated in Fig.\ \ref{knitting}}. However, as indicated in Fig.\ \ref{knitting} we want this 
process to be ongoing, such that the two points in the separated universes are identified, and even further, we 
want such identification for all points in the one-dimensional spaces. This is illustrated in Fig.\ \ref{torusknitting}
{for one-dimensional universes of flavour 1 and 2, say, in this way forming $T^2$. Let us label the points in 
the spatial universe of flavour 1 by coordinate $x^1$ and the points in the spatial universe of flavour 2 by coordinate 
$x^2$. All flavour 1 points are now assumed to be connected to all flavour 2 points, and the combined space can be labelled
by coordinates $(x^1,x^2)$ representing the torus $T^2$. It is important in the identification with $T^2$ that we have 
different flavours 1 and 2. The basic wormhole process as shown in Fig.\  \ref{hdynamics} also takes place between 
two spatial universes of the same flavour (as actually shown in Fig.\ \ref{hdynamics}), but it does not lead to the formation
of a space with topology $T^2$. Rather, when  two points  in the different $T^1$s are identified it leads to a merging 
of two $T^1$ spaces to a new $T^1$ space. This is not possible if the universes have different flavours. }
{%
In this way the concept of distance appears in the higher dimensional space created by the knitting mechanism.}
{Consider now the formation the universe $T^3$ from three spatial universes with flavour 1, 2 and 3 mediated by 
a ``wormhole'' of flavour 8. The dominant mechanism is shown in Fig.\ \ref{torusknitting3D}. The fact that $d_{888}$
is different from 0, as well as $d_{118}$, $d_{228}$ and $d_{338}$, allows for three universes of flavour 8, emerging from 
spatial universes of flavour 1, 2 and 3, to join in a single point with coupling $d_{888}$. As in the case with only two universes,
these interactions are strongest for wormholes of the smallest extension in space and time, and we obtain a gluing between 
points with three different flavours, labelled by three different coordinates $x^1$, $x^2$ and $x^3$, to a point in $T^3$, now
labelled by coordinate $(x^1,x^2,x^3)$.
 }
{%
The important term mediating  knitting is 
the propagator, related to the kinetic term ${\cal H}_0$  in \rf{Hdisk} and depicted in Fig.\ \ref{hdynamics}
as a kind of ``wormhole" between the two universes.
Since the process does not involve the tadpole term in the Hamiltonian we expect the knitting mechanism to respect 
time reversal invariance.
}

The situation is the same if we consider the other $\tH_3(\Dbl{Q})$, $\Dbl{Q} = \Dbl{R},
 \Dbl{H},  \Dbl{O}$.  In the case of $ \Dbl{R}$ we have two flavours, 1 and 3, where the universes
 can expand to infinite spatial volume, while universes with flavour  4,6 and 8 will stay at cut-off scale.
 For  $\Dbl{H}$ we five different flavoured universes which can expand to infinity and for $ \Dbl{O}$ we have 9 different 
 flavoured universes which can expand to infinity, where in all cases two universes of the same flavour can 
 merge into a flavour 8 universe which then can split in two universes of one of flavours of potential infinite spatial 
 extension. This follows from the tables of  $d_{abc}$ symbols given in Appendix C. 
 Again we conjecture that going beyond the perturbation picture shown in Fig.\ \ref{hdynamics}, 
 the knitting dynamics operates and creates extended spatial universes 
 of dimensions 2, 3, 5 and 9 for $ \Dbl{R},  \Dbl{C},  \Dbl{H}$ and $ \Dbl{O}$, respectively. If we add the time 
 extension the models we end up with 3, 4, 6 and 10  dimensions of extended spacetime. It is interesting 
 that these are precise the spacetime where classical supersymmetric string theories can be formulated.

 Recall that the automorphism groups for the Jordan algebras    $\tH_3(\Dbl{Q})$, $\Dbl{Q} =, \Dbl{R}, \Dbl{C},
 \Dbl{H},  \Dbl{O}$, are $SO(3)$, $SU(3)$, $U\!sp(6)$ and $F_4$ and breaking the symmetry in the 8-direction
 leave us with the unbroken subgroups $SO(2)$, $SO(3)$, $SO(5)$ and $SO(9)$. This picture is compatible
 with knitting procedure, since the structure constants  $d_{aa8}$ are the same in the unbroken
 directions. 
 
\subsection{A small cosmological constant}

The naive perturbative picture we have in our model is that universes of some flavours can be 
 created from nothing with zero spatial extension. They can  expand and split into universes 
 of the same or different flavours,  and the opposite process where two spatial universes join to one universe 
 is also possible. All these 
 changes are mediated (more or less) by the modes of flavour 8, 
 and our conjecture is that  if we consider the flavours which can expand to macroscopic spatial 
 extension in a finite time,  they will combine to a higher dimensional space, glued together 
 by the flavour 8 mode. This way of expanding to macroscopic size in a finite time of order 
 $1/\sqrt{-\mu}$ (i.e.\ in "Planckian time'') is clearly very different from the standard inflationary 
 expansion, since the model stops to exist for time larger than  $1/\sqrt{-\mu}$. However, we believe that 
 that is not really the ``end of time'' in  our model-universe. Wormholes and baby universes are an integral part 
 of our model, and the Coleman mechanism \cite{coleman} provides a way in which the effective cosmological constant can be very different from the ``bare'' cosmological constant. 
Thus it is possible to 
 imagine a scenario where the universes expand very fast, the scale factor growing like
 \beq\label{jk6}
 a(T) \propto \tan (\sqrt{-\mu} T)
 \eeq
 rather than the usual exponential growth encountered in a standard inflation scenario, but where the 
 mutual interactions between universes or part of the same universe eventually slows down this expansion 
 before it reaches the strict infinity at a finite time. An effective, modified Friedmann model inspired from CDT
 was suggested in \cite{aw2}.

\section{Discussion and Summary}\label{sec6}

We have proposed a model for the universe which tries to address the question of which kind of 
theory could create macroscopic universes starting from ``nothing''. ``Nothing'' in this context is a 
theory which has no obvious geometric interpretation, but which has the potential  to ``create'' a world 
with spacetime geometry. The basic idea is very simple and somehow inverting the usual logic of 
quantum field theory where one starts out with a quadratic non-interacting Hamiltonian and then adds 
a cubic term which provides the interaction. Here we start with the cubic part as fundamental, and by  
symmetry breaking we then obtain the quadratic, non-interacting part which allows us to talk 
about time and space. 
Motivated by CDT and non-critical string theory we were led to 
associate the cubic interactions with Jordan algebras. The  so-called ``magic" Jordan algebras 
consisting of Hermitian $3\times 3$ matrices with entries being real numbers, complex numbers, quarternions
and octonions, respectively, offered the possibility of an interesting symmetry breaking pattern where 
spacetime with macroscopic extensions could be 3,4 6 and 10 dimensional, {can all be derived from the expectation value 
of the operator
\beq
\exp\!\bigg\{
  \mbox{res}_{z=0} \Big( \tr\! :\! \frac{1}{3} J(z)^3 \!: \!\Big)
\bigg\},
\eeq
where $J(z) = \sum_\mu E_\mu J^\mu(z)$ 
is a bosonic current in complex plane $z$ 
with favours $\mu$ and where $E_\mu$ are $3 \times 3$ Hermitian matrices which belong to one of the ``magical'' Jordan algebras}.

 In particular the model based 
on the traceless part, $\tH_3(\Dbl{O})$, of the Jordan algebra $H_3(\Dbl{O})$, 
where $\Dbl{O}$ denotes the octonions, is intriguing, since the simplest 
symmetry breaking pattern leads to 10 extended dimensions and 16 compactified dimensions, very much like 
the situation present in the heterotic string model. In our models we do not have any concept of supersymmetry, 
which makes it remarkable that we encounter 3,4,6 and 10 dimensions in the simplest symmetry breaking patterns, 
since these are precisely the dimensions where classical superstrings can be formulated. 
{Again, this makes the 10 dimensional model especially interesting for the following reason: we know that 
only the 10 dimensional superstring theory can be consistently quantized. In particular, if one uses lightcone coordinates
in the description of the classical string theories, Lorentz invariant is not manifest in the quantized theory, and 
only in 10 dimensions can it be restored. Also in the CDT quantum gravity models, Lorentz invariance is not manifest.  
The starting point of 
the whole construction was precisely a CDT string field theory which was modelled after non-critical lightcone string field theory,
and maybe Lorentz invariance  can be restored for the model based on the Jordan algebra $H_3(\Dbl{O})$.}

It is also remarkably that a concept like inflation seems to be a natural consequence of our model, although 
the ``inflation'' we encounter is somewhat different from the standard inflation. In fact our main problem is 
to actually stop the universe from expanding to infinity in a finite, very short time. For this we appealed to the 
Coleman mechanism. In \cite{aw2} we developed a phenomenological model based on the assumption
that the Coleman mechanism would work in our model. The remarkable feature was that still some 
trace of the creation of baby universes, i.e.\ the splitting and joining of spatial universes, would survive 
in our present universe and actually result in an expansion of the universe, very much like what is observed, 
and this even if the Coleman mechanism had put the cosmological constant to zero. 
In this way the very origin of our universe gets connected to its eventual destiny. 
More details of this will be published elsewhere
\cite{to-appear}, where we will also argue that our model solves the horizon and flatness problem in cosmology
and that it makes explicit cosmological predictions which can be verified or falsified in the near future. 
{%
There are of course many aspects of Coleman's mechanism which are highly conjectural. One aspect is the concrete
existence of wormholes and precisely  how they enter in the functional integral and at which scale. Our model is well suited to turn
many of these conjectures into concrete calculations, since the starting point from where we were led to the present model,
namely CDT string field theory, {\it is} a theory of wormholes and not classical wormholes, but fully quantum wormholes. 
It was precisely constructed to have a mathematical well defined functional integral over geometries with a finite number of wormholes. In the theory outlined in this article the dominant effect of the  
wormholes is  at short distance to glue together one-dimensional spaces with different flavour 
and create higher dimensional tori and provide toroidal coordinates to these spaces by the knitting mechanism. 
On the other hand the wormholes can have any length, although long  wormholes are less probabilistic favoured, and
it is precisely the long wormholes which are used by Coleman.
The thickness of these wormholes are of Planck scale in our model
and this feature makes the Coleman mechanism much more realistic than some of the attempts to use  
Einstein-Rosen bridges as the wormholes in the Coleman mechanism. We have not yet managed to perform a detailed 
calculation leading to a concrete example of the Coleman mechanism, but we believe it is technically doable.}

There are many questions which still need to be addressed. It would nice and important to understand precisely 
which symmetry is present in the model. The $W^{(3)}$ algebra related to the Jordan algebra is only exact 
at the classical level. Clearly there are some intriguing algebraic relations involving the higher $W^{(n)}$'s, as is 
clear from the relations reported in Appendices D and E. However, a full understanding is still missing.
It is most important to understand in detail the splitting and merging of universes and the process of ``knitting''.
Perturbatively one can calculate the amplitudes for these processes to any order, but we need a non-perturbative 
solution to truly  understand how higher dimensional space emerges  from the one-dimensional spaces of different
flavour.  If one could understand this, one would also have a tool to discuss the Coleman mechanism in detail, 
something which to our knowledge never has been done before.  Also, it would be highly desirable to have a 
more explicit realisation of the symmetry breaking mechanism. Here it is postulated and implemented via the 
choice of a coherent state in the state space defined by the $W^{(3)}$ algebra, but the dynamics behind this 
is not understood. Of course the challenge is to develop what can be called ``dynamics'' before space and time 
have come into existence. 
 
\section*{Acknowledgment }
 YW thanks Akio Hosoya for discussions. JA and YW acknowledge the support from 
the Danish Research Council, via the grant ``Quantum Geometry'', grant no.  7014-00066B.
This work was partially supported by JSPS KAKENHI Grant no. JP18K03612.

\vspace{36pt}

%
\appendix
\setcounter{section}{0}
\setcounter{equation}{0}
\def\theequation{\Alph{section}.\arabic{equation}}
\def\thefigure{\Alph{section}.\arabic{figure}}
%

\section{
Mathematical properties of the Green function}
\label{app:GreenFun}

\subsection{The eigenvalue equation}

Let us consider the following eigenvalue  equation obtained in CDT, as described in Sec. 2:
\begin{equation}\label{DiffEqWithMarkedPoint}
H {\tinyspace} \psi(L)
\,=\,
E {\tinyspace} \psi(L)
\,,
\end{equation}
where
\begin{equation}\label{HamiltonianCylinderAmp}
H \,:=\, L \Big(\! - \frac{d^2}{d L^2} + \cc \Big)
\,,
\end{equation}
and $E$ is a constant and where the range of $L$ is from 0 to $\infty$. 
The inner product which makes $H$ an Hermitian operator  is 
\begin{equation}\label{InnerProductDef0}
\langle \psi^{\prime} | \psi \rangle
\,=\,
\int_0^\infty\! \frac{d L}{L} \!\>
\psi^{\prime\ast}(L) \psi(L)
\,.
\end{equation}

Changing  variable from $L$ to $z$ by
\begin{equation}
z \,:=\, 2 \sqrt{\cc}\!\; L
\,,
\end{equation}
and also changing  the wave function $\psi(L)$ to $\varphi(z)$ by
\begin{equation}
\psi(L) \,=\, z {\dbltinyspace} e^{-z /2} \varphi(z)
\,,
\end{equation}
we obtain
\begin{equation}\label{DifferentialEqPhi2}
\bigg\{
  z \frac{d^2}{d z^2} + ( 2 - z ) \frac{d}{d z}
  + \frac{E}{2 \sqrt{\cc}} - 1
\bigg\}
\varphi(z)
\,=\,
0
\,.
\end{equation}
The general solution of the differential equation 
\rf{DifferentialEqPhi2} is
\begin{equation}\label{WCDT_GeneralSolutions}
\varphi(z) \,=\,
c_1 M\!\!\>\Big( 1 - \frac{E}{2\sqrt{\cc}} , 2 , z \Big)
\,+\,
c_2 \!\; U\!\!\>\Big( 1 - \frac{E}{2\sqrt{\cc}} , 2 , z \Big)
\,.
\end{equation}
$M(\alpha,\beta,z)$ and $U(\alpha,\beta,z)$ are
the confluent hypergeometric functions 
of the first kind and the second kind, respectively.
If we demand that $\psi(L)$ is square integrable with respect to the scalar 
product \rf{InnerProductDef0}  we cannot use the $U$-function and 
the argument $\alpha =1-E/2\mu$ in $M(\alpha,2,z)$ has to be a non-positive integer $n$.
$M(n,2,z)$ is proportional to the  
associated Laguerre polynomial $L^1_n(z)$. Recall that the  
associated Laguerre equation is
\begin{equation}
\bigg\{
  z \frac{d^2}{d z^2} + ( m + 1 - z ) \frac{d}{d z} + n - m
\bigg\}
L^m_n(z)
\,=\,
0
\,,
\end{equation}
and the inner product, which makes the differential operator defining $L^m_n(z)$ Hermitian, is given by
\begin{equation}\label{firstdef}
\int_0^\infty\! d z \!\>
e^{-z} z^m L^m_k(z) L^m_n(z)
\,=\,
\frac{n!}{(n-m)!}\!\; \delta_{k,n}
\,,
\end{equation}
and the completeness is
\begin{equation}
\sum_{n=m}^\infty
\frac{(n\!-\!m)!}{n!}\!\> L^m_n(z) L^m_n(z')
\,=\,
\frac{e^z}{z^m}\!\; \delta(z-z')
\,.
\end{equation}
The definitions of associated Laguerre polynomials are\footnote{%
There are two major definitions for associated Laguerre polynomials.
One is the one given above, and the other is the following:
\begin{equation}
L_n^m(z) \,:=\,
\frac{e^z z^{-m}}{n!}\!\; \frac{d^n}{d z^n}\!\> (\!\> e^{-z} z^{m+n} )
\end{equation}
}
\begin{eqnarray}
&&
L_n^m(z) \,:=\, (-1)^m \frac{d^m}{d z^m}\!\> L_n(z)
\,,
\\
&&
L_n(z) \,:=\, \frac{e^z}{n!}\!\; \frac{d^n}{d z^n}\!\> (\!\> e^{-z} z^n )
\,,
\qquad\hbox{[\,$n \!=\! 0$, $1$, $2$,\ldots; $m \!=\! 0$, $1$, \ldots, $n$\,]}
\,.
\qquad
\end{eqnarray}
Therefore, the solution of the above differential equation is
\begin{equation}
\varphi = \varphi_n := \frac{1}{\sqrt{n}}\!\; L^1_n(z)
\,,
\qquad
E = E_n := 2 \sqrt{\cc}\, n
\,,
\qquad\hbox{[\,$n \!=\! 1$, $2$, \ldots\,]}
\,.
\quad
\end{equation}
Then, we find
\begin{equation}
\psi_n(L) \,=\,
\frac{A_n}{\sqrt{n}} {\dbltinyspace}
( 2 \sqrt{\cc} L ) {\dbltinyspace} e^{- \sqrt{\cc} L}
L^1_n\big(2 \sqrt{\cc}\!\;L\big)
\,,
\qquad\quad\hbox{[\,$n \!=\! 1$, $2$, \ldots\,]}
\,,
\end{equation}
where $A_n$ are the normalization factor.
The inner product becomes
\begin{eqnarray}
\langle \psi_m | \psi_n \rangle
\!&=&\!
\int_0^\infty\! \frac{d L}{L} \!\>
\psi^{\ast}_m(L) \psi_n(L)
\ = \ 
|A_n|^2 {\tinyspace} \delta_{m,n}
\,,
\end{eqnarray}
for $m$, $n \ge 1$.
So, $A_n = 1$. 
On the other hand,
the completeness becomes
\begin{eqnarray}
\sum_{n=1}^\infty
\psi_n(L') \psi_n^{\ast}(L)
\!&=&\!
|A_n|^2
{\dbltinyspace} L' {\tinyspace} \delta(L' - L)
\,=\,
L {\dbltinyspace}
\delta(L' - L)
\,.
\end{eqnarray}
In the last equation we used $A_n \!=\! 1$.
This result is consistent with the inner product 
(\ref{InnerProductDef0}).

One  should note that
there exists non-square-integrable solutions.
One of them is, for example,
\begin{equation}
\varphi = \varphi_0 := \frac{1}{z}
\qquad\quad\hspace{40pt}
E = E_0 := 0
\,.
\end{equation}
Then, the amplitude is
\begin{equation}
\psi_0(L) \,=\,
A_0
\!\;e^{- \sqrt{\cc} L}
\,.
\end{equation}
$\psi_0(L)$ is not square-integrable
and is not orthogonal to other $\psi_n(L)$
[\,$n \!\ge\! 1$\,]. Nevertheless it is an important solution: it is 
precisely the disk amplitude \rf{DiskAmpDef}, 
and thus the solution to the Wheeler-deWitt equation for 2d CDT. 

\subsection{The Green function}

By using the formula%
\begin{equation}
\sum_{n=m}^\infty
 \frac{z^n}{(n\!-\!m\!+\!1)_m}\!\> L_n^m(x) L_n^m(y)
\,=\,
\frac{z^{m/2} \,e^{-\, \frac{z}{1-z} (x+y)}}{(1\!-\!z) (x y)^{m/2}}\,
I_m\!\!\;\bigg(\frac{2\sqrt{z x y}}{1\!-\!z}\bigg)
\,,
\end{equation}
one finds
\begin{eqnarray}\label{CylinderAmplitude}
&&
L {\tinyspace} G(L_0,L;T)
\,:=\,
\sum_{n=1}^\infty
e^{- E_n T}
\psi_n(L_0) \psi_n^{\ast}(L)
\nonumber\\&&=\,
\frac{\sqrt{\cc L_0 L}}{\sinh(\sqrt{\cc}\, T)}
\,e^{- \sqrt{\cc}\,(L_0+L) \coth(\sqrt{\cc}\, T)}
I_1\!\!\;\bigg(\frac{2\sqrt{\cc L_0 L}}{\sinh(\sqrt{\cc}\, T)}\bigg)
\,.
\end{eqnarray}
For $L_0 \to 0$ one obtains
the amplitude of Big Bang, 
\begin{eqnarray}\label{BigBangAmplitude}
G_{\rm BB}(L;T) \!&:=&\!
\lim_{L_0 \to 0} \frac{1}{L_0}\!\; G(L_0,L;T)
\,=\,
\frac{\cc {\dbltinyspace} e^{- \sqrt{\cc}\,L \coth(\sqrt{\cc}\, T)}}
     {\sinh^2(\sqrt{\cc}\, T)}
\,.
\end{eqnarray}
We obtain the disk amplitude  \rf{DiskAmpDef}, i.e.\ a result 
proportional to $\psi_0(L)$, by integrating over $T$.\footnote{%
For $z := \coth(\sqrt{\cc}\, \T)$,
then we find
$d z = -\, \frac{\sqrt{\cc}\, d \T}{\sinh^2(\sqrt{\cc}\, \T)}$
and
$z \in [\infty,1] \,\leftrightarrow\, \T \in [0,\infty]$.
}

\subsection{The Green function and the average size of space}

The expectation value of the length of exit loop is 
\begin{equation}\label{AverageLengthUnderExpansion}
\langle L \rangle_{(L_0,T)}
\,:=\,
\frac{\int_0^\infty d L {\dbltinyspace} L {\tinyspace} G(L_0,L;T)}
     {\int_0^\infty d L {\dbltinyspace} G(L_0,L;T)}
\,,
\end{equation}
where $L_0$ is the length of entrance loop.  
The numerator and the denominator of the right-hand side are
\begin{eqnarray}
&&
\int\limits_0^\infty\! d L {\dbltinyspace} G(L_0,L;T)
\,=\,
\big(
1
-
e^{- \sqrt{\cc}\,L_0 / ( \sinh(\sqrt{\cc}\, T) \cosh(\sqrt{\cc}\, T) )}
\big)
\,e^{- \sqrt{\cc}\,L_0 \tanh(\sqrt{\cc}\, T)}
\,,
\\
&&
\int\limits_0^\infty\! d L {\dbltinyspace} L {\tinyspace} G(L_0,L;T)
\,=\,
\frac{L_0 \,e^{- \sqrt{\cc}\,L_0 \tanh(\sqrt{\cc}\, T)}}
     {\cosh^2(\sqrt{\cc}\, T)}
\,,
\\
&&
\int\limits_0^\infty\! d L {\dbltinyspace} L^2 G(L_0,L;T)
\,=\,
\bigg(
  \frac{2 L_0 \sinh(\sqrt{\cc}\, T)}
       {\sqrt{\cc}\,\cosh^3(\sqrt{\cc}\, T)}
  +
  \frac{L_0^2}{\cosh^4(\sqrt{\cc}\, T)}
\bigg)
\,e^{- \sqrt{\cc}\,L_0 \tanh(\sqrt{\cc}\, T)}
\,.
\qquad\quad
\end{eqnarray}
So, the above average becomes
\begin{eqnarray}\label{AverageLengthOfSpace}
\langle L \rangle_{(L_0,T)}
\!&=&\!
\frac{L_0}
     {\cosh^2(\sqrt{\cc}\, T) \big(
        1
        -
        e^{- 2 \sqrt{\cc}\,L_0 / \sinh(2 \sqrt{\cc}\, T)}
      \big)}
\,,
\end{eqnarray}
and the variance $\langle L^2 \rangle_{(L_0,T)}- \langle L \rangle^2_{(L_0,T)}$ becomes
\begin{eqnarray}\label{VarianceLengthOfSpace}
{\rm Var}_{(L_0,T)}(L)
\!&=&\!
\frac{2 L_0 \sinh(\sqrt{\cc}\, T)}
     {\sqrt{\cc} \cosh^3(\sqrt{\cc}\, T) \big(
        1
        -
        e^{- 2 \sqrt{\cc}\,L_0 / \sinh(2 \sqrt{\cc}\, T)}
      \big)}
\nonumber\\&&\!
-\,
\frac{L_0^2 \,e^{- 2 \sqrt{\cc}\,L_0 / \sinh(2 \sqrt{\cc}\, T)}}
     {\cosh^4(\sqrt{\cc}\, T) \big(
        1
        -
        e^{- 2 \sqrt{\cc}\,L_0 / \sinh(2 \sqrt{\cc}\, T)}
      \big)^2}
\,.
\end{eqnarray}
For $L_0 \!=\! 0$, we obtain
\beq\label{jk32}
\langle L \rangle_{T} = \frac{1}{\sqrt{\mu}} \, \tanh \sqrt{\mu} T,\qquad 
{\rm Var}_{T}(L) = \frac{1}{\mu} \, \tanh^2 \sqrt{\mu} T= \langle L \rangle^2_{T} 
\eeq
$\langle L \rangle_{T} $ is the average size of space at time $T$ after the Big Bang, as already mentioned
in the main text, \rf{AverageLengthOfSpaceZero1}. Already after a time not much longer that $T \sim 1/\sqrt{\mu}$
the size of space has almost reached its maximum  $L_{\rm cs} := \frac{1}{\sqrt{\cc}}$. It should also 
be noted that the standard deviation $\Delta L (T) = \sqrt{ {\rm Var}_{T}(L)} = \langle L \rangle_{T}$. 
So the standard deviation is quite large. 

If the spatial extent of the universe shrinks to zero, it will cease to exist and be absorbed in the vacuum.
The probability that this happens at time $T$, starting out with the Big Bang, is proportional to 
\begin{equation}\label{BubbleProbability}
\lim_{L_0 \to 0} \lim_{L \to 0}
\frac{1}{L_0} {\dbltinyspace} G(L_0,L;T)
\,=\,
\frac{\cc}{\sinh^2(\sqrt{\cc}\, T)}
\,.
\end{equation}

\subsection{The Green function with negative cosmological constant}

The differential equation \rf{DiffEqWithMarkedPoint}
is analytic with respect to the coefficient $\cc$.
Without explicit calculation one can make the analytic continuation
$\cc \to e^{i\theta} \cc$ [\,$0 \!\le\! \theta \!\le\! \pi$\,]
in the solutions which we already obtained\footnote{%
Even if $\sqrt{\mu}$ appears in the solution, there is no cut at $\mu =0$ in the solution for the Green function 
as one can explicitly check by making both analytic continuations
$\cc \to e^{i\theta} \cc$ [\,$0 \!\le\! \theta \!\le\! \pi$\,]
and 
$\cc \to e^{-i\theta} \cc$ [\,$0 \!\le\! \theta \!\le\! \pi$\,]. They give the same result (see e.g.\ footnote 
\ref{fo1} and \ref{fo2})
}.

When $\cc$ is negative, 
the Green function \rf{CylinderAmplitude} becomes\footnote{\label{fo1}%
The calculation is as follows:
\begin{eqnarray*}
&&
\frac{\sqrt{\cc L_0 L}}{\sinh(\sqrt{\cc}\, T)}
\ \to \ 
\frac{\pm i \sqrt{-\cc L_0 L}}{\sinh(\pm i \sqrt{-\cc}\, T)}
\,=\,
\frac{\sqrt{-\cc L_0 L}}{\sin(\sqrt{-\cc}\, T)}\, ,
\\
&&
\sqrt{\cc}\coth(\sqrt{\cc}\, T)
\ \to \ 
\pm i\sqrt{\cc}\coth(\pm i\sqrt{-\cc}\, T)
\,=\,
\sqrt{-\cc}\cot(\sqrt{-\cc}\, T)\, .
\end{eqnarray*}
\vspace{-12pt}
}
\begin{eqnarray}\label{CylinderAmplitudeNegativeMu}
&&
L {\tinyspace} G(L_0,L;T)
\,=\,
\frac{\sqrt{-\cc L_0 L}}{\sin(\sqrt{-\cc}\, T)}
\,e^{- \sqrt{-\cc}\,(L_0+L) \cot(\sqrt{-\cc}\, T)}
I_1\!\!\;\bigg(\frac{2\sqrt{-\cc L_0 L}}{\sin(\sqrt{-\cc}\, T)}\bigg)
\,,
\nn&&
\end{eqnarray}
and 
the Big Bang amplitude \rf{BigBangAmplitude} also becomes
\begin{eqnarray}\label{BigBangAmplitudeNegativeMu}
G_{\rm BB}(L;T) \!
&=&\!
\frac{-\cc {\dbltinyspace} e^{- \sqrt{-\cc}\,L \cot(\sqrt{-\cc}\, T)}}
     {\sin^2(\sqrt{-\cc}\, T)}
\,.
\end{eqnarray}

When $\cc$ is negative, 
the average length of the exit loop  when we start at $L_0 =0$, becomes
\footnote{\label{fo2}%
The calculation is as follows:
\begin{eqnarray*}
&&
\frac{\tanh(\sqrt{\cc}\, T)}{\sqrt{\cc}}
\ \to \ 
\frac{\tanh(\pm i \sqrt{-\cc}\, T)}{\pm i \sqrt{-\cc}}
\,=\,
\frac{\tan(\sqrt{-\cc}\, T)}{\sqrt{-\cc}}
\,.
\end{eqnarray*}
}
\beq\label{jk37}
\langle L \rangle_{T} = \frac{1}{\sqrt{-\mu}} \, \tan \sqrt{-\mu}\, T \,.
\eeq
The average length goes to infinity for  When $T \to T_{\rm ps} :=\frac{\pi}{2\sqrt{-\cc}}$.
Thus we have a kind of inflation, culminating  at $T_{\rm ps} $ where this is an infinitely fast expansion.
The ``inflation'' can be seen from \rf{BigBangAmplitudeNegativeMu}, too.
When $T$ is less than $T_{\rm ps}$,
$L_{\rm max}$, the peak of $G_{\rm bb}(L;T)$ with fixed $T$, is finite.
But, when $T$ approaches $T_{\rm ps}$,
$L_{\rm max}$ goes to infinity.

When $\cc$ is negative, the probability that a Big Bang universe disappears at time   $T < T_{\rm ps} $becomes
\begin{equation}
\lim_{L_0 \to 0} \lim_{L \to 0}
\frac{1}{L_0} {\dbltinyspace} G(L_0,L;T)
\,=\,
\frac{-\cc}{\sin^2(\sqrt{-\cc}\, T)}
\,.
\end{equation}

\subsection{The Green function without cosmological constant}

We can directly take the limit $\mu = 0$ in the above expressions and obtain the results for 
zero cosmological constant.
When $\cc$ is zero, 
the Green function \rf{CylinderAmplitude} becomes
\begin{eqnarray}\label{CylinderAmplitudeZeroMu}
&&
L {\tinyspace} G(L_0,L;T)
\,=\,
\frac{\sqrt{L_0 L}}{T}
\,e^{- (L_0+L) / T}
I_1\!\!\;\bigg(\frac{2\sqrt{L_0 L}}{T}\bigg)
\,,
\end{eqnarray}
and the Big Bang amplitude \rf{BigBangAmplitude} also becomes
\begin{eqnarray}\label{BigBangAmplitudeZeroMu}
G_{\rm BB}(L;T) \!
&=&\!
\frac{e^{- L / T}}
     {T^2}
\,.
\end{eqnarray}
Instead of \rf{jk32} we have 
\beq\label{jk42}
\langle L \rangle_{T} = \frac{1}{\sqrt{\mu}} \, \tanh \sqrt{\mu} T,\qquad 
{\rm Var}_{T}(L) = \frac{1}{\mu} \, \tanh^2 \sqrt{\mu} T= \langle L \rangle^2_{T} 
\eeq
and instead of \rf{BubbleProbability} 
\begin{equation}
\lim_{L_0 \to 0} \lim_{L \to 0}
\frac{1}{L_0} {\dbltinyspace} G(L_0,L;T)
\,=\,
\frac{1}{T^2}
\,.
\end{equation}

\section{
Real, complex, quaternion and octonion numbers}
\label{app:RCQOnumbers}


The set of 
real numbers, complex numbers, quaternions, octonions is 
denoted by $\Dbl{R}$, $\Dbl{C}$, $\Dbl{H}$, $\Dbl{O}$, 
respectively. 
One of 
real numbers, complex numbers, quaternions, octonions is 
\beq\label{RCHOnumbers}
x \,+ \sum_n e_n x_n
\,,
\eeq
where $x$, $x_1$, $x_2$, \ldots are real numbers, and 
the set of $e_n$, i.e.\ $E = \{ e_1, e_2, \ldots \}$ is 
\begin{align}\label{ImaginaryUnitRCHO}
&
E
\ \define \ 
\{\, \,\}
&\hspace{-60pt}\hbox{[\,for $\Dbl{R}$\,]}
\,,
\\
&
E
\ \define \ 
\{\, i \,\}
&\hspace{-60pt}\hbox{[\,for $\Dbl{C}$\,]}
\,,
\label{ImaginaryUnitC}
\\
&
E
\ \define \ 
\{\, i,\, j,\, k \,\}
&\hspace{-60pt}\hbox{[\,for $\Dbl{H}$\,]}
\,,
\label{ImaginaryUnitH}
\\
&
E
\ \define \ 
\{\, i,\, j,\, k ,\, \ell ,\, \bar{i} ,\, \bar{j} ,\, \bar{k} \,\}
&\hspace{-60pt}\hbox{[\,for $\Dbl{O}$\,]}
\,.
\label{ImaginaryUnitO}
\end{align}
The number of imaginary units, i.e.\ $|E{\tinyspace}|$, is 
$0$, $1$, $3$, $7$, 
for $\Dbl{R}$, $\Dbl{C}$, $\Dbl{H}$, $\Dbl{O}$, 
respectively. 
For real numbers, \rf{RCHOnumbers} reads $x$
[\,$x {\negdbltinyspace}\in{\negdbltinyspace} \Dbl{R}$\,].
For complex numbers, \rf{RCHOnumbers} reads
$x {\negdbltinyspace}+{\negdbltinyspace} i y$
[\,$x$, $y {\negdbltinyspace}\in{\negdbltinyspace} \Dbl{R}$\,].
For quaternions and octonions, 
we will explain the details in the following sections. 

\subsection{Quaternions}

According to \rf{RCHOnumbers} and \rf{ImaginaryUnitH}, 
the quaternions are numbers defined by
\begin{equation}\label{CartesianCoordinateQuaternion}
z  \ = \ 
x
 {\tinyspace}+{\tinyspace} i{\tinyspace}y
 {\tinyspace}+{\tinyspace} j{\tinyspace}u
 {\tinyspace}+{\tinyspace} k{\tinyspace}v
\qquad\quad\hbox{[\,$x$, $y$, $u$,
 $v {\negdbltinyspace}\in{\negdbltinyspace} \Dbl{R}$\,]}
\end{equation}
The quaternion conjugate of $z$ is defined by
\begin{equation}\label{QuaternionNumberQC}
z^\ast \ \define \ 
x
 {\tinyspace}-{\tinyspace} i{\tinyspace}y
 {\tinyspace}-{\tinyspace} j{\tinyspace}u
 {\tinyspace}-{\tinyspace} k{\tinyspace}v\, .
\end{equation}
The norm of a quaternion is defined by
\begin{equation}\label{QuaternionNumberABS}
|z| \ \define \  \sqrt{x^2 + y^2 + u^2 + v^2}\, .
\end{equation}
The multiplication of these imaginary units are as follows:
\begin{eqnarray}\label{QuaternionUnitSquare}
&&
i^2 \,=\, j^2 \,=\, k^2 \,=\, i j k \,=\, -1
\nonumber\\
&&
i{\tinyspace}j \,=\, - j{\tinyspace}i \,=\, k
\qquad
j{\tinyspace}k \,=\, - k{\tinyspace}j \,=\, i
\qquad
k{\tinyspace}i \,=\, - i{\tinyspace}k \,=\, j
\end{eqnarray}
Multiplications of quaternions do not necessarily do not necessarily commute, 
but satisfy associativity. 

\subsection{Octonions}

According to \rf{RCHOnumbers} and \rf{ImaginaryUnitO}, 
the octonions are numbers defined by
\begin{eqnarray}\label{CartesianCoordinateOctonion}
z  \!&=&\!
x {\dbltinyspace}+{\dbltinyspace} i{\halftinyspace}y
  {\dbltinyspace}+{\dbltinyspace} j{\halftinyspace}u
  {\dbltinyspace}+{\dbltinyspace} k{\halftinyspace}v
  {\dbltinyspace}+{\dbltinyspace} \ell{\tinyspace}\bar{x}
  {\dbltinyspace}+{\dbltinyspace} \bar{i}{\tinyspace}\bar{y}
  {\dbltinyspace}+{\dbltinyspace} \bar{j}{\tinyspace}\bar{u}
  {\dbltinyspace}+{\dbltinyspace} \bar{k}{\tinyspace}\bar{v}
\nonumber\\&=&\!
\xi {\negtinyspace}+{\negtinyspace}
 \eta{\tinyspace}\ell
\end{eqnarray}
where
$x$,{\negtinyspace} $y$,{\negtinyspace}
 $u$,{\negtinyspace} $v$,{\negtinyspace} $\bar{x}$,{\negtinyspace}
 $\bar{y}$,{\negtinyspace} $\bar{u}$,{\negtinyspace} 
 $\bar{v} {\negdbltinyspace}\in{\negdbltinyspace} \Dbl{R}$,
and
\begin{eqnarray}
\xi \!&=&\!
x {\dbltinyspace}+{\dbltinyspace} i{\halftinyspace}y
  {\dbltinyspace}+{\dbltinyspace} j{\halftinyspace}u
  {\dbltinyspace}+{\dbltinyspace} k{\halftinyspace}v
\in \Dbl{H}
\nonumber\\
\eta \!&=&\!
\bar{x}
  {\dbltinyspace}+{\dbltinyspace} i{\halftinyspace}\bar{y}
  {\dbltinyspace}+{\dbltinyspace} j{\halftinyspace}\bar{u}
  {\dbltinyspace}+{\dbltinyspace} k{\halftinyspace}\bar{v}
\in \Dbl{H}
\end{eqnarray}
The octonion conjugate of $z$ is defined by
\begin{eqnarray}\label{OctonionNumberOC}
z^\ast \!&\define&\!
x {\tinyspace}-{\tinyspace} i{\tinyspace}y
  {\tinyspace}-{\tinyspace} j{\tinyspace}u
  {\tinyspace}-{\tinyspace} k{\tinyspace}v
  {\tinyspace}-{\tinyspace} \ell{\tinyspace}\bar{x}
  {\tinyspace}-{\tinyspace} \bar{i}{\tinyspace}\bar{y}
  {\tinyspace}-{\tinyspace} \bar{j}{\tinyspace}\bar{u}
  {\tinyspace}-{\tinyspace} \bar{k}{\tinyspace}\bar{v}
\nonumber\\&=&\!
\xi^\ast {\negdbltinyspace}-{\negdbltinyspace}
 \eta{\tinyspace}\ell
\end{eqnarray}
The norm of a octonion is defined by
\begin{eqnarray}\label{OctonionNumberABS}
|z| \!&\define&\!
\sqrt{x^2 + y^2 + u^2 + v^2
      + \bar{x}^2 + \bar{y}^2 + \bar{u}^2 + \bar{v}^2}
\nonumber\\&=&\!
\sqrt{|\xi|^2 + |\eta|^2}
\end{eqnarray}
The multiplication of two octonions is defined by
\begin{equation}\label{OctonionMultiplication}
( \xi {\negtinyspace}+{\negtinyspace}
  \eta{\tinyspace}\ell {\tinyspace})
( \xi' {\negtinyspace}+{\negtinyspace}
  \eta'{\negtinyspace}\ell {\tinyspace})
\ = \ 
\xi \xi' {\negtinyspace}-{\negtinyspace} \eta^{\prime\ast} \eta
+ (
    \eta' \xi {\negtinyspace}+{\negtinyspace}
    \eta {\tinyspace} \xi^{\prime\ast}
  ){\trehalftinyspace}\ell
\end{equation}
where $\xi$, $\eta$, $\xi'$, $\eta'$ are quaternions.
The squares of imaginary units are all $-1$, 
and all imaginary units anticommute, 
in the same way as the imaginary units of quaternions do. 
So multiplication is not necessarily commutative, but also it is not necessarily associative. 

\section{
Symmetric and Hermitian Matrices}
\label{app:HermitianMatrices}

Let us consider the set of 
$\dim {\negdbltinyspace}\times{\negdbltinyspace} \dim$ 
Hemitian matrices.  
We classify the Hermitian matrices into four group, 
where the  matrix elements are real, complex, quaternion, octonion, respectively.
Hermiticity means that matrix elements satisfy $a_{ij} = a^*_{ji}$, where conjugation $a^*$ is defined above
(in the non-trial cases of quarternions and octonions).

The {\it real} vectorspace of these Hermitian matrices is a direct sum of the one-dimensional vector space spanned by the 
unit matrix and the vector space spanned by the traceless Hermitian matrices. We denote a basis of the 
traceless Hermitian matrices by $\{ \lambda_a | \,\hbox{$a \!=\! 1$, \ldots, $N$}\}$.
For the traceless matrices we can write
$S + E \!\times A$, 
where
$S$ is the set of real symmetric traceless matrices, 
$A$ is the set of real antisymmetric matrices, 
$E$ is the set of imaginary units defined in eqs.\ \rf{ImaginaryUnitRCHO}-\rf{ImaginaryUnitO}. 
The number of real symmetric traceless matrices is 
$|S| {\negdbltinyspace}={\negdbltinyspace}
 \frac{\dim(\dim+1)}{2} {\negtinyspace}-{\negtinyspace} 1$, 
and 
the number of real antisymmetric matrices is 
$|A| {\negdbltinyspace}={\negdbltinyspace}
 \frac{\dim(\dim-1)}{2}$. 
$E \!\times\! A$ is the set which consists of 
all possible multiplications of an element of $E$ and an element of $A$, 
so 
$|E \!\times\! A| = |E| \!\cdot\! |A|$. 
Let $N$ denote the dimension of the set $S + E \!\times\! A$.  Thus
\beq
N \,=\,
| S + E \!\times\! A |
\,=\,
|S| {\negdbltinyspace}+{\negdbltinyspace} |E| \!\cdot\! |A|
\eeq

\newcommand{\SUtwo}{%
\newcommand{\sig}{\lambda}

\subsection{$2$D Hermitian-matrix type}

In the case of $2 {\negdbltinyspace}\times{\negdbltinyspace} 2$ matrices, 
the set of symmetric traceless matices is
$S \!\define\! \{ S_{xy} , S_z \}$, where
\begin{eqnarray}\label{SU2SpinMatrixS}
&&
S_{xy} \define
  \left(\!\begin{array}{cc}
           0   &    1   \\
           1   &    0   \\
  \end{array}\!\right)
\,,
\qquad
S_z \define
  \left(\!\begin{array}{cc}
           1   &    0   \\
           0   &   -1   \\
  \end{array}\!\right)
\,,
\end{eqnarray}
and 
the set of antisymmetric matices is
$A \!\define\! \{ A_{xy} \}$, where
\begin{eqnarray}\label{SU2SpinMatrixA}
&&
A_{xy} \define
  \left(\!\begin{array}{cc}
           0   &   -1   \\
           1   &    0   \\
  \end{array}\!\right)
\,.
\end{eqnarray}
Since $|S| \!=\! 2$, $|A| \!=\! 1$ in the case of $\dim \!=\! 2$, 
$N \!=\! 2$, $3$, $5$, $9$, 
for $\Dbl{R}$, $\Dbl{C}$, $\Dbl{H}$, $\Dbl{O}$ matrix types, respectively. 
The concrete definitions are listed in the following subsections. 

\subsubsection{Real symmetric traceless matrices}

\beq
\sig_1 = S_{xy}
\,,
\qquad
\sig_3 = S_z
\,.
\eeq

\subsubsection{Hermite traceless matrices}

\beq
\sig_1 = S_{xy}
\,,
\qquad
\sig_3 = S_z
\,,
\qquad
\sig_2 = i A_{xy}
\,.
\eeq
In this case, $\sig_a$ are the Pauli matrices. 

\subsubsection{Quaternion Hermite traceless matrices}

\bea
&&
\sig_1 = S_{xy}
\,,
\qquad
\sig_3 = S_z
\,,
\qquad
\sig_2 = i A_{xy}
\,,
\qquad
\sig_{2'} = j A_{xy}
\,,
\qquad
\sig_{2''} = k A_{xy}
\,.
\qquad
\eea

\subsubsection{Octonion Hermite traceless matrices}

\bea
&&
\sig_1 = S_{xy}
\,,
\qquad
\sig_3 = S_z
\,,
\qquad
\sig_2 = i A_{xy}
\,,
\qquad
\sig_{2'} = j A_{xy}
\,,
\qquad
\sig_{2''} = k A_{xy}
\,,
\qquad
\nonumber\\
&&
\sig_{2^\circ} = \ell A_{xy}
\,,
\qquad
\sig_{\bar{2}} = \bar{i} A_{xy}
\,,
\qquad
\sig_{\bar{2}'} = \bar{j} A_{xy}
\,,
\qquad
\sig_{\bar{2}''} = \bar{k} A_{xy}
\,.
\qquad
\eea
}\SUtwo		

\subsection{$3$D Hermitian-matrix type}

In the case of $3 {\negdbltinyspace}\times{\negdbltinyspace} 3$ matrices, 
the set of symmetric traceless matices is 
\beq
S \,\define\,
\{ S_{xy} ,\, S_{yz} ,\, S_{zx} ,\, S_{x^2-y^2} ,\, S_{z^2} \}
\,,
\eeq
where
\begin{eqnarray}\label{SU3SpinMatrixS}
&&
S_{xy} \define
  \left(\!\begin{array}{ccc}
           0   &    1   &    0   \\
           1   &    0   &    0   \\
           0   &    0   &    0
  \end{array}\!\right)
\,,
\quad
S_{yz} \define
  \left(\!\begin{array}{ccc}
           0   &    0   &    0   \\
           0   &    0   &    1   \\
           0   &    1   &    0
  \end{array}\!\right)
\,,
\quad
S_{zx} \define
  \left(\!\begin{array}{ccc}
           0   &    0   &    1   \\
           0   &    0   &    0   \\
           1   &    0   &    0
  \end{array}\!\right)
\,,
\nonumber\\
&&
S_{x^2-y^2} \define
  \left(\!\begin{array}{ccc}
           1   &    0   &    0   \\
           0   &   -1   &    0   \\
           0   &    0   &    0
  \end{array}\!\right)
\,,
\quad
S_{z^2} \define
  \frac{1}{\sqrt{3}}\!
  \left(\!\begin{array}{ccc}
           1   &    0   &    0   \\
           0   &    1   &    0   \\
           0   &    0   &   -2
  \end{array}\!\right)
\,,
\qquad\ 
\end{eqnarray}
and 
the set of antisymmetric matices is 
\beq
A \,\define\, \{ A_{xy} , A_{yz} , A_{zx} \}
\,,
\eeq
where
\begin{eqnarray}\label{SU3SpinMatrixA}
&&
A_{xy} \define
  \left(\!\begin{array}{ccc}
           0   &   -1   &    0   \\
           1   &    0   &    0   \\
           0   &    0   &    0
  \end{array}\!\right)
\,,
\quad
A_{yz} \define
  \left(\!\begin{array}{ccc}
           0   &    0   &    0   \\
           0   &    0   &   -1   \\
           0   &    1   &    0
  \end{array}\!\right)
\,,
\quad
A_{zx} \define
  \left(\!\begin{array}{ccc}
           0   &    0   &    1   \\
           0   &    0   &    0   \\
          -1   &    0   &    0
  \end{array}\!\right)
\,.
\qquad\ 
\end{eqnarray}
Since $|S| \!=\! 5$, $|A| \!=\! 3$ in the case of $\dim \!=\! 3$, 
$N \!=\! 5$, $8$, $14$, $26$, 
for $\Dbl{R}$, $\Dbl{C}$, $\Dbl{H}$, $\Dbl{O}$ matrix types, respectively. 
The concrete definitions are listed in the following subsections. 

\subsubsection{Real symmetric traceless matrices}

\bea
&&
\lambda_1 = S_{xy}
\,,
\qquad
\lambda_6 = S_{yz}
\,,
\qquad
\lambda_4 = S_{zx}
\,,
\qquad
\lambda_3 = S_{x^2-y^2}
\,,
\qquad
\lambda_8 = S_{z^2}
\,.
\qquad
\eea
The non-zero $d_{abc}$ 
[\,$a$, $b$, $c \!=\! 1$, \ldots\,]
are in the following:
\begin{eqnarray}\label{SO3antistructureconstant2}
&&
d_{118} \,=\, 
d_{338} \,=\, \frac{1}{\sqrt{3}}
\,,
\qquad
d_{448} \,=\, 
d_{668} \,=\, -{\dbltinyspace}\frac{1}{2\sqrt{3}}
\,,
\nonumber\\
&&
d_{888} \,=\, -{\dbltinyspace}\frac{1}{\sqrt{3}}
\,,
\qquad
d_{344} \,=\, \frac{1}{2}
\,,
\qquad\quad
d_{366} \,=\, -{\dbltinyspace}\frac{1}{2}
\,,
\qquad\quad
d_{146} \,=\, \frac{1}{2}
\,.
\qquad
\end{eqnarray}
We here omit $d_{abc}$ obtained by the permutation of $a$, $b$, $c$. 

\subsubsection{Hermite traceless matrices}

\bea
&&
\lambda_1 = S_{xy}
\,,
\qquad
\lambda_6 = S_{yz}
\,,
\qquad
\lambda_4 = S_{zx}
\,,
\qquad
\lambda_3 = S_{x^2-y^2}
\,,
\qquad
\lambda_8 = S_{z^2}
\,,
\qquad
\nonumber\\
&&
\lambda_2 = i A_{xy}
\,,
\qquad
\lambda_7 = i A_{yz}
\,,
\qquad
\lambda_5 = -i A_{zx}
\,.
\eea
In this case, $\lambda_a$ are the Gell-Mann matrices. 
The non-zero $d_{abc}$ 
[\,$a$, $b$, $c \!=\! 1$, \ldots\,]
are in the following:
\begin{eqnarray}\label{SU3antistructureconstant2}
&&
d_{118} \,=\, 
d_{228} \,=\, 
d_{338} \,=\, \frac{1}{\sqrt{3}}
\,,
\qquad
d_{448} \,=\, 
d_{558} \,=\, 
d_{668} \,=\, 
d_{778} \,=\, -{\dbltinyspace}\frac{1}{2\sqrt{3}}
\,,
\nonumber\\
&&
d_{888} \,=\, -{\dbltinyspace}\frac{1}{\sqrt{3}}
\,,
\qquad
d_{344} \,=\, 
d_{355} \,=\, \frac{1}{2}
\,,
\qquad\quad
d_{366} \,=\, 
d_{377} \,=\, -{\dbltinyspace}\frac{1}{2}
\,,
\nonumber\\
&&
d_{146} \,=\, 
d_{157} \,=\, 
d_{256} \,=\, \frac{1}{2}
\,,
\qquad\quad
d_{247} \,=\, -{\dbltinyspace}\frac{1}{2}
\,.
\end{eqnarray}
Again we here omit $d_{abc}$ obtained by the permutation of $a$, $b$, $c$. 

\subsubsection{Quaternion Hermite traceless matrices}

\bea
&&
\lambda_1 = S_{xy}
\,,
\qquad
\lambda_6 = S_{yz}
\,,
\qquad
\lambda_4 = S_{zx}
\,,
\qquad
\lambda_3 = S_{x^2-y^2}
\,,
\qquad
\lambda_8 = S_{z^2}
\,,
\qquad
\nonumber\\
&&
\lambda_2 = i A_{xy}
\,,
\qquad
\lambda_7 = i A_{yz}
\,,
\qquad
\lambda_5 = -i A_{zx}
\,,
\qquad
\nonumber\\
&&
\lambda_{2'} = j A_{xy}
\,,
\qquad
\lambda_{7'} = j A_{yz}
\,,
\qquad
\lambda_{5'} = -j A_{zx}
\,,
\qquad
\nonumber\\
&&
\lambda_{2''} = k A_{xy}
\,,
\qquad
\lambda_{7''} = k A_{yz}
\,,
\qquad
\lambda_{5''} = -k A_{zx}
\,.
\eea
Non-zero $d_{abc}$ 
[\,$a$, $b$, $c \!=\! 1$, \ldots\,]
are in the following:
\begin{eqnarray}\label{Sp3antistructureconstant2}
&&
d_{118} \,=\, 
d_{228} \,=\, 
d_{2'2'8} \,=\, 
d_{2''2''8} \,=\, 
d_{338} \,=\, 
\frac{1}{\sqrt{3}}
\,,
\nonumber\\
&&
d_{448} \,=\, 
d_{558} \,=\, 
d_{5'5'8} \,=\, 
d_{5''5''8} \,=\, 
d_{668} \,=\, 
d_{778} \,=\, 
d_{7'7'8} \,=\, 
d_{7''7''8} \,=\, 
-{\dbltinyspace}\frac{1}{2\sqrt{3}}
\,,
\nonumber\\
&&
d_{888} \,=\, -{\dbltinyspace}\frac{1}{\sqrt{3}}
\,,
\nonumber\\
&&
d_{344} \,=\, 
d_{355} \,=\, 
d_{35'5'} \,=\, 
d_{35''5''} \,=\, 
\frac{1}{2}
\,,
\nonumber\\
&&
d_{366} \,=\, 
d_{377} \,=\, 
d_{37'7'} \,=\, 
d_{37''7''} \,=\, 
-{\dbltinyspace}\frac{1}{2}
\,,
\nonumber\\
&&
d_{146} \,=\, 
d_{157} \,=\, 
d_{15'7'} \,=\, 
d_{15''7''} \,=\, 
d_{256} \,=\, 
d_{2'5'6} \,=\, 
d_{2''5''6} \,=\, 
\frac{1}{2}
\,,
\nonumber\\
&&
d_{247} \,=\, 
d_{2'47'} \,=\, 
d_{2''47''} \,=\, 
-{\dbltinyspace}\frac{1}{2}
\,,
\nonumber\\
&&
d_{25'7''} \,=\, 
d_{2'5''7} \,=\, 
d_{2''57'} \,=\, 
\frac{1}{2}
\,,
\qquad\quad
d_{2'57''} \,=\, 
d_{2''5'7} \,=\, 
d_{25''7'} \,=\, 
-{\dbltinyspace}\frac{1}{2}
\,.
\end{eqnarray}
Also here we here omit $d_{abc}$ obtained by the permutation of $a$, $b$, $c$. 

\subsubsection{Octonion Hermite traceless matrices}

\bea
&&
\lambda_1 = S_{xy}
\,,
\qquad
\lambda_6 = S_{yz}
\,,
\qquad
\lambda_4 = S_{zx}
\,,
\qquad
\lambda_3 = S_{x^2-y^2}
\,,
\qquad
\lambda_8 = S_{z^2}
\,,
\qquad
\nonumber\\
&&
\lambda_2 = i A_{xy}
\,,
\qquad
\lambda_7 = i A_{yz}
\,,
\qquad
\lambda_5 = -i A_{zx}
\,,
\qquad
\nonumber\\
&&
\lambda_{2'} = j A_{xy}
\,,
\qquad
\lambda_{7'} = j A_{yz}
\,,
\qquad
\lambda_{5'} = -j A_{zx}
\,,
\qquad
\nonumber\\
&&
\lambda_{2''} = k A_{xy}
\,,
\qquad
\lambda_{7''} = k A_{yz}
\,,
\qquad
\lambda_{5''} = -k A_{zx}
\,,
\qquad
\nonumber\\
&&
\lambda_{2^\circ} = \ell A_{xy}
\,,
\qquad
\lambda_{7^\circ} = \ell A_{yz}
\,,
\qquad
\lambda_{5^\circ} = -\ell A_{zx}
\,,
\qquad
\nonumber\\
&&
\lambda_{\bar{2}} = \bar{i} A_{xy}
\,,
\qquad
\lambda_{\bar{7}} = \bar{i} A_{yz}
\,,
\qquad
\lambda_{\bar{5}} = -\bar{i} A_{zx}
\,,
\qquad
\nonumber\\
&&
\lambda_{\bar{2}'} = \bar{j} A_{xy}
\,,
\qquad
\lambda_{\bar{7}'} = \bar{j} A_{yz}
\,,
\qquad
\lambda_{\bar{5}'} = -\bar{j} A_{zx}
\,,
\qquad
\nonumber\\
&&
\lambda_{\bar{2}''} = \bar{k} A_{xy}
\,,
\qquad
\lambda_{\bar{7}''} = \bar{k} A_{yz}
\,,
\qquad
\lambda_{\bar{5}''} = -\bar{k} A_{zx}
\,.
\eea
The non-zero $d_{abc}$ 
[\,$a$, $b$, $c \!=\! 1$, \ldots\,]
are in the following:
\begin{eqnarray}\label{SOoct3antistructureconstant2}
&&
d_{118} \,=\, 
d_{228} \,=\, 
d_{2'2'8} \,=\, 
d_{2''2''8} \,=\, 
d_{2^\circ2^\circ8} \,=\, 
d_{\bar{2}\bar{2}8} \,=\, 
d_{\bar{2}'\bar{2}'8} \,=\, 
d_{\bar{2}''\bar{2}''8} \,=\, 
d_{338} \,=\, 
\frac{1}{\sqrt{3}}
\,,
\nonumber\\
&&
d_{448} \,=\, 
d_{558} \,=\, 
d_{5'5'8} \,=\, 
d_{5''5''8} \,=\, 
d_{5^\circ5^\circ8} \,=\, 
d_{\bar{5}\bar{5}8} \,=\, 
d_{\bar{5}'\bar{5}'8} \,=\, 
d_{\bar{5}''\bar{5}''8} \,=\, 
\nonumber\\&&
d_{668} \,=\, 
d_{778} \,=\, 
d_{7'7'8} \,=\, 
d_{7''7''8} \,=\, 
d_{7^\circ7^\circ8} \,=\, 
d_{\bar{7}\bar{7}8} \,=\, 
d_{\bar{7}'\bar{7}'8} \,=\, 
d_{\bar{7}''\bar{7}''8} \,=\, 
-{\dbltinyspace}\frac{1}{2\sqrt{3}}
\,,
\nonumber\\
&&
d_{888} \,=\, -{\dbltinyspace}\frac{1}{\sqrt{3}}
\,,
\nonumber\\
&&
d_{344} \,=\, 
d_{355} \,=\, 
d_{35'5'} \,=\, 
d_{35''5''} \,=\, 
d_{35^\circ5^\circ} \,=\, 
d_{3\bar{5}\bar{5}} \,=\, 
d_{3\bar{5}'\bar{5}'} \,=\, 
d_{3\bar{5}''\bar{5}''} \,=\, 
\frac{1}{2}
\,,
\nonumber\\
&&
d_{366} \,=\, 
d_{377} \,=\, 
d_{37'7'} \,=\, 
d_{37''7''} \,=\, 
d_{37^\circ7^\circ} \,=\, 
d_{3\bar{7}\bar{7}} \,=\, 
d_{3\bar{7}'\bar{7}'} \,=\, 
d_{3\bar{7}''\bar{7}''} \,=\, 
-{\dbltinyspace}\frac{1}{2}
\,,
\nonumber\\
&&
d_{146} \,=\, 
d_{157} \,=\, 
d_{15'7'} \,=\, 
d_{15''7''} \,=\, 
d_{15^\circ7^\circ} \,=\, 
d_{1\bar{5}\bar{7}} \,=\, 
d_{1\bar{5}'\bar{7}'} \,=\, 
d_{1\bar{5}''\bar{7}''} \,=\, 
\nonumber\\&&
d_{256} \,=\, 
d_{2'5'6} \,=\, 
d_{2''5''6} \,=\, 
d_{2^\circ5^\circ6} \,=\, 
d_{\bar{2}\bar{5}6} \,=\, 
d_{\bar{2}'\bar{5}'6} \,=\, 
d_{\bar{2}''\bar{5}''6} \,=\, 
\frac{1}{2}
\,,
\nonumber\\
&&
d_{247} \,=\, 
d_{2'47'} \,=\, 
d_{2''47''} \,=\, 
d_{2^\circ47^\circ} \,=\, 
d_{\bar{2}4\bar{7}} \,=\, 
d_{\bar{2}'4\bar{7}'} \,=\, 
d_{\bar{2}''4\bar{7}''} \,=\, 
-{\dbltinyspace}\frac{1}{2}
\,,
\nonumber\\
&&
d_{25'7''} \,=\, 
d_{2'5''7} \,=\, 
d_{2''57'} \,=\, 
d_{\bar{2}'\bar{5}7''} \,=\, 
d_{\bar{2}''\bar{5}'7} \,=\, 
d_{\bar{2}\bar{5}''7'} \,=\, 
\nonumber\\&&
d_{\bar{2}'5\bar{7}''} \,=\, 
d_{\bar{2}''5'\bar{7}} \,=\, 
d_{\bar{2}5''\bar{7}'} \,=\, 
d_{2'\bar{5}\bar{7}''} \,=\, 
d_{2''\bar{5}'\bar{7}} \,=\, 
d_{2\bar{5}''\bar{7}'} \,=\, 
\nonumber\\&&
d_{\bar{2}57^\circ} \,=\, 
d_{\bar{2}'5'7^\circ} \,=\, 
d_{\bar{2}''5''7^\circ} \,=\, 
d_{25^\circ\bar{7}} \,=\, 
d_{2'5^\circ\bar{7}'} \,=\, 
d_{2''5^\circ\bar{7}''} \,=\, 
\nonumber\\&&
d_{2^\circ\bar{5}7} \,=\, 
d_{2^\circ\bar{5}'7'} \,=\, 
d_{2^\circ\bar{5}''7''} \,=\, 
\frac{1}{2}
\,,
\nonumber\\
&&
d_{2'57''} \,=\, 
d_{2''5'7} \,=\, 
d_{25''7'} \,=\, 
d_{\bar{2}\bar{5}'7''} \,=\, 
d_{\bar{2}'\bar{5}''7} \,=\, 
d_{\bar{2}''\bar{5}7'} \,=\, 
\nonumber\\&&
d_{2\bar{5}'\bar{7}''} \,=\, 
d_{2'\bar{5}''\bar{7}} \,=\, 
d_{2''\bar{5}\bar{7}'} \,=\, 
d_{\bar{2}5'\bar{7}''} \,=\, 
d_{\bar{2}'5''\bar{7}} \,=\, 
d_{\bar{2}''5\bar{7}'} \,=\, 
\nonumber\\&&
d_{2\bar{5}7^\circ} \,=\, 
d_{2'\bar{5}'7^\circ} \,=\, 
d_{2''\bar{5}''7^\circ} \,=\, 
d_{2^\circ5\bar{7}} \,=\, 
d_{2^\circ5'\bar{7}'} \,=\, 
d_{2^\circ5''\bar{7}''} \,=\, 
\nonumber\\&&
d_{\bar{2}5^\circ7} \,=\, 
d_{\bar{2}'5^\circ7'} \,=\, 
d_{\bar{2}''5^\circ7''} \,=\, 
-{\dbltinyspace}\frac{1}{2}
\,.
\end{eqnarray}
As usual we here omit $d_{abc}$ obtained by the permutation of $a$, $b$, $c$.

\newcommand{\tsu}{\hspace{-0.75pt}}
\newcommand{\tsv}{\hspace{-0.50pt}}
\newcommand{\tsw}{\hspace{-0.25pt}}

\newcommand{\tsx}{\hspace{0.25pt}}
\newcommand{\tsy}{\hspace{0.50pt}}
\newcommand{\tsz}{\hspace{0.75pt}}

\newcommand{\Tr}{{\rm Tr}{\dbltinyspace}}

\section{
Properties of the trace}
\label{app:PropertiesTrace}

Below we will list a number of properties of traces of the generator-matrices for the Jordan algebras.
Underlined indicies means that we symmetrise with respect to the underlined indices.

\subsection{Spin-matrix type}

Let $X$ be an ${2^{{\tinyspace}[N/2]}} \!\times {2^{{\tinyspace}[N/2]}}$ matrix. 
In order to simplify the expressions we introduce the following trace:
\beq\label{TraceSpinMatrix}
\Tr X \,\define\,
\frac{1}{2^{{\tinyspace}[N/2]}}{\dbltinyspace}
\tr X
\eeq
$\tr$ is the standard trace to the matrix. 
Using this trace, 
one finds the following properties:
\bea
&&
\Tr
 ( 1 ) \,=\, 1
\\
&&
\Tr
 ( \gamma_a ) \,=\, 0
\\
&&
\Tr
 ( \gamma_a \!\circ\! \gamma_b ) \,=\, \delta_{ab}
\\
&&
\Tr
 \big( ( \gamma_a \!\circ\! \gamma_b ) \!\circ\! \gamma_c \big)
\,=\, 0
\\
&&
\Tr
 \big( ( ( \gamma_a \!\circ\! \gamma_b )
       \!\circ\! \gamma_c ) \!\circ\! \gamma_d \big)
\,=\,
\Tr
 \big( ( \gamma_a \!\circ\! \gamma_b ) \!\circ\!
       ( \gamma_c \!\circ\! \gamma_d ) \big)
\,=\, \delta_{ab} {\tinyspace} \delta_{cd}
\eea

\subsection{Hermitian-matrix type}

Let $X$ be a $K \times K$ Hermitian matrix and let $N$ be the dimension of the real vector space spanned 
by the traceless Hermitian matrices. We introduce the following trace:
\beq\label{TraceHermitianMatrix}
\Tr X \,\define\,
\frac{1}{K}{\dbltinyspace}
\tr X
\eeq
$\tr$ is the standard trace to the matrix. 
Using this trace, 
one finds the following propertyes:
\bea
&&\hspace{-20pt}
\Tr
 ( 1 ) \,=\, 1
\\
&&\hspace{-20pt}
\Tr
 ( \lambda_a ) \,=\, 0
\\
&&\hspace{-20pt}
\Tr
 ( \lambda_a \!\circ\! \lambda_b )
\,=\, \frac{2}{K} {\dbltinyspace} \delta_{ab}
\\
&&\hspace{-20pt}
\Tr
 \big( ( \lambda_a \!\circ\! \lambda_b ) \!\circ\! \lambda_c \big)
\,=\, \frac{2}{K} {\dbltinyspace} d_{abc}
\\
&&\hspace{-20pt}
\Tr
 \big( ( ( \lambda_a \!\circ\! \lambda_b )
       \!\circ\! \lambda_c ) \!\circ\! \lambda_d \big)
\,=\,
\Tr
 \big( ( \lambda_a \!\circ\! \lambda_b ) \!\circ\!
       ( \lambda_c \!\circ\! \lambda_d ) \big)
\,=\,
\Big( \frac{2}{K} \Big)^{\!2} \delta_{ab} {\tinyspace} \delta_{cd}
+
\frac{2}{K}
\sum_e d_{abe} d_{cde}
\qquad\quad
\eea

\subsubsection{$3$D Hermitian-matrix type}

\vspace{6pt}

\ {\bf Three basic loop diagrams:}
\begin{eqnarray}
&&
\sum_{i,{\hspace{1pt}}j}
  \delta_{i{\tsx}j}
  {\hspace{1pt}}
  d_{a{\tsx}i{\tsx}j}
\,=\,
0
\\
&&
\sum_{i,{\hspace{1pt}}j}
  d_{a{\tsx}i{\tsx}j}
  {\hspace{1pt}}
  d_{{\tsx}b{\tsx}i{\tsx}j}
\,=\,
\frac{N + 2}{6}\!\;
\delta_{ab}
\\
&&
\sum_{i,{\hspace{1pt}}j,{\hspace{1pt}}k}
  d_{a{\tsx}i{\tsx}j}
  {\hspace{1pt}}
  d_{{\tsx}b{\tsx}jk}
  {\hspace{1pt}}
  d_{c{\tsz}k{\tsy}i}
\,=\,
-\,
\frac{N - 2}{12}\!\;
  d_{ab{\tsx}c}
\end{eqnarray}
\ {\bf Definitions of diagrams:}
\begin{eqnarray}
&&
M_{a{\tsx}b{\tsy}c{\tsy}d}^{(402)}
\,\define\,
\sum_{i}
  d_{a{\tsx}b{\tsy}i}
  {\hspace{1pt}}
  d_{c{\tsy}d{\tsz}i}
\\
&&
M_{a{\tsx}b{\tsy}c{\tsy}d{\tsy}e}^{(503)}
\,\define\,
\sum_{i,{\hspace{1pt}}j}
  d_{a{\tsx}b{\tsy}i}
  {\hspace{1pt}}
  d_{c{\tsz}i{\tsy}j}
  {\hspace{1pt}}
  d_{d{\tsy}e{\tsy}j}
\\
&&
M_{a{\tsx}b{\tsy}c{\tsy}d{\tsy}ef}^{(604)}
\,\define\,
\sum_{i,{\hspace{1pt}}j,{\hspace{1pt}}k}
  d_{a{\tsx}b{\tsy}i}
  {\hspace{1pt}}
  d_{c{\tsz}i{\tsy}j}
  {\hspace{1pt}}
  d_{d{\tsy}j{\tsx}k}
  {\hspace{1pt}}
  d_{ef{\tsx}k}
\\
&&
M_{a{\tsx}b{\tsy}c{\tsy}d{\tsy}ef}^{(604)\prime}
\,\define\,
\sum_{i,{\hspace{1pt}}j,{\hspace{1pt}}k}
  d_{i{\tsx}j{\tsx}k}
  {\hspace{1pt}}
  d_{a{\tsx}b{\tsy}i}
  {\hspace{1pt}}
  d_{c{\tsy}d{\tsx}j}
  {\hspace{1pt}}
  d_{ef{\tsx}k}
\\
&&
M_{a{\tsx}b{\tsy}c{\tsx}d}^{(414)}
\,\define\,
\sum_{i,{\hspace{1pt}}j,{\hspace{1pt}}k,{\hspace{1pt}}l}
  d_{a{\tsx}i{\tsy}j}
  {\hspace{1pt}}
  d_{{\tsx}b{\tsx}j{\tsx}k}
  {\hspace{1pt}}
  d_{c{\tsy}k{\tsy}l}
  {\hspace{1pt}}
  d_{d{\tsy}l{\tsy}i}
\\
&&
M_{a{\tsx}b{\tsy}c{\tsx}d{\tsx}e}^{(515)}
\,\define\,
\sum_{i,{\hspace{1pt}}j,{\hspace{1pt}}k,{\hspace{1pt}}l,{\hspace{1pt}}m}
  d_{a{\tsx}i{\tsy}j}
  {\hspace{1pt}}
  d_{{\tsx}b{\tsx}j{\tsx}k}
  {\hspace{1pt}}
  d_{c{\tsy}k{\tsy}l}
  {\hspace{1pt}}
  d_{d{\tsy}l{\tsy}m}
  {\hspace{1pt}}
  d_{e{\tsy}m{\tsy}i}
\\
&&
M_{a{\tsx}b{\tsy}c{\tsx}d{\tsx}e{\tsv}f}^{(616)}
\,\define\,
\sum_{i,{\hspace{1pt}}j,{\hspace{1pt}}k,{\hspace{1pt}}l,{\hspace{1pt}}m,{\hspace{1pt}}n}
  d_{a{\tsx}i{\tsy}j}
  {\hspace{1pt}}
  d_{{\tsx}b{\tsx}j{\tsx}k}
  {\hspace{1pt}}
  d_{c{\tsy}k{\tsy}l}
  {\hspace{1pt}}
  d_{d{\tsy}l{\tsy}m}
  {\hspace{1pt}}
  d_{e{\tsy}m{\tsy}n}
  {\hspace{1pt}}
  d_{f{\tsx}n{\tsy}i}
\end{eqnarray}

\ {\bf Total symmetric tree diagrams:}
\begin{eqnarray}
&&
M_{\underline{a{\tsx}b{\tsy}c{\tsy}d}}^{(402)}
\,=\,
\frac{1}{3}\!\>
  \delta_{\underline{a{\tsx}b}}
  {\hspace{1pt}}
  \delta_{\underline{c{\tsy}d}}
\\
&&
M_{\underline{a{\tsx}b{\tsy}c{\tsy}d{\tsy}e}}^{(503)}
\,=\,
\frac{1}{3}\!\>
  \delta_{\underline{a{\tsx}b}}
  {\hspace{1pt}}
  d_{\underline{c{\tsy}d{\tsx}e}}
\\
&&
M_{\underline{a{\tsx}b{\tsy}c{\tsy}d{\tsy}e{\tsv}f}}^{(604)}
\,=\,
\frac{1}{9}\!\>
  \delta_{\underline{a{\tsx}b}}
  {\hspace{1pt}}
  \delta_{\underline{c{\tsy}d}}
  {\hspace{1pt}}
  \delta_{\underline{e{\tsv}f}}
\\
&&
M_{\underline{a{\tsx}b{\tsy}c{\tsy}d{\tsy}e{\tsv}f}}^{(604)\prime}
\,=\,
\frac{2}{3}\!\>
  d_{\underline{a{\tsx}b{\tsx}c}}
  {\hspace{1pt}}
  d_{\underline{d{\tsy}e{\tsv}f}}
-
\frac{1}{9}\!\>
  \delta_{\underline{a{\tsx}b}}
  {\hspace{1pt}}
  \delta_{\underline{c{\tsy}d}}
  {\hspace{1pt}}
  \delta_{\underline{e{\tsv}f}}
\end{eqnarray}
\ {\bf Partial symmetric tree diagrams:}
\begin{eqnarray}
&&
M_{\underline{b{\tsy}c}{\tsz}a{\tsy}\underline{d{\tsy}e}}^{(503)}
\,=\,
\frac{2}{3}{\hspace{1pt}}
  \delta_{a{\tsx}\underline{b}}
  {\hspace{1pt}}
  d_{{\tsx}\underline{c{\tsy}d{\tsy}e}}
-
\frac{1}{3}{\hspace{1pt}}
  \delta_{{\tsx}\underline{b{\tsy}c}}
  {\hspace{1pt}}
  d_{a{\tsx}\underline{d{\tsy}e}}
\\
&&
M_{a{\tsx}b{\tsz}c{\tsy}d{\tsy}e}^{(503)}
+
M_{b{\tsx}c{\tsz}d{\tsy}e{\tsy}a}^{(503)}
+
M_{c{\tsx}d{\tsz}e{\tsy}a{\tsx}b}^{(503)}
+
M_{d{\tsx}e{\tsy}a{\tsx}b{\tsy}c}^{(503)}
+
M_{e{\tsy}a{\tsx}b{\tsy}c{\tsy}d}^{(503)}
\nonumber\\&&
\phantom{%
}\hspace{-2.5pt}%
-
M_{e{\tsy}b{\tsy}c{\tsy}d{\tsy}a}^{(503)}
-
M_{a{\tsx}c{\tsy}d{\tsy}e{\tsy}b}^{(503)}
-
M_{b{\tsx}d{\tsy}e{\tsx}a{\tsx}c}^{(503)}
-
M_{c{\tsz}e{\tsz}a{\tsx}b{\tsy}d}^{(503)}
-
M_{d{\tsy}a{\tsx}b{\tsy}c{\tsy}e}^{(503)}
\nonumber\\&&
=\,
\frac{1}{6}{\hspace{1pt}}
 \big({\tsz}
  \delta_{a{\tsx}b} {\hspace{1pt}} d_{c{\tsy}d{\tsy}e}
  +
  \delta_{{\tsx}b{\tsy}c} {\hspace{1pt}} d_{d{\tsy}e{\tsy}a}
  +
  \delta_{c{\tsy}d} {\hspace{1pt}} d_{e{\tsy}a{\tsx}b}
  +
  \delta_{d{\tsy}e} {\hspace{1pt}} d_{a{\tsx}b{\tsx}c}
  +
  \delta_{e{\tsy}a} {\hspace{1pt}} d_{{\tsx}b{\tsy}c{\tsy}d}
\nonumber\\&&
\phantom{%
=\,
\frac{1}{6}{\hspace{1pt}}
 \big({\tsz}
}\hspace{-2.5pt}%
  -
  \delta_{e{\tsx}b} {\hspace{1pt}} d_{c{\tsy}d{\tsy}a}
  -
  \delta_{a{\tsx}c} {\hspace{1pt}} d_{d{\tsy}e{\tsy}b}
  -
  \delta_{{\tsx}b{\tsy}d} {\hspace{1pt}} d_{e{\tsy}a{\tsx}c}
  -
  \delta_{c{\tsy}e} {\hspace{1pt}} d_{a{\tsx}b{\tsy}d}
  -
  \delta_{d{\tsy}a} {\hspace{1pt}} d_{{\tsx}b{\tsy}c{\tsy}e}
 {\tsz}\big)
\\
&&
  M_{\underline{c{\tsx}f}{\tsx}a{\tsx}b{\tsy}\underline{d{\tsy}e}}^{(604)}
  +
  M_{\underline{d{\tsy}e}{\tsy}c{\tsx}f{\tsx}\underline{a{\tsx}b}}^{(604)}
  +
  M_{\underline{a{\tsx}b}{\tsy}d{\tsy}e{\tsy}\underline{c{\tsx}f}}^{(604)}
+
\frac{1}{2}
\big(
  M_{a{\tsx}b{\tsy}\underline{c{\tsx}f{\tsv}d{\tsy}e}}^{(604)}
  +
  M_{c{\tsx}f{\tsx}\underline{d{\tsy}e{\tsy}a{\tsx}b}}^{(604)}
  +
  M_{d{\tsy}e{\tsy}\underline{a{\tsx}b{\tsy}c{\tsx}f}}^{(604)}
{\tsz}\big)
\nonumber\\&&
=\,
\frac{1}{6}
\big(
  d_{a{\tsx}c{\tsy}e}
  {\hspace{1pt}}
  d_{{\tsx}b{\tsy}d{\tsv}f}
  +
  d_{a{\tsx}b{\tsy}c}
  {\hspace{1pt}}
  d_{f{\tsv}e{\tsy}d}
  +
  d_{d{\tsy}a{\tsx}b}
  {\hspace{1pt}}
  d_{c{\tsx}f{\tsv}e}
  +
  d_{e{\tsy}d{\tsy}a}
  {\hspace{1pt}}
  d_{{\tsx}b{\tsy}c{\tsx}f}
\big)
\nonumber\\&&\phantom{%
=\,}%
+ \frac{1}{18}
\big(
  d_{a{\tsx}b{\tsv}f}
  {\hspace{1pt}}
  d_{e{\tsy}d{\tsy}c}
  +
  d_{{\tsx}b{\tsy}c{\tsy}e}
  {\hspace{1pt}}
  d_{d{\tsy}a{\tsv}f}
  +
  d_{c{\tsx}f{\tsv}d}
  {\hspace{1pt}}
  d_{a{\tsx}b{\tsy}e}
  +
  d_{f{\tsv}e{\tsy}a}
  {\hspace{1pt}}
  d_{{\tsx}b{\tsy}c{\tsy}d}
  +
  d_{e{\tsy}d{\tsy}b}
  {\hspace{1pt}}
  d_{c{\tsx}f{\tsv}a}
  +
  d_{d{\tsy}a{\tsy}c}
  {\hspace{1pt}}
  d_{f{\tsv}e{\tsx}b}
\big)
\nonumber\\&&\phantom{%
=\,}%
+ \frac{1}{54}
\big(
  \delta_{a{\tsx}b}
  {\hspace{1pt}}
  \delta_{e{\tsv}f}
  {\hspace{1pt}}
  \delta_{d{\tsy}c}
  +
  \delta_{{\tsx}b{\tsy}c}
  {\hspace{1pt}}
  \delta_{d{\tsy}e}
  {\hspace{1pt}}
  \delta_{a{\tsv}f}
  +
  \delta_{c{\tsx}f}
  {\hspace{1pt}}
  \delta_{a{\tsx}d}
  {\hspace{1pt}}
  \delta_{{\tsx}b{\tsy}e}
  +
  \delta_{a{\tsv}f}
  {\hspace{1pt}}
  \delta_{d{\tsy}b}
  {\hspace{1pt}}
  \delta_{e{\tsy}c}
  +
  \delta_{{\tsx}b{\tsy}e}
  {\hspace{1pt}}
  \delta_{a{\tsx}c}
  {\hspace{1pt}}
  \delta_{d{\tsv}f}
  +
  \delta_{c{\tsy}d}
  {\hspace{1pt}}
  \delta_{{\tsx}b{\tsv}f}
  {\hspace{1pt}}
  \delta_{a{\tsy}e}
\big)
\nonumber\\&&\phantom{%
=\,}%
+ \frac{1}{18}
  {\hspace{1pt}}
  \delta_{a{\tsv}f}
  {\hspace{1pt}}
  \delta_{{\tsx}b{\tsy}e}
  {\hspace{1pt}}
  \delta_{c{\tsy}d}
\\
&&
M_{c{\tsy}d{\tsy}a{\tsx}b{\tsy}e{\tsv}f}^{(604)}
+
M_{e{\tsv}f{\tsv}a{\tsx}b{\tsy}c{\tsy}d}^{(604)}
+
M_{f{\tsv}d{\tsy}a{\tsx}b{\tsy}e{\tsy}c}^{(604)}
+
M_{e{\tsy}c{\tsy}a{\tsx}b{\tsv}f{\tsv}d}^{(604)}
\nonumber\\
&&
+\,
M_{e{\tsy}b{\tsy}c{\tsx}f{\tsx}a{\tsx}d}^{(604)}
+
M_{a{\tsx}d{\tsy}c{\tsx}f{\tsx}e{\tsy}b}^{(604)}
+
M_{d{\tsy}b{\tsy}c{\tsx}f{\tsx}a{\tsx}e}^{(604)}
+
M_{a{\tsx}e{\tsy}c{\tsx}f{\tsx}d{\tsy}b}^{(604)}
\nonumber\\
&&
+\,
M_{b{\tsy}c{\tsy}d{\tsx}e{\tsv}f{\tsv}a}^{(604)}
+
M_{f{\tsv}a{\tsx}d{\tsx}e{\tsy}b{\tsy}c}^{(604)}
+
M_{a{\tsx}c{\tsy}d{\tsx}e{\tsv}f{\tsx}b}^{(604)}
+
M_{f{\tsx}b{\tsy}d{\tsx}e{\tsy}a{\tsx}c}^{(604)}
\nonumber\\
&&
+\,
\frac{1}{2}
\big(
  M_{d{\tsy}e{\tsy}a{\tsx}b{\tsv}f{\tsv}c}^{(604)}
  -
  M_{d{\tsy}e{\tsy}a{\tsv}f{\tsx}b{\tsy}c}^{(604)}
  -
  M_{d{\tsy}e{\tsy}a{\tsx}c{\tsx}f{\tsx}b}^{(604)}
  +
  M_{f{\tsv}c{\tsy}a{\tsx}b{\tsy}d{\tsy}e}^{(604)}
  -
  M_{a{\tsv}c{\tsy}f{\tsx}b{\tsy}d{\tsy}e}^{(604)}
  -
  M_{f{\tsv}a{\tsx}c{\tsy}b{\tsy}d{\tsy}e}^{(604)}
{\tsz}\big)
\nonumber\\
&&
+\,
\frac{1}{2}
\big(
  M_{b{\tsy}a{\tsx}c{\tsx}f{\tsv}d{\tsy}e}^{(604)}
  -
  M_{b{\tsy}a{\tsx}c{\tsy}d{\tsv}f{\tsv}e}^{(604)}
  -
  M_{b{\tsy}a{\tsx}c{\tsx}e{\tsy}d{\tsv}f}^{(604)}
  +
  M_{d{\tsy}e{\tsy}c{\tsx}f{\tsy}b{\tsy}a}^{(604)}
  -
  M_{c{\tsy}e{\tsy}d{\tsx}f{\tsy}b{\tsy}a}^{(604)}
  -
  M_{d{\tsy}c{\tsy}e{\tsv}f{\tsx}b{\tsy}a}^{(604)}
{\tsz}\big)
\nonumber\\
&&
+\,
\frac{1}{2}
\big(
  M_{c{\tsx}f{\tsv}d{\tsy}e{\tsy}a{\tsx}b}^{(604)}
  -
  M_{c{\tsx}f{\tsv}d{\tsy}a{\tsx}e{\tsy}b}^{(604)}
  -
  M_{c{\tsx}f{\tsv}d{\tsx}b{\tsy}a{\tsx}e}^{(604)}
  +
  M_{a{\tsx}b{\tsy}d{\tsx}e{\tsy}c{\tsx}f}^{(604)}
  -
  M_{d{\tsy}b{\tsy}a{\tsx}e{\tsy}c{\tsx}f}^{(604)}
  -
  M_{a{\tsx}d{\tsy}b{\tsy}e{\tsy}c{\tsx}f}^{(604)}
{\tsz}\big)
\nonumber\\&&
=\,
  d_{a{\tsx}c{\tsy}e}
  {\hspace{1pt}}
  d_{{\tsx}b{\tsy}d{\tsv}f}
  +
  d_{a{\tsx}c{\tsy}d}
  {\hspace{1pt}}
  d_{{\tsx}b{\tsy}e{\tsv}f}
  +
  d_{{\tsx}b{\tsy}c{\tsy}d}
  {\hspace{1pt}}
  d_{a{\tsy}e{\tsv}f}
  +
  d_{{\tsx}b{\tsy}c{\tsy}e}
  {\hspace{1pt}}
  d_{a{\tsx}d{\tsv}f}
\nonumber\\&&\phantom{%
=\,}%
- \frac{2}{3}
\big(
  d_{a{\tsx}b{\tsy}e}
  {\hspace{1pt}}
  d_{c{\tsy}d{\tsv}f}
  +
  d_{a{\tsx}d{\tsy}e}
  {\hspace{1pt}}
  d_{{\tsx}b{\tsy}c{\tsx}f}
  +
  d_{c{\tsx}d{\tsy}e}
  {\hspace{1pt}}
  d_{a{\tsx}b{\tsv}f}
  +
  d_{a{\tsx}b{\tsy}d}
  {\hspace{1pt}}
  d_{c{\tsy}e{\tsv}f}
  +
  d_{{\tsx}b{\tsx}d{\tsy}e}
  {\hspace{1pt}}
  d_{a{\tsx}c{\tsx}f}
  +
  d_{a{\tsx}b{\tsy}c}
  {\hspace{1pt}}
  d_{d{\tsy}e{\tsv}f}
\big)
\nonumber\\&&\phantom{%
=\,}%
+ \frac{1}{9}
\big(
  \delta_{a{\tsx}b}
  {\hspace{1pt}}
  \delta_{c{\tsx}d}
  {\hspace{1pt}}
  \delta_{e{\tsv}f}
  +
  \delta_{a{\tsx}d}
  {\hspace{1pt}}
  \delta_{{\tsx}b{\tsx}e}
  {\hspace{1pt}}
  \delta_{c{\tsx}f}
  +
  \delta_{{\tsx}b{\tsx}c}
  {\hspace{1pt}}
  \delta_{d{\tsx}e}
  {\hspace{1pt}}
  \delta_{a{\tsv}f}
  +
  \delta_{a{\tsx}e}
  {\hspace{1pt}}
  \delta_{{\tsx}b{\tsx}d}
  {\hspace{1pt}}
  \delta_{c{\tsx}f}
  +
  \delta_{a{\tsx}c}
  {\hspace{1pt}}
  \delta_{d{\tsx}e}
  {\hspace{1pt}}
  \delta_{{\tsx}b{\tsv}f}
  +
  \delta_{a{\tsx}b}
  {\hspace{1pt}}
  \delta_{c{\tsx}e}
  {\hspace{1pt}}
  \delta_{d{\tsv}f}
\big)
\nonumber\\&&\phantom{%
=\,}%
+ \frac{1}{3}
  {\hspace{1pt}}
  \delta_{a{\tsx}b}
  {\hspace{1pt}}
  \delta_{d{\tsx}e}
  {\hspace{1pt}}
  \delta_{c{\tsx}f}
\\
&&
M_{a{\tsx}b{\tsy}c{\tsy}d{\tsy}e{\tsv}f}^{(604)}
+
M_{b{\tsy}c{\tsy}d{\tsy}e{\tsv}f{\tsv}a}^{(604)}
+
M_{c{\tsy}d{\tsy}e{\tsv}f{\tsv}a{\tsx}b}^{(604)}
+
M_{d{\tsy}e{\tsv}f{\tsv}a{\tsx}b{\tsy}c}^{(604)}
+
M_{e{\tsv}f{\tsv}a{\tsx}b{\tsy}c{\tsy}d}^{(604)}
+
M_{f{\tsv}a{\tsx}b{\tsy}c{\tsy}d{\tsy}e}^{(604)}
\nonumber\\
&&
-\,
\big(
  M_{a{\tsv}f{\tsv}c{\tsy}e{\tsy}d{\tsy}b}^{(604)}
  +
  M_{f{\tsv}e{\tsy}b{\tsy}d{\tsy}c{\tsy}a}^{(604)}
  +
  M_{e{\tsy}d{\tsy}a{\tsx}c{\tsy}b{\tsv}f}^{(604)}
  +
  M_{d{\tsy}c{\tsx}f{\tsx}b{\tsy}a{\tsy}e}^{(604)}
  +
  M_{c{\tsy}b{\tsy}e{\tsy}a{\tsv}f{\tsv}d}^{(604)}
  +
  M_{b{\tsy}a{\tsx}d{\tsv}f{\tsv}e{\tsy}c}^{(604)}
{\tsz}\big)
\nonumber\\
&&
+\,
\frac{1}{2}
\big(
  M_{a{\tsx}b{\tsy}c{\tsy}e{\tsy}d{\tsv}f}^{(604)}
  +
  M_{f{\tsv}a{\tsx}b{\tsy}d{\tsy}c{\tsy}e}^{(604)}
  + 
  M_{e{\tsv}f{\tsv}a{\tsx}c{\tsy}b{\tsy}d}^{(604)}
  +
  M_{d{\tsy}e{\tsv}f{\tsx}b{\tsy}a{\tsx}c}^{(604)}
  + 
  M_{c{\tsy}d{\tsy}e{\tsy}a{\tsv}f{\tsx}b}^{(604)}
  +
  M_{b{\tsy}c{\tsy}d{\tsv}f{\tsv}e{\tsy}a}^{(604)}
\nonumber\\
&&\phantom{%
+\,
\frac{1}{2}
\big(}%
  +
  M_{b{\tsy}e{\tsy}c{\tsx}f{\tsv}d{\tsy}a}^{(604)}
  +
  M_{a{\tsx}d{\tsy}b{\tsy}e{\tsy}c{\tsx}f}^{(604)}
  +
  M_{f{\tsv}c{\tsy}a{\tsx}d{\tsy}b{\tsy}e}^{(604)}
  +
  M_{b{\tsy}e{\tsv}f{\tsv}c{\tsy}d{\tsy}a}^{(604)}
  +
  M_{a{\tsx}d{\tsy}e{\tsy}b{\tsy}c{\tsx}f}^{(604)}
  +
  M_{f{\tsv}c{\tsy}d{\tsy}a{\tsx}b{\tsy}e}^{(604)}
{\tsz}\big)
\nonumber\\
&&
-\,
\frac{1}{2}
\big(
  M_{f{\tsx}b{\tsy}c{\tsy}d{\tsy}e{\tsy}a}^{(604)}
  +
  M_{e{\tsy}a{\tsx}b{\tsy}c{\tsy}d{\tsv}f}^{(604)}
  + 
  M_{d{\tsv}f{\tsv}a{\tsx}b{\tsy}c{\tsy}e}^{(604)}
  +
  M_{c{\tsy}e{\tsv}f{\tsv}a{\tsx}b{\tsy}d}^{(604)}
  + 
  M_{b{\tsy}d{\tsy}e{\tsv}f{\tsv}a{\tsx}c}^{(604)}
  +
  M_{a{\tsx}c{\tsy}d{\tsy}e{\tsv}f{\tsy}b}^{(604)}
\nonumber\\
&&\phantom{%
-\,
\frac{1}{2}
\big(}%
  +
  M_{f{\tsy}b{\tsy}c{\tsy}e{\tsy}d{\tsy}a}^{(604)}
  +
  M_{e{\tsy}a{\tsx}b{\tsy}d{\tsy}c{\tsx}f}^{(604)}
  + 
  M_{d{\tsv}f{\tsv}a{\tsx}c{\tsy}b{\tsy}e}^{(604)}
  +
  M_{c{\tsy}e{\tsv}f{\tsy}b{\tsy}a{\tsx}d}^{(604)}
  + 
  M_{b{\tsy}d{\tsy}e{\tsy}a{\tsv}f{\tsv}c}^{(604)}
  +
  M_{a{\tsx}c{\tsy}d{\tsv}f{\tsv}e{\tsy}b}^{(604)}
{\tsz}\big)
\nonumber\\
&&
+\,
\frac{3}{2}
\big(
  M_{a{\tsx}b{\tsy}c{\tsx}f{\tsv}e{\tsy}d}^{(604)}
  +
  M_{f{\tsv}a{\tsx}b{\tsy}e{\tsy}d{\tsy}c}^{(604)}
  + 
  M_{e{\tsv}f{\tsv}a{\tsx}d{\tsy}c{\tsy}b}^{(604)}
  +
  M_{a{\tsx}b{\tsv}f{\tsv}c{\tsy}e{\tsy}d}^{(604)}
  + 
  M_{f{\tsv}a{\tsy}e{\tsy}b{\tsy}d{\tsy}c}^{(604)}
  +
  M_{e{\tsv}f{\tsv}d{\tsy}a{\tsx}c{\tsy}b}^{(604)}
{\tsz}\big)
\nonumber\\
&&
-\,
\frac{1}{2}
\big(
  M_{a{\tsx}b{\tsy}c{\tsy}e{\tsv}f{\tsv}d}^{(604)\prime}
  +
  M_{b{\tsy}c{\tsy}d{\tsv}f{\tsv}a{\tsy}e}^{(604)\prime}
  +
  M_{c{\tsy}d{\tsy}e{\tsy}a{\tsx}b{\tsv}f}^{(604)\prime}
  +
  M_{d{\tsy}e{\tsv}f{\tsx}b{\tsy}c{\tsy}a}^{(604)\prime}
  +
  M_{e{\tsv}f{\tsv}a{\tsx}c{\tsy}d{\tsy}b}^{(604)\prime}
  +
  M_{f{\tsv}a{\tsx}b{\tsy}d{\tsy}e{\tsy}c}^{(604)\prime}
{\tsz}\big)
\nonumber\\
&&
+\,
\frac{7}{2}
\big(
  M_{a{\tsx}b{\tsy}e{\tsy}d{\tsv}f{\tsv}c}^{(604)\prime}
  +
  M_{b{\tsy}c{\tsx}f{\tsv}e{\tsy}a{\tsx}d}^{(604)\prime}
  +
  M_{c{\tsy}d{\tsy}a{\tsv}f{\tsy}b{\tsy}e}^{(604)\prime}
{\tsz}\big)
\,+\,
3
\big(
  M_{a{\tsx}b{\tsy}c{\tsy}d{\tsy}e{\tsv}f}^{(604)\prime}
  +
  M_{f{\tsv}a{\tsx}b{\tsy}c{\tsy}d{\tsy}e}^{(604)\prime}
{\tsz}\big)
\,+\,
\frac{9}{2}
  M_{a{\tsx}d{\tsy}b{\tsy}e{\tsy}c{\tsx}f}^{(604)\prime}
\nonumber\\&&
=\,
\frac{3}{2}
\big(
  d_{a{\tsx}b{\tsy}c}
  {\hspace{1pt}}
  d_{d{\tsy}e{\tsv}f}
  +
  d_{f{\tsv}a{\tsx}b}
  {\hspace{1pt}}
  d_{c{\tsy}d{\tsy}e}
  +
  d_{e{\tsv}f{\tsv}a}
  {\hspace{1pt}}
  d_{{\tsx}b{\tsy}c{\tsy}d}
  +
  d_{a{\tsx}b{\tsy}d}
  {\hspace{1pt}}
  d_{e{\tsv}f{\tsv}c}
  +
  d_{{\tsx}b{\tsy}c{\tsy}e}
  {\hspace{1pt}}
  d_{f{\tsv}a{\tsx}d}
\nonumber\\&&\phantom{%
=\,
\frac{3}{2}
\big(}\hspace{-2.5pt}%
  +
  d_{c{\tsy}d{\tsv}f}
  {\hspace{1pt}}
  d_{a{\tsx}b{\tsy}e}
  +
  d_{d{\tsy}e{\tsy}a}
  {\hspace{1pt}}
  d_{{\tsx}b{\tsy}c{\tsx}f}
  +
  d_{e{\tsv}f{\tsy}b}
  {\hspace{1pt}}
  d_{c{\tsy}d{\tsy}a}
  +
  d_{f{\tsv}a{\tsx}c}
  {\hspace{1pt}}
  d_{d{\tsy}e{\tsy}b}
  -
  d_{a{\tsx}c{\tsy}e}
  {\hspace{1pt}}
  d_{{\tsx}b{\tsy}d{\tsx}f}
{\tsx}\big)
\nonumber\\&&\phantom{%
=\,}%
-
\frac{1}{18}
\big(
  \delta_{a{\tsx}b}
  {\hspace{1pt}}
  \delta_{e{\tsy}d}
  {\hspace{1pt}}
  \delta_{{\tsv}f{\tsy}c}
  +
  \delta_{{\tsx}b{\tsy}c}
  {\hspace{1pt}}
  \delta_{{\tsv}f{\tsv}e}
  {\hspace{1pt}}
  \delta_{a{\tsx}d}
  +
  \delta_{c{\tsy}d}
  {\hspace{1pt}}
  \delta_{a{\tsv}f}
  {\hspace{1pt}}
  \delta_{{\tsx}b{\tsy}e}
  +
  \delta_{a{\tsx}b}
  {\hspace{1pt}}
  \delta_{c{\tsy}e}
  {\hspace{1pt}}
  \delta_{{\tsv}f{\tsv}d}
  +
  \delta_{{\tsx}b{\tsy}c}
  {\hspace{1pt}}
  \delta_{d{\tsv}f}
  {\hspace{1pt}}
  \delta_{a{\tsy}e}
\nonumber\\&&\phantom{%
=\,
-
\frac{1}{18}
\big(}\hspace{-1.5pt}%
  +
  \delta_{c{\tsy}d}
  {\hspace{1pt}}
  \delta_{e{\tsy}a}
  {\hspace{1pt}}
  \delta_{{\tsx}b{\tsv}f}
  +
  \delta_{d{\tsy}e}
  {\hspace{1pt}}
  \delta_{{\tsv}f{\tsx}b}
  {\hspace{1pt}}
  \delta_{c{\tsy}a}
  +
  \delta_{e{\tsv}f}
  {\hspace{1pt}}
  \delta_{a{\tsx}c}
  {\hspace{1pt}}
  \delta_{d{\tsy}b}
  +
  \delta_{{\tsv}f{\tsv}a}
  {\hspace{1pt}}
  \delta_{{\tsx}b{\tsy}d}
  {\hspace{1pt}}
  \delta_{e{\tsy}c}
{\tsy}\big)
\nonumber\\&&\phantom{%
=\,}
-
\frac{1}{9}
\big(
  \delta_{a{\tsx}d}
  {\hspace{1pt}}
  \delta_{{\tsv}f{\tsx}b}
  {\hspace{1pt}}
  \delta_{e{\tsy}c}
  +
  \delta_{{\tsx}b{\tsy}e}
  {\hspace{1pt}}
  \delta_{a{\tsx}c}
  {\hspace{1pt}}
  \delta_{{\tsv}f{\tsv}d}
  +
  \delta_{c{\tsx}f}
  {\hspace{1pt}}
  \delta_{{\tsx}b{\tsy}d}
  {\hspace{1pt}}
  \delta_{a{\tsy}e}
{\tsy}\big)
-
\frac{1}{6}{\hspace{1pt}}
  \delta_{a{\tsx}d}
  {\hspace{1pt}}
  \delta_{{\tsx}b{\tsy}e}
  {\hspace{1pt}}
  \delta_{c{\tsx}f}
\end{eqnarray}
\ {\bf Total symmetric loop diagrams:}
\begin{eqnarray}
&&
M_{\underline{a{\tsx}b{\tsx}c{\tsy}d}}^{(414)}
\,=\,
\frac{3N + 10}{72}\!\;
  \delta_{\underline{a{\tsx}b}}
  {\hspace{1pt}}
  \delta_{\underline{c{\tsy}d}}
\\
&&
M_{\underline{a{\tsx}b{\tsx}c{\tsy}d{\tsy}e}}^{(515)}
\,=\,
-\,
\frac{5N - 10}{144}\!\;
  \delta_{\underline{a{\tsx}b}}
  {\hspace{1pt}}
  d_{\underline{c{\tsy}d{\tsx}e}}
\\
&&
M_{\underline{a{\tsx}b{\tsx}c{\tsy}d{\tsy}e{\tsv}f}}^{(616)}
\,=\,
\frac{N-2}{144}\!\;
  d_{\underline{a{\tsx}b{\tsx}c}}
  {\hspace{1pt}}
  d_{\underline{d{\tsy}e{\tsv}f}}
+
\frac{9N+46}{864}\!\>
  \delta_{\underline{a{\tsx}b}}
  {\hspace{1pt}}
  \delta_{\underline{c{\tsy}d}}
  {\hspace{1pt}}
  \delta_{\underline{e{\tsv}f}}
\qquad\quad
\end{eqnarray}
\ {\bf Partial symmetric loop diagrams:}
\begin{eqnarray}
&&
M_{a{\tsx}b{\tsx}c{\tsy}d}^{(414)}
+
\frac{N-6}{24}
\big(
  M_{a{\tsx}c{\tsy}b{\tsy}d}^{(404)}
  -
  M_{a{\tsx}b{\tsy}c{\tsy}d}^{(404)}
  -
  M_{a{\tsx}d{\tsy}b{\tsy}c}^{(404)}
{\tsz}\big)
\nonumber\\&&
=\,
\frac{N+4}{36}{\hspace{1pt}}
 \big({\tsz}
  \delta_{a{\tsx}b} {\hspace{1pt}} \delta_{c{\tsy}d}
  +
  \delta_{a{\tsx}d} {\hspace{1pt}} \delta_{{\tsx}b{\tsy}c}
 {\tsz}\big)
-
\frac{N}{36}{\hspace{2pt}}
  \delta_{a{\tsy}c} {\hspace{1pt}} \delta_{{\tsx}b{\tsy}d}
\\
&&
M_{a{\tsx}b{\tsx}c{\tsy}\underline{d{\tsy}e}}^{(515)}
\,+
\frac{N-10}{24}
\big(
  2
  M_{a{\tsx}c{\tsy}\underline{d{\tsy}e}{\tsy}b}^{(503)}
  +
  M_{\underline{d}{\tsx}a{\tsx}b{\tsy}c{\tsy}\underline{e}}^{(503)}
  +
  M_{\underline{d}{\tsy}b{\tsx}a{\tsy}c{\tsy}\underline{e}}^{(503)}
  +
  M_{\underline{d}{\tsy}b{\tsy}c{\tsy}a{\tsy}\underline{e}}^{(503)}
{\tsz}\big)
\nonumber\\
&&=\,
-\hspace{3pt}
\frac{3N+10}{288}{\hspace{1pt}}
 \big({\tsz}
  \delta_{a{\tsx}b} {\hspace{1pt}} d_{c{\tsy}d{\tsx}e}
  +
  \delta_{c{\tsy}b} {\hspace{1pt}} d_{a{\tsx}d{\tsy}e}
  +
  \delta_{d{\tsx}e} {\hspace{1pt}} d_{a{\tsx}b{\tsx}c}
 {\tsz}\big)
\nonumber\\
&&\phantom{=}\,\hspace{1pt}
+
\frac{5N-26}{288}{\hspace{1pt}}
 \big({\tsz}
  \delta_{a{\tsx}c} {\hspace{1pt}} d_{{\tsx}b{\tsx}d{\tsy}e}
  +
  \delta_{{\tsx}b{\tsx}d} {\hspace{1pt}} d_{a{\tsx}c{\tsy}e}
  +
  \delta_{{\tsx}b{\tsx}e} {\hspace{1pt}} d_{a{\tsx}c{\tsx}d}
 {\tsz}\big)
\nonumber\\
&&\phantom{=}\,\hspace{1pt}
+
\frac{N-18}{288}{\hspace{1pt}}
 \big({\tsz}
  \delta_{a{\tsx}d} {\hspace{1pt}} d_{c{\tsy}b{\tsy}e}
  +
  \delta_{a{\tsx}e} {\hspace{1pt}} d_{c{\tsy}b{\tsx}d}
  +
  \delta_{c{\tsy}d} {\hspace{1pt}} d_{a{\tsx}b{\tsy}e}
  +
  \delta_{c{\tsy}e} {\hspace{1pt}} d_{a{\tsx}b{\tsy}d}
 {\tsz}\big)
\\
&&
M_{a{\tsx}b{\tsx}c{\tsz}\underline{d{\tsy}e{\tsv}f}}^{(616)}
\,-
\frac{N-2}{24}
\big(
  M_{\underline{d}{\tsy}b{\tsy}c{\tsy}\underline{e{\tsv}f}{\tsv}a}^{(604)}
  +
  M_{c{\tsy}a{\tsx}b{\tsy}\underline{d{\tsy}e{\tsv}f}}^{(604)}
  +
  M_{\underline{d}{\tsy}b{\tsy}a{\tsy}\underline{e{\tsv}f}{\tsv}c}^{(604)}
  +
  M_{\underline{d}{\tsy}a{\tsy}c{\tsy}\underline{e{\tsv}f}{\tsx}b}^{(604)}
  +
  M_{c{\tsy}a{\tsx}\underline{d}{\tsy}b{\tsy}\underline{e{\tsv}f}}^{(604)}
  +
  M_{\underline{d}{\tsy}c{\tsy}a{\tsx}\underline{e{\tsv}f}{\tsx}b}^{(604)}
{\tsz}\big)
\nonumber\\
&&\phantom{%
M_{a{\tsx}b{\tsx}c{\tsz}\underline{d{\tsy}e{\tsv}f}}^{(616)}
}\hspace{0pt}%
\,+
\frac{N-5}{6}{\hspace{1pt}}
  M_{a{\tsy}\underline{d}{\tsy}b{\tsy}\underline{e}{\tsy}c{\tsx}\underline{f}}^{(604)\prime}
\,+
\frac{N-4}{6}
\big(
  M_{a{\tsx}b{\tsy}c{\tsy}\underline{d{\tsy}e{\tsv}f}}^{(604)\prime}
  +
  M_{c{\tsy}b{\tsy}a{\tsx}\underline{d{\tsy}e{\tsv}f}}^{(604)\prime}
{\tsz}\big)
\nonumber\\&&
=\,
\frac{5N - 22}{144}{\hspace{1pt}}
  d_{a{\tsx}b{\tsy}c}
  {\hspace{1pt}}
  d_{d{\tsy}e{\tsv}f}
+
\frac{N - 4}{24}{\hspace{1pt}}
  d_{a{\tsx}c{\tsy}\underline{d}}
  {\hspace{1pt}}
  d_{{\tsx}b{\tsy}\underline{e{\tsv}f}}
+
\frac{19N - 82}{144}
\big({\tsz}
  d_{a{\tsx}b{\tsy}\underline{d}}
  {\hspace{1pt}}
  d_{c{\tsy}\underline{e{\tsv}f}}
  +
  d_{{\tsx}b{\tsy}c{\tsy}\underline{d}}
  {\hspace{1pt}}
  d_{a{\tsy}\underline{e{\tsv}f}}
{\tsz}\big)
\nonumber\\&&\phantom{=\,
}%
-
\frac{2N - 7}{54}
  \delta_{a{\tsx}\underline{d}}
  {\hspace{1pt}}
  \delta_{{\tsx}b{\tsz}\underline{e}}
  {\hspace{1pt}}
  \delta_{c{\tsx}\underline{f}}
-
\frac{17(N - 2)}{864}{\hspace{1pt}}
  \delta_{a{\tsx}c}
  {\hspace{1pt}}
  \delta_{{\tsx}b{\tsz}\underline{d}}
  {\hspace{1pt}}
  \delta_{\underline{e{\tsv}f}}
-
\frac{7N - 78}{864}
\big(
  \delta_{a{\tsx}b}
  {\hspace{1pt}}
  \delta_{c{\tsz}\underline{d}}
  {\hspace{1pt}}
  \delta_{\underline{e{\tsv}f}}
  +
  \delta_{c{\tsy}b}
  {\hspace{1pt}}
  \delta_{a{\tsx}\underline{d}}
  {\hspace{1pt}}
  \delta_{\underline{e{\tsv}f}}
\big)
\nonumber\\
&&
\end{eqnarray}
\ {\bf Non-symmetric loop diagrams:} (for $N \!=\! 5$)
\begin{eqnarray}
&&
M_{a{\tsx}b{\tsx}c{\tsy}d{\tsy}e}^{(515)}
\,-
\frac{5}{24}
\big(
  M_{e{\tsx}c{\tsx}a{\tsx}b{\tsy}d}^{(503)}
  +
  M_{d{\tsx}a{\tsx}b{\tsy}c{\tsy}e}^{(503)}
  +
  M_{b{\tsx}e{\tsx}c{\tsy}d{\tsy}a}^{(503)}
  +
  M_{c{\tsx}a{\tsx}d{\tsy}b{\tsy}e}^{(503)}
  +
  M_{d{\tsx}b{\tsx}e{\tsy}a{\tsy}c}^{(503)}
{\tsz}\big)
\nonumber\\
&&=\,
-\,
\frac{25}{288}{\hspace{1pt}}
 \big({\tsz}
  \delta_{a{\tsx}b} {\hspace{1pt}} d_{c{\tsy}d{\tsy}e}
  +
  \delta_{{\tsx}b{\tsx}c} {\hspace{1pt}} d_{d{\tsy}e{\tsy}a}
  +
  \delta_{c{\tsx}d} {\hspace{1pt}} d_{e{\tsy}a{\tsx}b}
  +
  \delta_{d{\tsx}e} {\hspace{1pt}} d_{a{\tsx}b{\tsx}c}
  +
  \delta_{e{\tsx}a} {\hspace{1pt}} d_{{\tsx}b{\tsy}c{\tsy}d}
 {\tsz}\big)
\nonumber\\
&&\phantom{=}\,\hspace{1pt}
-
\frac{1}{288}{\hspace{1pt}}
 \big({\tsz}
  \delta_{e{\tsx}b} {\hspace{1pt}} d_{a{\tsx}c{\tsy}d}
  +
  \delta_{a{\tsx}c} {\hspace{1pt}} d_{{\tsx}b{\tsy}d{\tsy}e}
  +
  \delta_{{\tsx}b{\tsx}d} {\hspace{1pt}} d_{c{\tsz}e{\tsx}a}
  +
  \delta_{c{\tsx}e} {\hspace{1pt}} d_{d{\tsx}a{\tsx}b}
  +
  \delta_{d{\tsx}a} {\hspace{1pt}} d_{e{\tsy}b{\tsx}c}
 {\tsz}\big)
\nonumber\\
&&\phantom{=}\,\hspace{1pt}
+
\frac{7}{64\sqrt{3}}{\hspace{2pt}}
  \ep_{a{\tsx}b{\tsx}c{\tsy}d{\tsx}e}
\\
&&
M_{a{\tsx}b{\tsx}c{\tsy}d{\tsy}e{\tsv}f}^{(616)}
\,-
\frac{13}{168}
\big(
  M_{a{\tsx}b{\tsy}c{\tsy}d{\tsy}e{\tsv}f}^{(604)}
  +
  M_{f{\tsv}a{\tsx}b{\tsy}c{\tsy}d{\tsy}e}^{(604)}
  +
  M_{e{\tsv}f{\tsv}a{\tsx}b{\tsy}c{\tsy}d}^{(604)}
  +
  M_{d{\tsy}e{\tsv}f{\tsv}a{\tsx}b{\tsy}c}^{(604)}
  +
  M_{c{\tsy}d{\tsy}e{\tsv}f{\tsv}a{\tsx}b}^{(604)}
  +
  M_{b{\tsy}c{\tsy}d{\tsy}e{\tsv}f{\tsv}a}^{(604)}
{\tsz}\big)
\nonumber\\
&&\phantom{%
M_{a{\tsx}b{\tsx}c{\tsy}d{\tsy}e{\tsv}f}^{(616)}
}%
\,+
\frac{11}{84}
\big(
  M_{a{\tsv}f{\tsv}c{\tsy}e{\tsy}d{\tsy}b}^{(604)}
  +
  M_{f{\tsv}e{\tsy}b{\tsy}d{\tsy}c{\tsy}a}^{(604)}
  +
  M_{e{\tsy}d{\tsy}a{\tsy}c{\tsy}b{\tsv}f}^{(604)}
  +
  M_{d{\tsy}c{\tsx}f{\tsx}b{\tsy}a{\tsy}e}^{(604)}
  +
  M_{c{\tsy}b{\tsy}e{\tsy}a{\tsv}f{\tsv}d}^{(604)}
  +
  M_{b{\tsy}a{\tsx}d{\tsv}f{\tsv}e{\tsy}c}^{(604)}
{\tsz}\big)
\nonumber\\
&&\phantom{%
M_{a{\tsx}b{\tsx}c{\tsy}d{\tsy}e{\tsv}f}^{(616)}
}%
\,-
\frac{11}{84}
\big(
  M_{f{\tsx}b{\tsy}c{\tsy}d{\tsy}e{\tsy}a}^{(604)}
  +
  M_{e{\tsy}a{\tsx}b{\tsy}c{\tsy}d{\tsv}f}^{(604)}
  +
  M_{d{\tsv}f{\tsv}a{\tsx}b{\tsy}c{\tsy}e}^{(604)}
  +
  M_{c{\tsy}e{\tsv}f{\tsv}a{\tsx}b{\tsy}d}^{(604)}
  +
  M_{b{\tsy}d{\tsy}e{\tsv}f{\tsv}a{\tsy}c}^{(604)}
  +
  M_{a{\tsy}c{\tsy}d{\tsy}e{\tsv}f{\tsx}b}^{(604)}
{\tsz}\big)
\nonumber\\
&&\phantom{%
M_{a{\tsx}b{\tsx}c{\tsy}d{\tsy}e{\tsv}f}^{(616)}
}%
\,+
\frac{11}{336}
\big(
  M_{f{\tsx}b{\tsy}c{\tsy}e{\tsy}d{\tsy}a}^{(604)}
  +
  M_{e{\tsy}a{\tsx}b{\tsy}d{\tsy}c{\tsx}f}^{(604)}
  +
  M_{d{\tsv}f{\tsv}a{\tsy}c{\tsy}b{\tsy}e}^{(604)}
  +
  M_{c{\tsy}e{\tsv}f{\tsx}b{\tsy}a{\tsx}d}^{(604)}
  +
  M_{b{\tsy}d{\tsy}e{\tsy}a{\tsv}f{\tsv}c}^{(604)}
  +
  M_{a{\tsy}c{\tsy}d{\tsv}f{\tsv}e{\tsy}b}^{(604)}
{\tsz}\big)
\nonumber\\
&&\phantom{%
M_{a{\tsx}b{\tsx}c{\tsy}d{\tsy}e{\tsv}f}^{(616)}
}%
\,-
\frac{11}{336}
\big(
  M_{e{\tsy}b{\tsy}c{\tsy}d{\tsy}a{\tsv}f}^{(604)}
  +
  M_{d{\tsy}a{\tsx}b{\tsy}c{\tsx}f{\tsv}e}^{(604)}
  +
  M_{c{\tsx}f{\tsv}a{\tsx}b{\tsy}e{\tsy}d}^{(604)}
  +
  M_{b{\tsy}e{\tsv}f{\tsv}a{\tsx}d{\tsy}c}^{(604)}
  +
  M_{a{\tsx}d{\tsy}e{\tsv}f{\tsv}c{\tsy}b}^{(604)}
  +
  M_{f{\tsv}c{\tsy}d{\tsy}e{\tsy}b{\tsy}a}^{(604)}
{\tsz}\big)
\nonumber\\
&&\phantom{%
M_{a{\tsx}b{\tsx}c{\tsy}d{\tsy}e{\tsv}f}^{(616)}
}%
\,-
\frac{11}{168}
\big(
  M_{a{\tsx}b{\tsy}c{\tsy}e{\tsy}d{\tsv}f}^{(604)}
  +
  M_{f{\tsv}a{\tsx}b{\tsy}d{\tsy}c{\tsy}e}^{(604)}
  +
  M_{e{\tsv}f{\tsv}a{\tsy}c{\tsy}b{\tsy}d}^{(604)}
  +
  M_{d{\tsy}e{\tsv}f{\tsx}b{\tsy}a{\tsy}c}^{(604)}
  +
  M_{c{\tsy}d{\tsy}e{\tsy}a{\tsv}f{\tsx}b}^{(604)}
  +
  M_{b{\tsy}c{\tsy}d{\tsv}f{\tsv}e{\tsy}a}^{(604)}
{\tsz}\big)
\nonumber\\
&&\phantom{%
M_{a{\tsx}b{\tsx}c{\tsy}d{\tsy}e{\tsv}f}^{(616)}
}%
\,+
\frac{1}{6}
\big(
  M_{a{\tsx}b{\tsy}c{\tsx}f{\tsv}e{\tsy}d}^{(604)}
  +
  M_{f{\tsv}a{\tsx}b{\tsy}e{\tsy}d{\tsy}c}^{(604)}
  +
  M_{e{\tsv}f{\tsv}a{\tsx}d{\tsy}c{\tsy}b}^{(604)}
  +
  M_{a{\tsx}b{\tsv}f{\tsv}c{\tsy}e{\tsy}d}^{(604)}
  +
  M_{f{\tsv}a{\tsy}e{\tsy}b{\tsy}d{\tsy}c}^{(604)}
  +
  M_{e{\tsv}f{\tsv}d{\tsy}a{\tsy}c{\tsy}b}^{(604)}
{\tsz}\big)
\nonumber\\
&&\phantom{%
M_{a{\tsx}b{\tsx}c{\tsy}d{\tsy}e{\tsv}f}^{(616)}
}%
\,+
\frac{23}{84}
\big(
  M_{d{\tsy}b{\tsy}c{\tsx}f{\tsv}e{\tsy}a}^{(604)}
  +
  M_{c{\tsy}a{\tsx}b{\tsy}e{\tsy}d{\tsv}f}^{(604)}
  +
  M_{b{\tsv}f{\tsv}a{\tsx}d{\tsy}c{\tsy}e}^{(604)}
  +
  M_{d{\tsy}b{\tsv}f{\tsv}c{\tsy}e{\tsy}a}^{(604)}
  +
  M_{c{\tsy}a{\tsy}e{\tsy}b{\tsy}d{\tsv}f}^{(604)}
  +
  M_{b{\tsv}f{\tsv}d{\tsy}a{\tsy}c{\tsy}e}^{(604)}
{\tsz}\big)
\nonumber\\
&&\phantom{%
M_{a{\tsx}b{\tsx}c{\tsy}d{\tsy}e{\tsv}f}^{(616)}
}%
\,+
\frac{9}{112}
\big(
  M_{b{\tsy}e{\tsy}c{\tsx}f{\tsv}d{\tsy}a}^{(604)}
  +
  M_{a{\tsx}d{\tsy}b{\tsy}e{\tsy}c{\tsx}f}^{(604)}
  +
  M_{f{\tsv}c{\tsy}a{\tsx}d{\tsy}b{\tsy}e}^{(604)}
  +
  M_{b{\tsy}e{\tsv}f{\tsv}c{\tsy}d{\tsy}a}^{(604)}
  +
  M_{a{\tsx}d{\tsy}e{\tsy}b{\tsy}c{\tsx}f}^{(604)}
  +
  M_{f{\tsv}c{\tsy}d{\tsy}a{\tsx}b{\tsy}e}^{(604)}
{\tsz}\big)
\nonumber\\
&&=\,
\frac{3}{56}\hspace{1pt}
  d_{a{\tsy}c{\tsy}e}
  {\hspace{1pt}}
  d_{{\tsx}b{\tsy}d{\tsv}f}
+
\frac{295}{1008}
\big(
  d_{a{\tsx}b{\tsy}c}
  {\hspace{1pt}}
  d_{d{\tsy}e{\tsv}f}
  +
  d_{f{\tsv}a{\tsx}b}
  {\hspace{1pt}}
  d_{c{\tsy}d{\tsy}e}
  +
  d_{e{\tsv}f{\tsv}a}
  {\hspace{1pt}}
  d_{{\tsx}b{\tsy}c{\tsy}d}
\big)
\nonumber\\&&\phantom{%
=\,}\hspace{-1.5pt}%
-
\frac{17}{112}
\big(
  d_{a{\tsx}b{\tsy}d}
  {\hspace{1pt}}
  d_{e{\tsv}f{\tsv}c}
  +
  d_{{\tsx}b{\tsy}c{\tsy}e}
  {\hspace{1pt}}
  d_{f{\tsv}a{\tsx}d}
  +
  d_{c{\tsy}d{\tsv}f}
  {\hspace{1pt}}
  d_{a{\tsx}b{\tsy}e}
  +
  d_{d{\tsy}e{\tsy}a}
  {\hspace{1pt}}
  d_{{\tsx}b{\tsy}c{\tsx}f}
  +
  d_{e{\tsv}f{\tsx}b}
  {\hspace{1pt}}
  d_{c{\tsy}d{\tsy}a}
  +
  d_{f{\tsv}a{\tsy}c}
  {\hspace{1pt}}
  d_{d{\tsy}e{\tsy}b}
\big)
\nonumber\\&&\phantom{%
=\,}\hspace{-1.5pt}%
+
\frac{7}{96}{\hspace{1pt}}
  \delta_{a{\tsx}d}
  {\hspace{1pt}}
  \delta_{{\tsx}b{\tsy}e}
  {\hspace{1pt}}
  \delta_{c{\tsx}f}
+
\frac{13}{224}
\big(
  \delta_{a{\tsx}b}
  {\hspace{1pt}}
  \delta_{e{\tsy}d}
  {\hspace{1pt}}
  \delta_{{\tsv}f{\tsv}c}
  +
  \delta_{{\tsx}b{\tsy}c}
  {\hspace{1pt}}
  \delta_{{\tsv}f{\tsv}e}
  {\hspace{1pt}}
  \delta_{a{\tsx}d}
  +
  \delta_{c{\tsy}d}
  {\hspace{1pt}}
  \delta_{a{\tsv}f}
  {\hspace{1pt}}
  \delta_{{\tsx}b{\tsy}e}
\big)
\nonumber\\&&\phantom{%
=\,}\hspace{-1.5pt}%
+
\frac{131}{6048}
\big(
  \delta_{a{\tsx}b}
  {\hspace{1pt}}
  \delta_{c{\tsy}d}
  {\hspace{1pt}}
  \delta_{e{\tsv}f}
  +
  \delta_{{\tsv}f{\tsv}a}
  {\hspace{1pt}}
  \delta_{{\tsx}b{\tsy}c}
  {\hspace{1pt}}
  \delta_{d{\tsy}e}
\big)
+
\frac{313}{6048}
\big(
  \delta_{a{\tsx}d}
  {\hspace{1pt}}
  \delta_{{\tsv}f{\tsx}b}
  {\hspace{1pt}}
  \delta_{e{\tsy}c}
  +
  \delta_{{\tsx}b{\tsy}e}
  {\hspace{1pt}}
  \delta_{a{\tsy}c}
  {\hspace{1pt}}
  \delta_{{\tsv}f{\tsv}d}
  +
  \delta_{c{\tsx}f}
  {\hspace{1pt}}
  \delta_{{\tsx}b{\tsy}d}
  {\hspace{1pt}}
  \delta_{a{\tsy}e}
\big)
\nonumber\\&&\phantom{%
=\,}\hspace{-1.5pt}%
-
\frac{89}{6048}
\big(
  \delta_{a{\tsx}b}
  {\hspace{1pt}}
  \delta_{c{\tsy}e}
  {\hspace{1pt}}
  \delta_{{\tsv}f{\tsv}d}
  +
  \delta_{{\tsx}b{\tsy}c}
  {\hspace{1pt}}
  \delta_{d{\tsv}f}
  {\hspace{1pt}}
  \delta_{a{\tsy}e}
  +
  \delta_{c{\tsy}d}
  {\hspace{1pt}}
  \delta_{e{\tsy}a}
  {\hspace{1pt}}
  \delta_{{\tsx}b{\tsv}f}
  +
  \delta_{d{\tsy}e}
  {\hspace{1pt}}
  \delta_{{\tsv}f{\tsx}b}
  {\hspace{1pt}}
  \delta_{c{\tsy}a}
  +
  \delta_{e{\tsv}f}
  {\hspace{1pt}}
  \delta_{a{\tsy}c}
  {\hspace{1pt}}
  \delta_{d{\tsy}b}
  +
  \delta_{{\tsv}f{\tsv}a}
  {\hspace{1pt}}
  \delta_{{\tsx}b{\tsy}d}
  {\hspace{1pt}}
  \delta_{e{\tsy}c}
\big)
\nonumber\\&&
\end{eqnarray}

\newpage

\newcommand{\commutator}[2]{[\, #1 {\dbltinyspace}, #2 \,]}
\newcommand{\commutatorbig}[2]{\big[\, #1 \,,\, #2 \,\big]}
\newcommand{\commutatorBig}[2]{\Big[\, #1 \,,\, #2 \,\Big]}
\newcommand{\commutatorbigg}[2]{\bigg[\tinyspace #1 \>,\;\! #2 \bigg]}
\newcommand{\anticommutator}[2]{\{\tinyspace #1 \;\!,\>\! #2 \tinyspace\}}
\newcommand{\anticommutatorbig}[2]{\big\{\tinyspace #1 \,,\, #2 \tinyspace\big\}}
\newcommand{\anticommutatorBig}[2]{\Big\{\, #1 \,,\, #2 \,\Big\}}
\newcommand{\anticommutatorbigg}[2]{\bigg\{\negtinyspace #1 \>,\;\! #2 \bigg\}}

\section{
Higher order commutators of the $W$-algebra}
\label{app:Walgebra}

The commutation relation of $\widetilde{W}^{(4)}(z)$ is 
\begin{eqnarray}
\commutator{\widetilde{W}^{(4)}(z)}{\widetilde{W}^{(4)}(w)}
\!&=&\!
\frac{6{\tinyspace} \widetilde{W}^{(6)}(w)}
     {(z{\negdbltinyspace}-{\negdbltinyspace}w)^2}
\,+\,
\frac{3{\tinyspace} \partial \widetilde{W}^{(6)}(w)}
     {z{\negdbltinyspace}-{\negdbltinyspace}w}
\nonumber\\&&\!
+\;
\frac{114{\tinyspace}C_3{\tinyspace} \widetilde{W}^{\prime(4)}(w)}
     {5{\tinyspace} (z{\negdbltinyspace}-{\negdbltinyspace}w)^4}
\,+\,
\frac{57{\tinyspace}C_3{\tinyspace} \partial \widetilde{W}^{\prime(4)}(w)}
     {5{\tinyspace} (z{\negdbltinyspace}-{\negdbltinyspace}w)^3}
\nonumber\\&&\!
+\;
\frac{19{\tinyspace}C_3{\tinyspace} \partial^2 \widetilde{W}^{\prime(4)}(w)}
     {6{\tinyspace} (z{\negdbltinyspace}-{\negdbltinyspace}w)^2}
\,+\,
\frac{19{\tinyspace}C_3{\tinyspace} \partial^3 \widetilde{W}^{\prime(4)}(w)}
     {30{\tinyspace} (z{\negdbltinyspace}-{\negdbltinyspace}w)}
\nonumber\\&&\!
+\;
\frac{72{\tinyspace}C_4{\tinyspace} \widetilde{W}^{(2)}(w)}
     {5(z{\negdbltinyspace}-{\negdbltinyspace}w)^6}
\,+\,
\frac{36{\tinyspace}C_4{\tinyspace} \partial \widetilde{W}^{(2)}(w)}
     {5{\tinyspace} (z{\negdbltinyspace}-{\negdbltinyspace}w)^5}
\nonumber\\&&\!
+\;
\frac{54{\tinyspace}C_4{\tinyspace} \partial^2 \widetilde{W}^{(2)}(w)}
     {25{\tinyspace} (z{\negdbltinyspace}-{\negdbltinyspace}w)^4}
\,+\,
\frac{12{\tinyspace}C_4{\tinyspace} \partial^3 \widetilde{W}^{(2)}(w)}
     {25{\tinyspace} (z{\negdbltinyspace}-{\negdbltinyspace}w)^3}
\nonumber\\&&\!
+\;
\frac{3{\tinyspace}C_4{\tinyspace} \partial^4 \widetilde{W}^{(2)}(w)}
     {35{\tinyspace} (z{\negdbltinyspace}-{\negdbltinyspace}w)^2}
\,+\,
\frac{9{\tinyspace}C_4{\tinyspace} \partial^5 \widetilde{W}^{(2)}(w)}
     {700{\tinyspace} (z{\negdbltinyspace}-{\negdbltinyspace}w)}
\,+\,
\frac{9 N{\tinyspace}C_4}
     {5{\tinyspace} (z{\negdbltinyspace}-{\negdbltinyspace}w)^8}
\,,
\end{eqnarray}
where
\begin{equation}
C_3 \,\define\, \frac{3N \!+\! 26}{108}
\,,
\qquad\quad
C_4 \,\define\, C_2 C_3
\,,
\end{equation}
The operator $\widetilde{W}^{\prime(4)}(z)$ 
is not an independent operator, 
is expressed by $\widetilde{W}^{(4)}(z)$ and a new operator 
$\widetilde{W}^{(2,2)}(z)$ as 
\begin{eqnarray}
\widetilde{W}^{\prime(4)}(z)
\!&=&\!
\frac{1}{9 {\tinyspace} C_3}
\bigg(
  \frac{11 N {\negdbltinyspace}+{\negdbltinyspace} 52}{19}
  {\dbltinyspace} \widetilde{W}^{(4)}(z)
  +
  \frac{5 ( N {\negdbltinyspace}-{\negdbltinyspace} 2 )}{108}
  {\dbltinyspace} \widetilde{W}^{(2,2)}(z)
\bigg)
\,.
\end{eqnarray}
It should be noted that 
the structure of above commutatin relations are 
very similar to the structure of $W^{(1+\infty)}$ algebra. 

Now one has to incorporate this new operator 
into the commutation relations. 
\begin{eqnarray}
\commutator{\widetilde{W}^{(2)}(z)}{\widetilde{W}^{(2,2)}(w)}
\!&=&\!
\frac{4{\tinyspace} \widetilde{W}^{(2,2)}(w)}
     {(z{\negdbltinyspace}-{\negdbltinyspace}w)^2}
\,+\,
\frac{\partial \widetilde{W}^{(2,2)}(w)}
     {z{\negdbltinyspace}-{\negdbltinyspace}w}
\,+\,
\frac{324{\tinyspace}C_2{\tinyspace} \widetilde{W}^{(2)}(w)}
     {95{\tinyspace} (z{\negdbltinyspace}-{\negdbltinyspace}w)^4}
\,,
\\
\commutator{\widetilde{W}^{(3)}(z)}{\widetilde{W}^{(2,2)}(w)}
\!&=&\!
\frac{9{\tinyspace}C_2}{19{\tinyspace}C_3}
\bigg(
 \frac{5{\tinyspace} \widetilde{W}^{(3,2)}(w)}
      {(z{\negdbltinyspace}-{\negdbltinyspace}w)^2}
 \,+\,
 \frac{2{\tinyspace} \partial \widetilde{W}^{(3,2)}(w)}
      {(z{\negdbltinyspace}-{\negdbltinyspace}w)}
 \,+\,
\frac{\widetilde{W}^{(3,3)}(w)}
     {z{\negdbltinyspace}-{\negdbltinyspace}w}
\nonumber\\&&\!\phantom{%
\frac{9{\tinyspace}C_2}{19{\tinyspace}C_3}
\bigg(
}
 +\,
 \frac{54{\tinyspace}C_3{\tinyspace} \widetilde{W}^{(3)}(w)}
      {5{\tinyspace} (z{\negdbltinyspace}-{\negdbltinyspace}w)^4}
 \,+\,
 \frac{18{\tinyspace}C_3{\tinyspace} \partial \widetilde{W}^{(3)}(w)}
      {5{\tinyspace} (z{\negdbltinyspace}-{\negdbltinyspace}w)^3}
\nonumber\\&&\!\phantom{%
\frac{9{\tinyspace}C_2}{19{\tinyspace}C_3}
\bigg(
}
 +\,
 \frac{27{\tinyspace}C_3{\tinyspace} \partial^2 \widetilde{W}^{(3)}(w)}
      {35{\tinyspace} (z{\negdbltinyspace}-{\negdbltinyspace}w)^2}
 \,+\,
 \frac{9{\tinyspace}C_3{\tinyspace} \partial^3 \widetilde{W}^{(3)}(w)}
      {70{\tinyspace} (z{\negdbltinyspace}-{\negdbltinyspace}w)}
\bigg)
\,,
\\
\commutator{\widetilde{W}^{(4)}(z)}{\widetilde{W}^{(2,2)}(w)}
\!&=&\!
\frac{10{\tinyspace}C_2}{19{\tinyspace}C_3}
\bigg(
 \frac{6{\tinyspace} \widetilde{W}^{\prime(6)}(w)}
      {(z{\negdbltinyspace}-{\negdbltinyspace}w)^2}
 \,+\,
 \frac{3{\tinyspace} \partial \widetilde{W}^{\prime(6)}(w)}
      {z{\negdbltinyspace}-{\negdbltinyspace}w}
 \,+\,
 \frac{\widetilde{W}^{(4,3)}(w)}
      {z{\negdbltinyspace}-{\negdbltinyspace}w}
\nonumber\\&&\!\phantom{%
\frac{10{\tinyspace}C_2}{19{\tinyspace}C_3}
\bigg(
}
 +\,
 \frac{114{\tinyspace}C_3{\tinyspace} \widetilde{W}^{\prime\prime(4)}(w)}
      {5{\tinyspace} (z{\negdbltinyspace}-{\negdbltinyspace}w)^4}
 \,+\,
 \frac{57{\tinyspace}C_3{\tinyspace}
       \partial \widetilde{W}^{\prime\prime(4)}(w)}
      {5{\tinyspace} (z{\negdbltinyspace}-{\negdbltinyspace}w)^3}
\nonumber\\&&\!\phantom{%
\frac{10{\tinyspace}C_2}{19{\tinyspace}C_3}
\bigg(
}
 +\,
 \frac{19{\tinyspace}C_3{\tinyspace}
       \partial^2 \widetilde{W}^{\prime\prime(4)}(w)}
     {6{\tinyspace} (z{\negdbltinyspace}-{\negdbltinyspace}w)^2}
 \,+\,
 \frac{19{\tinyspace}C_3{\tinyspace}
       \partial^3 \widetilde{W}^{\prime\prime(4)}(w)}
     {30{\tinyspace} (z{\negdbltinyspace}-{\negdbltinyspace}w)}
\nonumber\\&&\!\phantom{%
\frac{10{\tinyspace}C_2}{19{\tinyspace}C_3}
\bigg(
}
 +\,
 \frac{72{\tinyspace}C_4{\tinyspace} \widetilde{W}^{(2)}(w)}
      {5{\tinyspace} (z{\negdbltinyspace}-{\negdbltinyspace}w)^6}
 \,+\,
 \frac{36{\tinyspace}C_4{\tinyspace} \partial \widetilde{W}^{(2)}(w)}
      {5{\tinyspace} (z{\negdbltinyspace}-{\negdbltinyspace}w)^5}
\nonumber\\&&\!\phantom{%
\frac{10{\tinyspace}C_2}{19{\tinyspace}C_3}
\bigg(
}
 +\,
 \frac{54{\tinyspace}C_4{\tinyspace} \partial^2 \widetilde{W}^{(2)}(w)}
      {25{\tinyspace} (z{\negdbltinyspace}-{\negdbltinyspace}w)^4}
 \,+\,
 \frac{12{\tinyspace}C_4{\tinyspace} \partial^3 \widetilde{W}^{(2)}(w)}
      {25{\tinyspace} (z{\negdbltinyspace}-{\negdbltinyspace}w)^3}
\nonumber\\&&\!\phantom{%
\frac{10{\tinyspace}C_2}{19{\tinyspace}C_3}
\bigg(
}
 +\,
 \frac{3{\tinyspace}C_4{\tinyspace} \partial^4 \widetilde{W}^{(2)}(w)}
      {35{\tinyspace} (z{\negdbltinyspace}-{\negdbltinyspace}w)^2}
 \,+\,
 \frac{9{\tinyspace}C_4{\tinyspace} \partial^5 \widetilde{W}^{(2)}(w)}
      {700{\tinyspace} (z{\negdbltinyspace}-{\negdbltinyspace}w)}
\nonumber\\&&\!\phantom{%
\frac{10{\tinyspace}C_2}{19{\tinyspace}C_3}
\bigg(
}
 +\,
 \frac{9 N {\tinyspace} C_4}
      {5{\tinyspace} (z{\negdbltinyspace}-{\negdbltinyspace}w)^8}
\bigg)
\,,
\\
\commutator{\widetilde{W}^{(2,2)}(z)}{\widetilde{W}^{(2,2)}(w)}
\!&=&\!
\frac{486{\tinyspace}C_2}{361{\tinyspace}C_3}
\bigg(
 \frac{6{\tinyspace} \widetilde{W}^{(2,4)}(w)}
      {(z{\negdbltinyspace}-{\negdbltinyspace}w)^2}
 \,+\,
 \frac{3{\tinyspace} \partial \widetilde{W}^{(2,4)}(w)}
      {z{\negdbltinyspace}-{\negdbltinyspace}w}
\nonumber\\&&\!\phantom{%
\frac{486{\tinyspace}C_2}{361{\tinyspace}C_3}
\bigg(
}
 +\,
 \frac{114{\tinyspace}C_3{\tinyspace} \widetilde{W}^{(2,2)}(w)}
      {5{\tinyspace} (z{\negdbltinyspace}-{\negdbltinyspace}w)^4}
 \,+\,
 \frac{57{\tinyspace}C_3{\tinyspace}
       \partial \widetilde{W}^{(2,2)}(w)}
      {5{\tinyspace} (z{\negdbltinyspace}-{\negdbltinyspace}w)^3}
\nonumber\\&&\!\phantom{%
\frac{486{\tinyspace}C_2}{361{\tinyspace}C_3}
\bigg(
}
 +\,
 \frac{19{\tinyspace}C_3{\tinyspace}
       \partial^2 \widetilde{W}^{(2,2)}(w)}
     {6{\tinyspace} (z{\negdbltinyspace}-{\negdbltinyspace}w)^2}
 \,+\,
 \frac{19{\tinyspace}C_3{\tinyspace}
       \partial^3 \widetilde{W}^{(2,2)}(w)}
     {30{\tinyspace} (z{\negdbltinyspace}-{\negdbltinyspace}w)}
\nonumber\\&&\!\phantom{%
\frac{486{\tinyspace}C_2}{361{\tinyspace}C_3}
\bigg(
}
 +\,
 \frac{72{\tinyspace}C_4{\tinyspace} \widetilde{W}^{(2)}(w)}
      {5{\tinyspace} (z{\negdbltinyspace}-{\negdbltinyspace}w)^6}
 \,+\,
 \frac{36{\tinyspace}C_4{\tinyspace} \partial \widetilde{W}^{(2)}(w)}
      {5{\tinyspace} (z{\negdbltinyspace}-{\negdbltinyspace}w)^5}
\nonumber\\&&\!\phantom{%
\frac{486{\tinyspace}C_2}{361{\tinyspace}C_3}
\bigg(
}
 +\,
 \frac{54{\tinyspace}C_4{\tinyspace} \partial^2 \widetilde{W}^{(2)}(w)}
      {25{\tinyspace} (z{\negdbltinyspace}-{\negdbltinyspace}w)^4}
 \,+\,
 \frac{12{\tinyspace}C_4{\tinyspace} \partial^3 \widetilde{W}^{(2)}(w)}
      {25{\tinyspace} (z{\negdbltinyspace}-{\negdbltinyspace}w)^3}
\nonumber\\&&\!\phantom{%
\frac{486{\tinyspace}C_2}{361{\tinyspace}C_3}
\bigg(
}
 +\,
 \frac{3{\tinyspace}C_4{\tinyspace} \partial^4 \widetilde{W}^{(2)}(w)}
      {35{\tinyspace} (z{\negdbltinyspace}-{\negdbltinyspace}w)^2}
 \,+\,
 \frac{9{\tinyspace}C_4{\tinyspace} \partial^5 \widetilde{W}^{(2)}(w)}
      {700{\tinyspace} (z{\negdbltinyspace}-{\negdbltinyspace}w)}
\nonumber\\&&\!\phantom{%
\frac{486{\tinyspace}C_2}{361{\tinyspace}C_3}
\bigg(
}
 +\,
 \frac{9 N {\tinyspace} C_4}
      {5{\tinyspace} (z{\negdbltinyspace}-{\negdbltinyspace}w)^8}
\bigg)
\,.
\end{eqnarray}
The operators
$\widetilde{W}^{\prime\prime(4)}(z)$ 
and 
$\widetilde{W}^{\prime(6)}(z)$ 
are not independent operators, 
are expressed by 
$\widetilde{W}^{(4)}(z)$, 
$\widetilde{W}^{(2,2)}(z)$, 
$\widetilde{W}^{(6)}(z)$ 
and new operators 
$\widetilde{W}^{(2,4)}(z)$, 
$\widetilde{W}^{(4,2)}(z)$ as
\begin{eqnarray}
\widetilde{W}^{\prime\prime(4)}(z)
\!&=&\!
\frac{2}{5}
\bigg(
  \frac{3 ( 4 N {\negdbltinyspace}+{\negdbltinyspace} 23)}{19}
  {\dbltinyspace} \widetilde{W}^{(4)}(z)
  -
  \frac{4 N {\negdbltinyspace}+{\negdbltinyspace} 13}{18}
  {\dbltinyspace} \widetilde{W}^{(2,2)}(z)
\bigg)
\,,
\\
\widetilde{W}^{\prime(6)}(z)
\!&=&\!
 \frac{27{\tinyspace}C_3}{5}
 {\dbltinyspace} \widetilde{W}^{(6)}(z)
 -
 \frac{9{\tinyspace}C_2}{10}
 {\dbltinyspace} \widetilde{W}^{(2,4)}(z)
\nonumber\\&&\!
-\,
\frac{C_4}{25} \bigg(
  18 {\tinyspace} \widetilde{W}^{(4,2)}(z)
  +
  6 {\tinyspace} \partial^2 \widetilde{W}^{(4)}(z)
  -
  \frac{19}{9} {\dbltinyspace} \partial^2 \widetilde{W}^{(2,2)}(z)
\bigg)
\,.
\end{eqnarray}

The above operators can be expressed by using the current $J^a(z)$. 
The operators 
$\widetilde{W}^{(2,2)}(z)$, 
are expressed by 
\begin{eqnarray}
\widetilde{W}^{(2,2)}(z)
\!&=&\!
\frac{54{\tinyspace}C_2}{19} {\dbltinyspace} W^{(4B)}(z)
\,,
\\
\widetilde{W}^{(3,2)}(z) \!&=&\!
\frac{36{\tinyspace}C_3}{5} {\dbltinyspace} W^{(5B)}(z)
\,,
\\
\widetilde{W}^{(3,3)}(z) \!&=&\!
3{\tinyspace}C_3 {\tinyspace} W^{\prime(6B)}(z)
\,,
\\
\widetilde{W}^{(6)}(z) \!&=&\!
W^{(6A)}
\,+\,
\frac{2 ( N {\negdbltinyspace}+{\negdbltinyspace} 4 )}{45}
 {\dbltinyspace} W^{(6B\alpha)}(z)
\,+\,
\frac{9 N \!+\! 22}{324 {\tinyspace} C_3}
 {\dbltinyspace} \widetilde{W}^{(2,4)}(z)
\nonumber\\
&&\!
   +\>
\frac{11N\!+\!34}{135}
\Big(
 W^{(6B\beta)}(z)
 \,+\,
 \frac{1}{3} {\dbltinyspace}
 \partial^2 W^{(4A)}(z)
\Big)
\,,
\\
\widetilde{W}^{(4,2)}(z) \!&=&\!
W^{(6B\alpha)}(z)
\,+\,
W^{(6B\beta)}(z)
\,,
\\
\widetilde{W}^{(2,4)}(z) \!&=&\!
\frac{3 {\tinyspace} C_4}{5}
\Big(
 W^{(6C)}(z)
 \,+\,
 \frac{2}{3}
 {\dbltinyspace} \partial^2 W^{(4B)}(z)
 \,-\,
 \frac{1}{140}
 {\dbltinyspace} \partial^4 \widetilde{W}^{(2)}(z)
\Big)
\,,
\\
\widetilde{W}^{(4,3)}(z) \!&=&\!
\frac{162 {\tinyspace} C_4}{15}
\Big(
 W^{\prime(7B)}(z)
 \,-\,
 \frac{1}{10}
 {\dbltinyspace} \partial \widetilde{W}^{(4,2)}(z)
\Big)
\,.
\end{eqnarray}

\begin{eqnarray}
W^{(4A)}(z) \!&=&\!
\frac{1}{12} \sum_{a,b}
 :\!
    J^a(z) J^a(z) J^b(z) J^b(z)
 \!:
\,,
\\
W^{(4B)}(z) \!&=&\!
\frac{1}{20} \sum_{a}
 :\!
    2{\tinyspace} \partial^2{\negdbltinyspace} J^a(z) J^a(z)
  - 3{\tinyspace} \partial J^a(z) \partial J^a(z)
 \!:
\,,
\\
W^{(5A)}(z) \!&=&\!
\frac{1}{15} \sum_{f,a,b,c} d_{abc}{\negdbltinyspace}
 :\!
    J^f(z) J^f(z) J^a(z) J^b(z) J^c(z)
 \!:
\,,
\\
W^{(5B)}(z) \!&=&\!
\frac{1}{21} \sum_{a,b,c} d_{abc}{\negdbltinyspace}
 :\!
    2{\tinyspace} \partial^2{\negdbltinyspace} J^a(z) J^b(z) J^c(z)
\nonumber\\&&\!\phantom{%
\frac{1}{21} \sum_{a,b,c} d_{abc}{\negdbltinyspace}
 :\!
}\!
  - 3{\tinyspace} \partial J^a(z) \partial J^b(z) J^c(z)
 \!:
\,,
\\
%
%
W^{\prime(6B)}(z) \!&=&\!
\frac{1}{10} \sum_{a,b,c} d_{abc}{\negdbltinyspace}
 :\!
    6{\tinyspace} \partial^2{\negdbltinyspace} J^a(z) \partial J^b(z) J^c(z)
\nonumber\\&&\!\phantom{%
\frac{1}{10} \sum_{a,b,c} d_{abc}{\negdbltinyspace}
 :\!
}\!
 - \partial^3{\negdbltinyspace} J^a(z) J^b(z) J^c(z)
\nonumber\\&&\!\phantom{%
\frac{1}{10} \sum_{a,b,c} d_{abc}{\negdbltinyspace}
 :\!
}\!
 - 6{\tinyspace} \partial J^a(z) \partial J^b(z) \partial J^c(z)
\!:
\,,
\\
W^{(6A)}(z) \!&=&\!
\frac{1}{54} \sum_{a,b,c}
 :\!
    J^a(z) J^a(z) J^b(z) J^b(z) J^c(z) J^c(z)
 \!:
\,,
\\
W^{(6B\alpha)}(z) \!&=&\!
\frac{1}{12} \sum_{a,b}
 :\!
    2{\tinyspace} J^a(z) J^a(z) \partial^2{\negdbltinyspace} J^b(z) J^b(z)
\nonumber\\&&\!\phantom{%
\frac{1}{12} \sum_{a,b}
 :\!
}\!
  - 3{\tinyspace} J^a(z) J^a(z) \partial J^b(z) \partial J^b(z)
 \!:
\,,
\\
W^{(6B\beta)}(z) \!&=&\!
-\,
\frac{1}{2} \sum_{a,b}
 :\!
    \partial J^a(z) J^a(z) \partial J^b(z) J^b(z)
 \!:
\,,
\\
W^{\prime(7B)}(z) \!&=&\!
-\,
\frac{1}{4} \sum_{a,b}
 :\!
    \partial J^a(z) \partial J^a(z) \partial J^b(z) J^b(z)
 \!:
\,,
\\
W^{(6C)}(z) \!&=&\!
-\,
\frac{1}{4} \sum_{a}
 :\!
    2{\tinyspace} \partial^3{\negdbltinyspace} J^a(z) \partial J^a(z)
  - 3{\tinyspace} \partial^2{\negdbltinyspace} J^a(z)
                  \partial^2{\negdbltinyspace} J^a(z)
 \!:
\,.
\end{eqnarray}

\end{document}